\RequirePackage{snapshot}
\documentclass[english,3p]{elsarticle}
\usepackage[T1]{fontenc}
\usepackage[utf8]{inputenc}
\usepackage{float}
\usepackage{amsmath}
\usepackage{amsthm}
\usepackage{graphicx}
\usepackage{subscript}

\makeatletter

\newcommand*\LyXZeroWidthSpace{\hspace{0pt}}
\providecommand{\tabularnewline}{\\}

\numberwithin{equation}{section}
\numberwithin{figure}{section}

\usepackage{algpseudocode,algorithm,algorithmicx}
\usepackage{tikz}
\usetikzlibrary{shapes,arrows,calc,decorations.pathreplacing}

\@ifundefined{showcaptionsetup}{}{%
 \PassOptionsToPackage{caption=false}{subfig}}
\usepackage{subfig}
\makeatother

\usepackage{babel}
\begin{document}

\title{Effect of ozone sensitization on the reflection patterns and stabilization
of standing detonation waves induced by curved ramps}

\author[NRL]{Eric J. Ching}
\author[NRL]{Ryan F. Johnson}

\address[NRL]{Laboratories for Computational Physics and Fluid Dynamics, U.S. Naval Research Laboratory, 4555 Overlook Ave SW, Washington, DC 20375}

\begin{abstract}
Standing detonation engines are a promising detonation-based propulsion
technology. The most commonly studied standing detonation configuration
involves a straight-sided wedge that induces an oblique detonation
wave. A recently introduced standing-detonation-engine concept entails
a curved ramp that leads to formation of a curved detonation wave.
The continuous compression or expansion induced by the ramp curvature
can have significant influence on the flow characteristics and wave
patterns of the detonation wave, offering greater flexibility in engine
design than conventional wedge geometries. This study aims to further
explore this relatively new standing-detonation-engine concept by
examining the effect of ignition promoters, namely ozone, on the flow
characteristics and reflection patterns of curved detonation waves
induced by convex or concave ramps inside a confined combustion chamber.
Simulations are performed using a positivity-preserving and entropy-bounded
discontinuous Galerkin method with curved elements to exactly represent
the ramp geometries. In the context of wedge-induced oblique detonation
waves, ozone addition has been found to decrease the initiation length
and lead to a smoother shock-detonation transition. This can then
attenuate the detonation and reduce stagnation-pressure losses, thus
improving the potential propulsion performance. In the context of
detonation waves induced by curved ramps, although ozone addition
similarly shortens the initiation zone, the curvature of the ramp
introduces additional effects that can amplify or counteract both
the ozone-induced contraction of the initiation zone and the aforementioned
detonation attenuation. For example, in the case of convex walls,
the reduced initiation length causes shock-detonation transition to
occur at a steeper ramp angle, where the curved leading shock is stronger.
As a result, the curved detonation wave can actually be strengthened,
which is detrimental to detonation stabilization (in the case of a
Mach reflection pattern) and stagnation-pressure recovery. However,
the initiation zone is shortened to a greater degree than in the case
of concave ramps since it no longer spans the shallower downstream
portion of the ramp. Conversely, in the case of concave walls, the
smaller initiation zone causes transition to occur at a shallower
ramp angle, where the leading shock is weaker. The attenuation of
the detonation can then be magnified, further reducing stagnation-pressure
losses and improving detonation stabilization (in the case of a Mach
reflection pattern), although the initiation length is decreased to
a smaller extent than in the case of convex ramps since the initiation
zone becomes restricted to shallower ramp angles. Finally, we present
specific examples wherein ozone addition changes the type of reflection
pattern (e.g., regular reflection, stationary Mach reflection, and
non-stationary Mach reflection).
\end{abstract}
\maketitle
\global\long\def\middlebar{\,\middle|\,}%
\global\long\def\average#1{\left\{  \!\!\left\{  #1\right\}  \!\!\right\}  }%
\global\long\def\expnumber#1#2{{#1}\mathrm{e}{#2}}%
 \newcommand*{\horzbar}{\rule[.5ex]{2.5ex}{0.5pt}}

\global\long\def\revisionmath#1{\textcolor{red}{#1}}%

\makeatletter \def\ps@pprintTitle{  \let\@oddhead\@empty  \let\@evenhead\@empty  \def\@oddfoot{\centerline{\thepage}}  \let\@evenfoot\@oddfoot} \makeatother

\let\svthefootnote\thefootnote\let\thefootnote\relax\footnotetext{\\ \hspace*{65pt}DISTRIBUTION STATEMENT A. Approved for public release:            distribution is unlimited.}\addtocounter{footnote}{0}\let\thefootnote\svthefootnote

\section{Introduction}

\label{sec:Introduction}

Detonation waves are supersonic combustion phenomena characterized
by a shock wave followed by a smooth, closely coupled combustion region.
The high temperature behind the shock ignites the mixture, which reacts
rapidly to the equilibrium state and leads to an energy release that
in turn drives the motion of the shock. Propulsion and power systems
based on detonation have emerged as a promising technology for hypersonic
flight, space travel, and reduced emissions due to higher thermodynamic
and combustion efficiencies and simpler geometric structure than traditional
deflagration-based systems~\citep{Kai00,Kai03}. A number of detonation-based
engine concepts exist, including rotating detonation engines~\citep{Kai11,Gup18},
pulse detonation engines~\citep{Kai03,Hei02}, and standing detonation
engines~\citep{Wol11,Ros21}. Standing detonation engines are appealing
due to their simple and economical design and their theoretical applicability
to a wide range of flight Mach numbers.

The most conventional standing-detonation-engine concept involves
high-speed flow past a fixed wedge, resulting in the formation of
an oblique detonation wave (ODW). For such wedge-induced ODWs, an
oblique shock wave (OSW) first forms at the leading edge, compressing
and heating the mixture before transitioning to an ODW. However, a
major challenge associated with this engine concept is reliable detonation
initiation and stabilization. In order to achieve a stable ODW, the
wedge angle must be large enough to obtain a complete detonation but
small enough to prevent detachment of the ODW~\citep{Pra91}. The
difficulty of sustaining a detonation is exacerbated in experiments
due to the limited run times of most facilities. In an important breakthrough
in 2021, Rosato et al.~\citep{Ros21} experimentally stabilized an
ODW at relevant conditions for approximately three seconds by using
a converging-diverging nozzle to sustain hypersonic flow at Mach 5.

The majority of studies in the literature that investigate ODW initiation
and stabilization or the underlying physics of ODWs rely on numerical
simulations. Early analyses of ODWs approximated the detonation wave
as an OSW coupled with instantaneous post-shock heat release~\citep{Pra91}.
However, it has since been determined that there exists a nonreactive
initiation zone between the OSW and ODW~\citep{Li94,Vig96}. There
are two types of OSW-to-ODW transition mechanisms: \emph{abrupt} transition
and \emph{smooth} transition~\citep{Wan11,Ten12,Wan15}, each associated
with a complex wave structure. In the former, which occurs at lower
Mach numbers, the leading OSW and ODW intersect at a multi-wave point.
In the latter, which is favorable for detonation stabilization, the
multi-wave point is replaced with a curved-shock region.  Hysteresis
associated with OSW-to-ODW transition has also been investigated~\citep{Liu18}.
Other studies have examined the cellular structure characterizing
the ODW front~\citep{Cho07,Ver13,Ten15}. In particular, the ODW
front comprises three zones: the first is characterized by ZND-type
detonation without any cellular structure, the second contains left-running
transverse waves, and the third consists of both left-running and
right-running transverse waves, giving rise to a cellular structure
similar to that observed in multi-headed normal detonations. The existence
of microscopic hypersonic jets (``micro-jets'') in the third zone
has also been observed~\citep{Ram24}.

The addition of ozone and other ignition promoters to the incoming
fuel-air mixture has recently been explored as a means to improve
the stabilization characteristics of ODWs. This is motivated by the
fact that ozone addition can accelerate ignition and reduce the induction
length in ZND calculations due to increased production of O radicals~\citep{Mag98}.
Additionally, experiments performed by Crane et al.~\citep{Cra19}
demonstrated that ozone doping can significantly reduce detonation
cell size, and the thermodynamic properties of the mixture (e.g.,
Chapman-Jouguet speed, post-shock state, and equilibrium state) are
largely unaffected. In the context of ODWs, Teng et al.~\citep{Ten24}
found that ozone addition can significantly decrease initiation length
and change abrupt ODW transitions to smooth transitions, improving
ODW stabilization characteristics and allowing for reduced combustor
sizes. The decrease in initiation length is more pronounced at lower
Mach numbers. The reason for the change in transition mechanism is
that the shorter initiation length suppresses the convergence of compression
waves into shocks, promoting smooth transitions. Analyses of total-pressure
losses revealed that ozonated ODWs exhibit improved propulsion performance.
Vashishtha et al.~\citep{Vas22} performed similar studies and reached
the same qualitative conclusions, though they also examined the effect
of H\textsubscript{2}O\textsubscript{2} (another ignition promoter)
addition and found that H\textsubscript{2}O\textsubscript{2} doping
is more effective than O\textsubscript{3} doping at higher Mach numbers
associated with smooth ODW transitions.

Apart from ODWs induced by straight-sided wedges, ODWs induced by
other geometries, such as cones~\citep{Ver12,Yan17,Han19}, truncated
cones~\citep{Zho23}, and double wedges~\citep{Hon22}, have been
investigated. Xiang et al.~\citep{Xia22} considered a wedge that
then transitions into a curved surface, and Xiong et al.~\citep{Xio23}
and Yan et al.~\citep{Yan24} examined curved detonation waves (CDWs)
induced by curved ramps\footnote{Although straight-sided wedges can give rise to detonations with a
curved shape~\citep{Yan24}, in this paper, a curved detonation wave
(CDW) refers specifically to a detonation induced by a curved wall.}. In~\citep{Yan24}, the stabilization characteristics and reflection
patterns of hydrogen-oxygen CDWs in two types of confined combustion
chambers were investigated: the first with a convex ramp and the second
with a concave ramp. Note that in computational studies of ODWs, chamber
confinement is often neglected (i.e., a semi-infinite wedge is assumed),
although a number of studies have considered a confined chamber~\citep{Wan20,Zha22}.
The effect of the degree of curvature was of particular interest.
A simple quantitative criterion for obtaining a stationary CDW was
also developed. Three broad types of reflection patterns were observed:
(a) non-stationary Mach reflection, which occurs if the curvature
is too high due to the presence of high-pressure, subsonic regions
that interact and merge; (b) stationary Mach reflection, which occurs
as the curvature is decreased; (c) stationary regular reflection,
which is the most favorable in terms of stability. The use of curved
ramps can significantly affect detonation initiation and stabilization.
For example, in the case of convex walls, the steepness of the leading
edge can reduce the initiation length compared to the case of a straight-sided
wedge with the same endpoints. In addition, the continuous expansion
caused by the convex surface can attenuate the detonation, potentially
decreasing the likelihood of a non-stationary Mach reflection and
reducing stagnation-pressure losses~\citep{Xio23}. On the other
hand, continuous compression induced by concave walls can facilitate
combustion. A ramp that changes in convexity can exploit both the
expansion and compression caused by convex and concave surfaces, respectively,
further improving detonation initiation and stabilization~\citep{Xio23,Yan24}
(though this concept has yet to be explored in detail). 

\begin{figure}[H]
\subfloat[\label{fig:convex-schematic}Convex wall.]{\includegraphics[width=0.96\columnwidth]{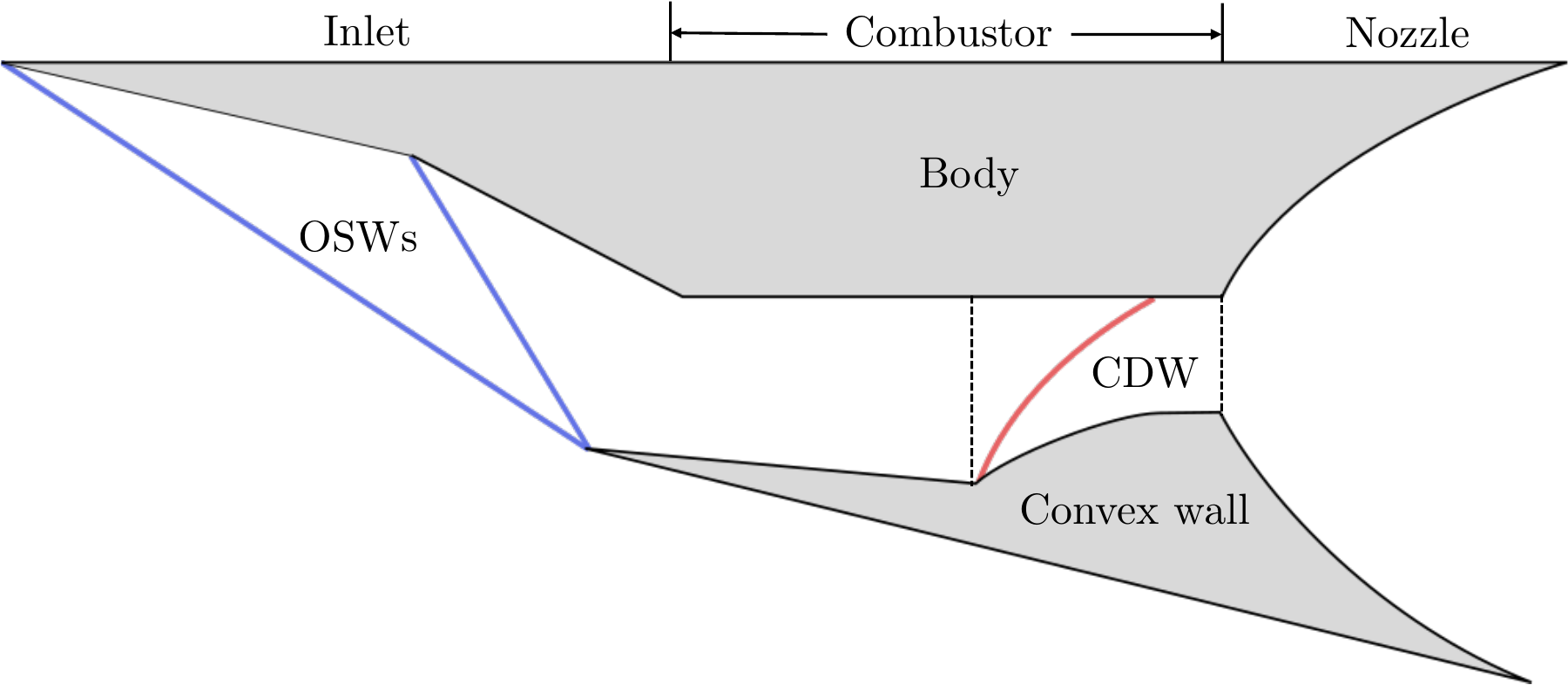}}\hfill{}\subfloat[\label{fig:concave-schematic}Concave wall.]{\includegraphics[width=0.96\columnwidth]{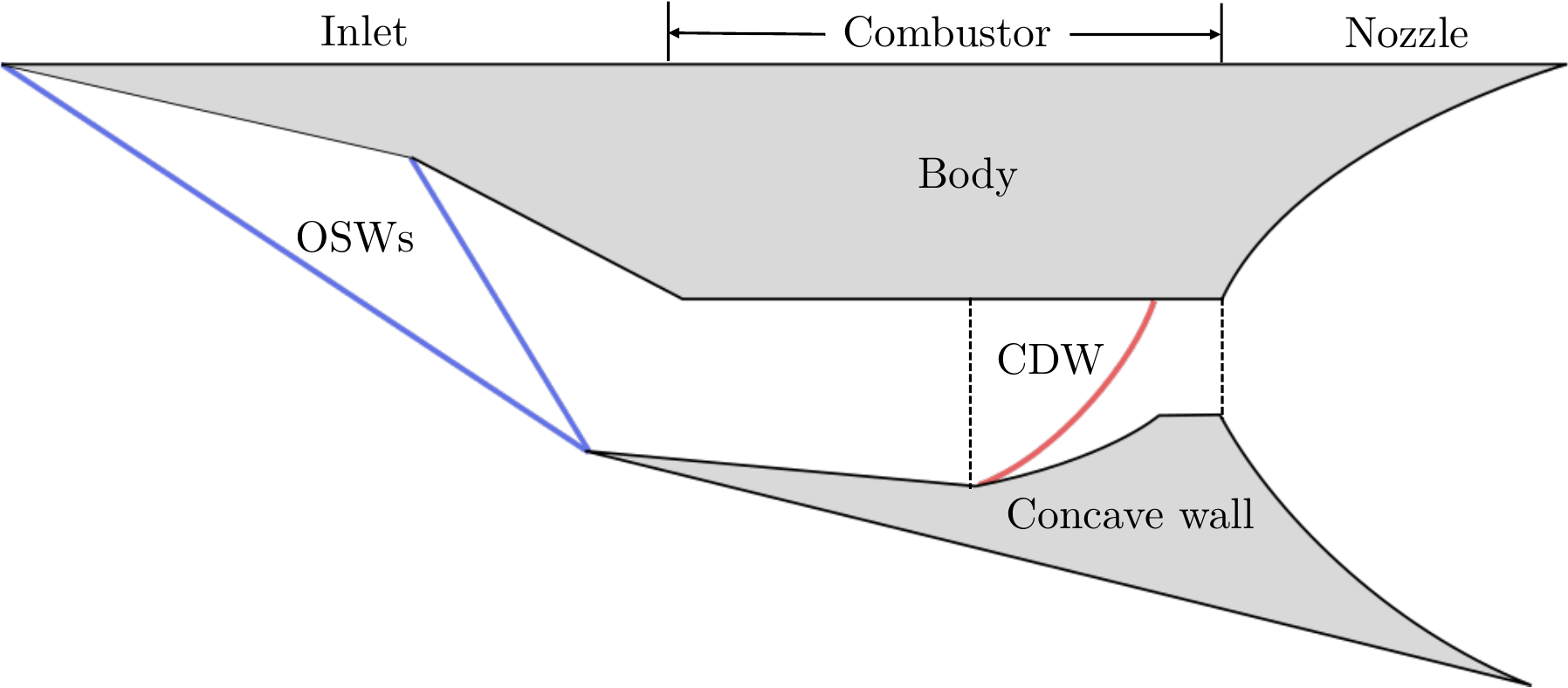}}

\caption{\label{fig:convex-concave-schematics}Representative schematic of
a standing detonation engine with a convex ramp and a concave ramp.
The region enclosed by the dashed lines represents the computational
domain. OSW: oblique shock wave. CDW: curved detonation wave.}
\end{figure}

Given the potential of both ozone doping and the use of curved walls,
this work aims to investigate the effect of ozone addition on the
reflection patterns and flow characteristics of standing hydrogen-oxygen
CDWs induced by curved ramps. In particular, we consider a similar
geometric configuration to that in~\citep{Yan24} (i.e., a confined
combustion chamber with either a convex ramp or a concave ramp), as
displayed in Figure~\ref{fig:convex-concave-schematics}. We examine
how ozone sensitization affects detonation initiation and the resulting
Mach-reflection and regular-reflection patterns at various curvatures.
The effect of ozone addition on stagnation-pressure losses is also
of interest. The findings of this study will improve understanding
of the effects of ignition promoters on this relatively new standing-detonation-engine
concept.

The remainder of this article is organized as follows. The governing
equations, associated physical models, and discretization techniques
are discussed in Section~\ref{sec:Physical-and-mathematical-modeling}.
The following section presents the results and corresponding discussion.
The paper then concludes with some final remarks.

\section{Physical and mathematical modeling}

\label{sec:Physical-and-mathematical-modeling}

\subsection{Governing equations and numerical methods}

The governing equations are the compressible, multicomponent, chemically
reacting Euler equations in two spatial dimensions. As in related
studies~\citep{Vas22,Ten24,Xio23,Yan24}, viscous effects are neglected.
These equations are written as
\begin{equation}
\frac{\partial y}{\partial t}+\nabla\cdot\mathcal{F}\left(y,\nabla y\right)-\mathcal{S}\left(y\right)=0,\label{eq:conservation-law-strong-form}
\end{equation}
where $y$ is the state vector, $t$ is time, $\mathcal{F}$ is the
convective flux, and $\mathcal{S}=\left(0,\ldots,0,0,\omega_{1},\ldots,\omega_{n_{s}}\right)^{T}$
is the chemical source term, with $\omega_{i}$ corresponding to the
production rate of the $i$th species. The physical coordinates are
denoted by $x=(x_{1},x_{2})$. The vector of state variables is expanded
as

\begin{equation}
y=\left(\rho v_{1},\rho v_{2},\rho e_{t},C_{1},\ldots,C_{n_{s}}\right)^{T},\label{eq:reacting-navier-stokes-state}
\end{equation}
where $\rho$ is density, $v=\left(v_{1},v_{2}\right)$ is the velocity
vector, $e_{t}$ is the specific total energy, $C=\left(C_{1},\ldots,C_{n_{s}}\right)$
is the vector of molar concentrations, and $n_{s}$ is the number
of species. The partial density of the $i$th species is defined as
\[
\rho_{i}=W_{i}C_{i},
\]
where $W_{i}$ is the molecular weight of the $i$th species, from
which the density can be computed as

\[
\rho=\sum_{i=1}^{n_{s}}\rho_{i}.
\]

\noindent The mole and mass fractions of the $i$th species are given
by
\[
X_{i}=\frac{C_{i}}{\sum_{i=1}^{n_{s}}C_{i}},\quad Y_{i}=\frac{\rho_{i}}{\rho}.
\]
The equation of state for the mixture is written as
\begin{equation}
P=R^{0}T\sum_{i=1}^{n_{s}}C_{i},\label{eq:EOS}
\end{equation}
where $P$ is the pressure, $T$ is the temperature, and $R^{0}$
is the universal gas constant. The specific total energy is the sum
of the mixture-averaged specific internal energy, $u$, and the specific
kinetic energy, written as

\[
e_{t}=u+\frac{1}{2}\sum_{k=1}^{d}v_{k}v_{k},
\]
where the former is the mass-weighted sum of the specific internal
energies of each species, given by
\[
u=\sum_{i=1}^{n_{s}}Y_{i}u_{i}.
\]
Assuming a thermally perfect gas, $u_{i}$ is defined as 
\[
u_{i}=h_{i}-R_{i}T=h_{\mathrm{ref},i}+\int_{T_{\mathrm{ref}}}^{T}c_{p,i}(\tau)d\tau-R_{i}T,
\]
where $h_{i}$ is the specific enthalpy of the $i$th species, $R_{i}=R^{0}/W_{i}$,
$T_{\mathrm{ref}}$ is the reference temperature of 298.15 K, $h_{\mathrm{ref},i}$
is the reference-state species formation enthalpy, and $c_{p,i}$
is the specific heat at constant pressure of the $i$th species, which
is approximated with a polynomial as a function of temperature based
on the NASA coefficients~\citep{Mcb93,Mcb02}. The $k$th spatial
component of the convective flux is written as
\begin{equation}
\mathcal{F}_{k}^{c}\left(y\right)=\left(\rho v_{k}v_{1}+P\delta_{k1},\ldots,\rho v_{k}v_{d}+P\delta_{kd},v_{k}\left(\rho e_{t}+P\right),v_{k}C_{1},\ldots,v_{k}C_{n_{s}}\right)^{T}.\label{eq:reacting-navier-stokes-spatial-convective-flux-component}
\end{equation}
We employ detailed chemical kinetics based on the H\textsubscript{2}
sub-model of the Foundational Fuel Chemistry Model Version 1.0 (FFCM-1)~\citep{FFCM1}
combined with the Princeton ozone sub-model~\citep{Zha16}.

The governing equations are spatially discretized using a positivity-preserving
and entropy-bounded discontinuous Galerkin method that can guarantee
nonnegative species concentrations, positive density, positive temperature,
and bounded specific entropy (from below)~\citep{Chi22,Chi22_2}.
Overintegration techniques to reduce spurious pressure oscillations
in smooth regions of the flow are applied~\citep{Joh20}. Strang
splitting is employed to deal with the stiff chemical source term,
where second-order strong-stability-preserving Runge-Kutta time integration~\citep{Got01,Ket08}
is used for the transport step and an implicit DG discretization in
time is used for the reaction step. All simulations in this work are
performed using a $p=1$ solution approximation, where $p$ denotes
the polynomial degree (with an order of accuracy of $p+1$ in smooth
regions), and a $p=2$ geometric approximation. To minimize numerical
artifacts at discontinuities, an artificial-viscosity term
\begin{equation}
\nabla\cdot\left(\nu_{\mathrm{AV}}\nabla y\right)\label{eq:artificial-viscosity-term}
\end{equation}
is added to the RHS of Equation~(\ref{eq:conservation-law-strong-form}),
where the artificial viscosity, $\nu_{\mathrm{AV}}$, is computed
as
\[
\nu_{\mathrm{AV}}=C_{\mathrm{AV}}S_{\mathrm{AV}}\left(\frac{h^{2}}{p+1}\right).
\]
$C_{\mathrm{AV}}$ is a user-defined parameter, $S_{\mathrm{AV}}$
is a shock sensor based on pressure variations inside a given element~\citep{Chi19},
and $h$ is a length scale associated with the element. $\nu_{\mathrm{AV}}$
is then made $C^{0}$-continuous as in~\citep{Per13}. The artificial-viscosity
term~(\ref{eq:artificial-viscosity-term}) is discretized using the
SIPG viscous flux function~\citep{Har08}.

\subsection{Computational setup}

\label{subsec:Computational-setup}

\begin{figure}[H]
\subfloat[\label{fig:concave-domain}Concave wall. $\mathrm{A}=(-5,100),\mathrm{B}=(0,100),\mathrm{D}=(210,62),\mathrm{E}=(210,0),\mathrm{F}=(125,0),\mathrm{G}=(125,15),\mathrm{H}=(60,70)$.]{\includegraphics[width=0.48\columnwidth]{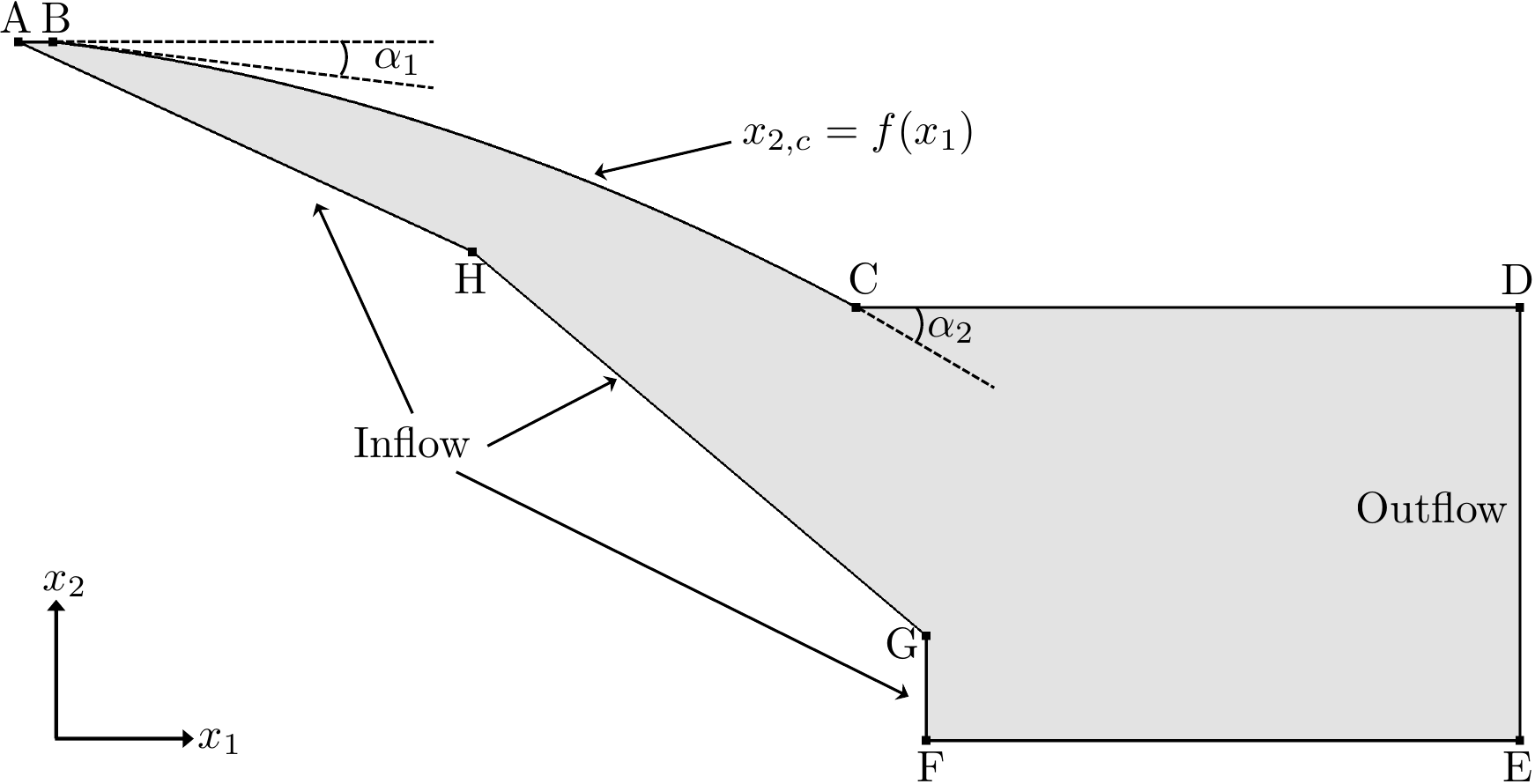}}\hfill{}\subfloat[\label{fig:convex-domain}Convex wall. $\mathrm{A}=(-5,100),\mathrm{B}=(0,100),\mathrm{D}=(210,62),\mathrm{E}=(210,0),\mathrm{F}=(120,0)$.]{\includegraphics[width=0.48\columnwidth]{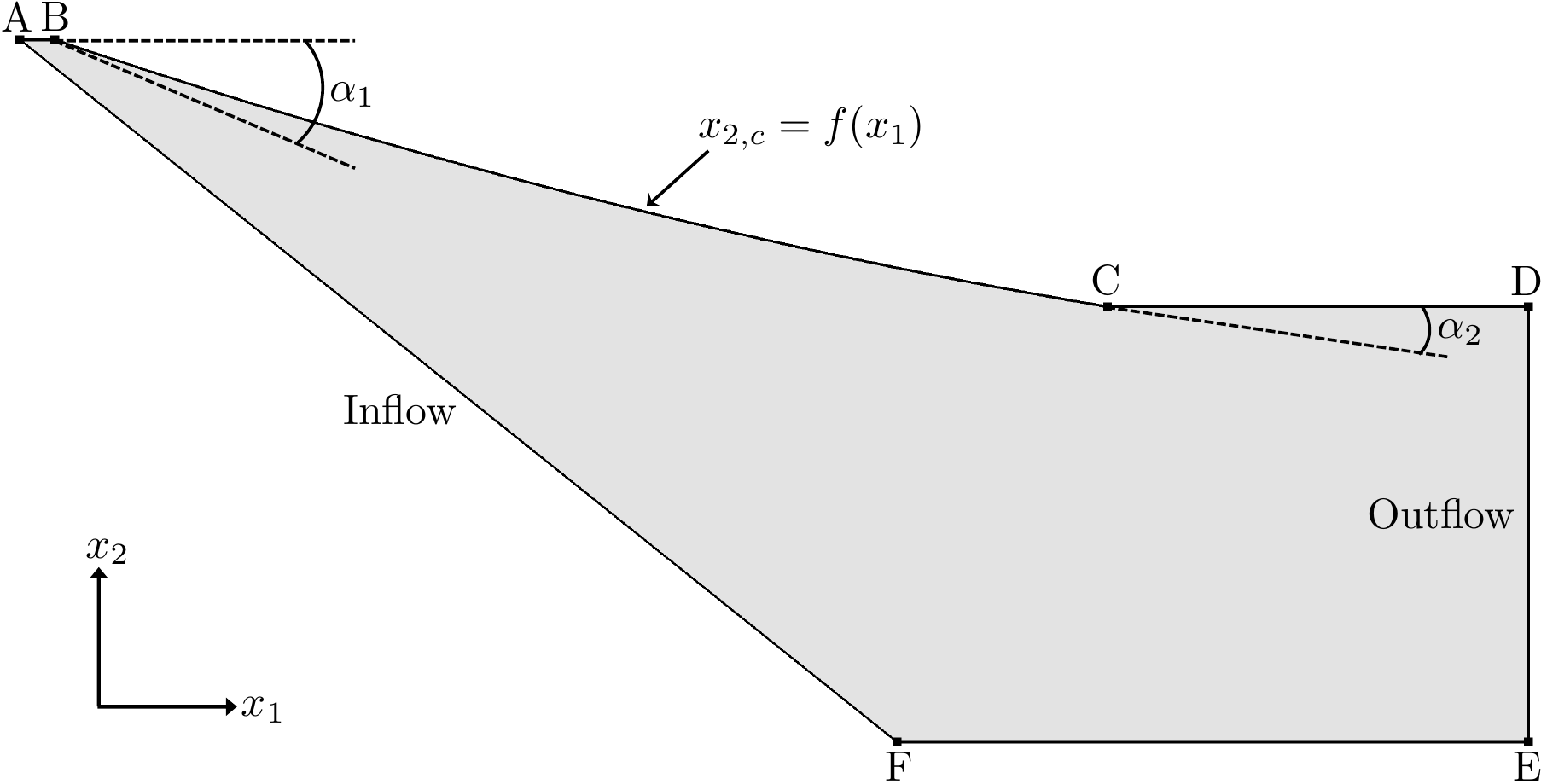}}

\caption{\label{fig:convex-concave-domains}Computational domains with (a)
concave wall and (b) convex wall. $\alpha_{1}$ and $\alpha_{2}$
are the initial and final angles, respectively, of the wall. Point
C is varied to obtain different mean curvatures. The point coordinates
are in units of mm. }
\end{figure}

Two types of computational domains, as shown in Figure~\ref{fig:convex-concave-domains}
are considered in this work. The overall dimensions of the combustion
chamber are similar to those in~\citep{Yan24}, which were based
on~\citep{Zha22} and~\citep{Wan20}, although a significant portion
of the freestream region is neglected here for computational savings.
For consistency with~\citep{Yan24}, the chamber is inverted in the
vertical direction, and $x_{2,c}$ is the ramp profile, which is a
quadratic function of $x_{1}$. Figure~\ref{fig:concave-domain}
displays a concave ramp, while Figure~\ref{fig:convex-domain} displays
a convex ramp. In Section~\ref{sec:Results-and-discussion}, Points
A and B are fixed, Point C is varied in the $x_{1}$-direction to
obtain different curvatures, and the remaining points are modified
to enlarge/reduce the domain as needed. $\alpha_{1}$ and $\alpha_{2}$
in Figure~\ref{fig:convex-concave-domains} are the initial and final
angles of the ramp, respectively. Following~\citep{Yan24}, the mean
curvature is defined as
\[
\kappa=\frac{\Delta\alpha}{s}=\frac{\alpha_{2}-\alpha_{1}}{s},
\]
where $s$ is the arc length of the curved ramp (i.e., Curve BC).
Note, however, that $\kappa$, while an important parameter, does
not by itself capture all relevant information about the ramp geometry.

As in~\citep{Yan24}, the state at the inflow boundaries in Figure~\ref{fig:convex-concave-domains}
is fully prescribed as $v_{1}=2495\text{ m/s}$, $P=1$ bar, $T=700\text{ K}$,
and a lean hydrogen-air mixture at an equivalence ratio of 0.34. These
conditions roughly correspond to Mach 9 flow at a temperature and
pressure of 200 K and 6 kPa, respectively, after compression by a
$25^{\circ}$ OSW. The inflow state is the initial condition as well.
Extrapolation is applied at the outflow boundary, and slip-wall conditions
are imposed on the remaining boundaries in Figure~\ref{fig:convex-concave-domains}.
Similar to~\citep{Yan24}, ozone is added to the mixture in the following
amounts: $0$, $1000$, and $10,000$ ppm (by mole). Gmsh~\citep{Geu09}
is used to generate unstructured triangular meshes with a characteristic
element size of $h=0.1$ mm. Information on grid convergence is provided
in~\ref{sec:grid-convergence}. Note that with a $p=1$ solution
approximation, each triangular element has three degrees of freedom
(per state variable), and with a $p=2$ geometric approximation, the
walls are exactly represented since $x_{2,c}$ is a quadratic function
of $x_{1}$.

\section{Results and discussion}

\label{sec:Results-and-discussion}

ZND calculations performed with the Shock Detonation Toolbox~\citep{sdtoolbox}
are first presented in order to qualitatively examine the effect of
ozone addition in the simplified one-dimensional setting. Results
for concave walls are then discussed, followed by those for convex
walls. All simulations in Sections~\ref{subsec:Concave-walls} and~\ref{subsec:Convex-walls}
are performed using a modified version of the JENRE\textregistered~Multiphysics
Framework~\citep{Joh20} that incorporates the extensions described
in~\citep{Chi22,Chi22_2} and in Section~\ref{sec:Physical-and-mathematical-modeling}.

\subsection{ZND calculations}

\label{subsec:ZND-calculations}

\begin{figure}[H]
\subfloat[\label{fig:ZND-T}Temperature.]{\includegraphics[width=0.32\columnwidth]{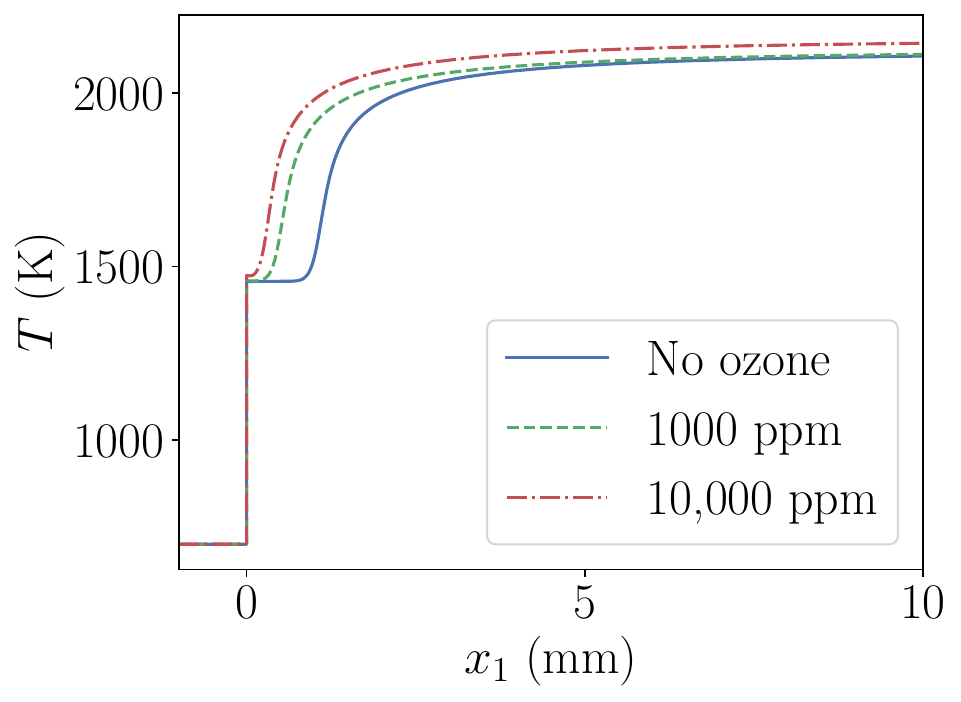}}\hfill{}\subfloat[\label{fig:ZND-P}Pressure.]{\includegraphics[width=0.32\columnwidth]{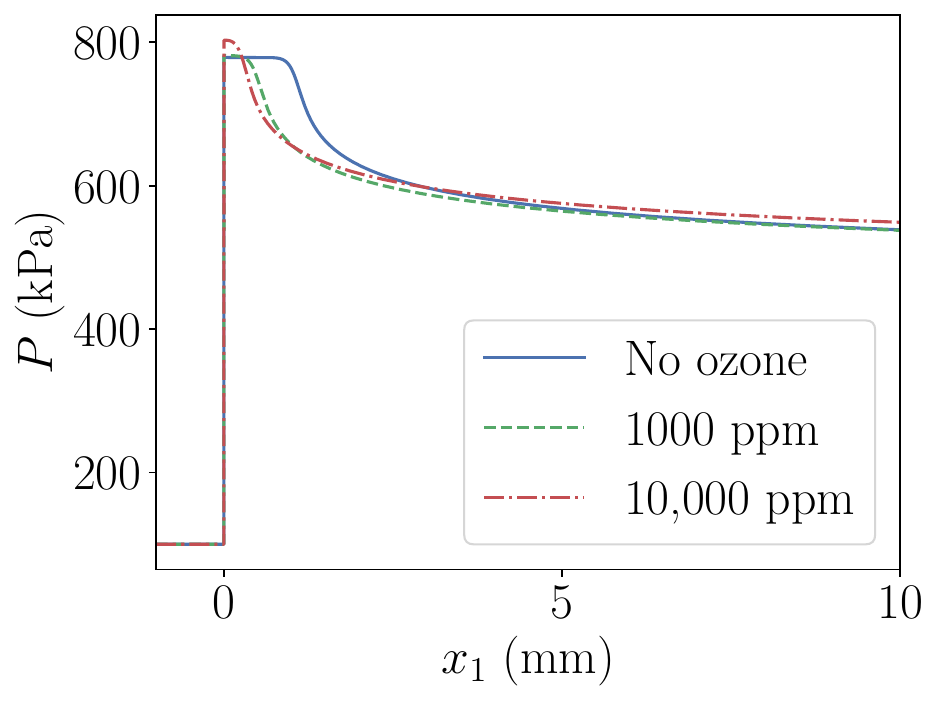}}\hfill{}\subfloat[\label{fig:ZND-SPR}Stagnation-pressure recovery (SPR).]{\includegraphics[width=0.32\columnwidth]{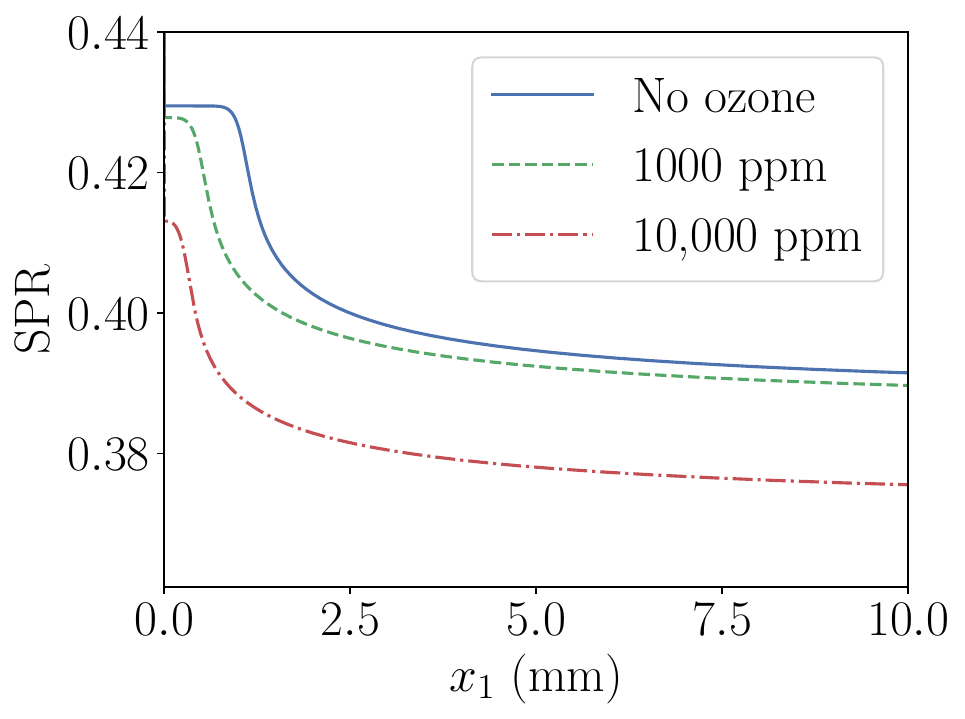}}

\caption{\label{fig:ZND}Temperature, pressure, and SPR profiles from ZND calculations
at Chapman-Jouguet conditions.}
\end{figure}

The initial, pre-shock state is set as the inflow state (with no ozone,
1000 ppm ozone, or 10,000 ppm ozone) described in Section~\ref{subsec:Computational-setup}.
Chapman-Jouguet conditions are assumed. Figure~\ref{fig:ZND} presents
the profiles of temperature, pressure, and stagnation-pressure recovery
(SPR), which is defined as the ratio between the (local) stagnation
pressure and the inflow stagnation pressure. This quantity is an approximate
measure of the propulsion performance~\citep{Xio23,Ten24}. The region
$x<0\;\mathrm{mm}$ corresponds to the initial, pre-shock state, and
the shock is located at $x=0\;\mathrm{mm}$. In Figure~\ref{fig:ZND-SPR},
the initial state, where SPR is simply unity, is not shown. Ozone
addition reduces the induction length and increases the temperature
and pressure behind the shock. The ozone-induced increase in temperature
behind the induction zone is more noticeable than the ozone-induced
increase in pressure. SPR is decreased as a result of ozone sensitization.
The overall changes in post-shock temperature, pressure, and SPR due
to ozone, however, are not significant. The difference between 1000
ppm ozone and 10,000 ppm ozone is greater than that between no ozone
and 1000 ppm.

\subsection{Concave walls}

\label{subsec:Concave-walls}

This section presents results for four concave cases, as listed in
Table~\ref{tab:concave-cases}. $\alpha_{1}$ is fixed across all
cases, and Point C in Figure~\ref{fig:concave-domain} is varied
in the $x_{1}$-direction to obtain different curvatures. As indicated
in the fifth column of Table~\ref{tab:concave-cases}, the letter
refers to the ozone amount (e.g., Case 1A corresponds to $\kappa=3.05\;\mathrm{km}^{-1}$
and no ozone). Stationary Mach reflections are observed in all cases
except Case 2C, which corresponds to a regular reflection, and Case
4A, which corresponds to a non-stationary Mach reflection. 

\begin{table}[h]
\begin{centering}
\caption{Geometric information for concave ramp. Units are in mm unless otherwise
specified.\label{tab:concave-cases}}
\par\end{centering}
\centering{}%
\begin{tabular}{cccccc}
\hline 
\noalign{\vskip\doublerulesep}
Case & Point C & Wall profile & $\kappa\;\left(\mathrm{km}^{-1}\right)$ & Ozone  & Reflection pattern\tabularnewline[\doublerulesep]
\hline 
\noalign{\vskip\doublerulesep}
\hline 
\noalign{\vskip\doublerulesep}
1 & $\left(115,62\right)$ & $x_{2,c}=-0.001805x_{1}^{2}-0.1229x_{1}+100$ & $3.05$ & %
\begin{tabular}{c}
1A: 0 ppm\tabularnewline
1B: 1000 ppm\tabularnewline
1C: 10,000 ppm\tabularnewline
\end{tabular} & %
\begin{tabular}{c}
1A: Mach reflection\tabularnewline
1B: Mach reflection\tabularnewline
1C: Mach reflection\tabularnewline
\end{tabular}\tabularnewline[\doublerulesep]
\noalign{\vskip\doublerulesep}
\hline 
\noalign{\vskip\doublerulesep}
2 & $\left(134,62\right)$ & $x_{2,c}=-0.001199x_{1}^{2}-0.1229x_{1}+100$ & $2.12$ & %
\begin{tabular}{c}
2A: 0 ppm\tabularnewline
2B: 1000 ppm\tabularnewline
2C: 10,000 ppm\tabularnewline
\end{tabular} & %
\begin{tabular}{c}
2A: Mach reflection\tabularnewline
2B: Mach reflection\tabularnewline
2C: Regular reflection\tabularnewline
\end{tabular}\tabularnewline[\doublerulesep]
\noalign{\vskip\doublerulesep}
\hline 
\noalign{\vskip\doublerulesep}
3 & $\left(90,62\right)$ & $x_{2,c}=-0.003326x_{1}^{2}-0.1229x_{1}+100$ & $5.09$ & %
\begin{tabular}{c}
3A: 0 ppm\tabularnewline
3B: 1000 ppm\tabularnewline
3C: 10,000 ppm\tabularnewline
\end{tabular} & %
\begin{tabular}{c}
3A: Mach reflection\tabularnewline
3B: Mach reflection\tabularnewline
3C: Mach reflection\tabularnewline
\end{tabular}\tabularnewline[\doublerulesep]
\noalign{\vskip\doublerulesep}
\hline 
\noalign{\vskip\doublerulesep}
4 & $\left(84,62\right)$ & $x_{2,c}=-0.003922x_{1}^{2}-0.1229x_{1}+100$ & $5.80$ & %
\begin{tabular}{c}
4A: 0 ppm\tabularnewline
4B: 1000 ppm\tabularnewline
4C: 10,000 ppm\tabularnewline
\end{tabular} & %
\begin{tabular}{c}
4A: Non-stationary\tabularnewline
4B: Mach reflection\tabularnewline
4C: Mach reflection\tabularnewline
\end{tabular}\tabularnewline[\doublerulesep]
\hline 
\noalign{\vskip\doublerulesep}
\end{tabular}
\end{table}

\subsubsection{Case 1}

\label{subsec:Concave-case-1}

\begin{figure}[H]
\subfloat[Case 1A.]{\includegraphics[width=0.32\columnwidth]{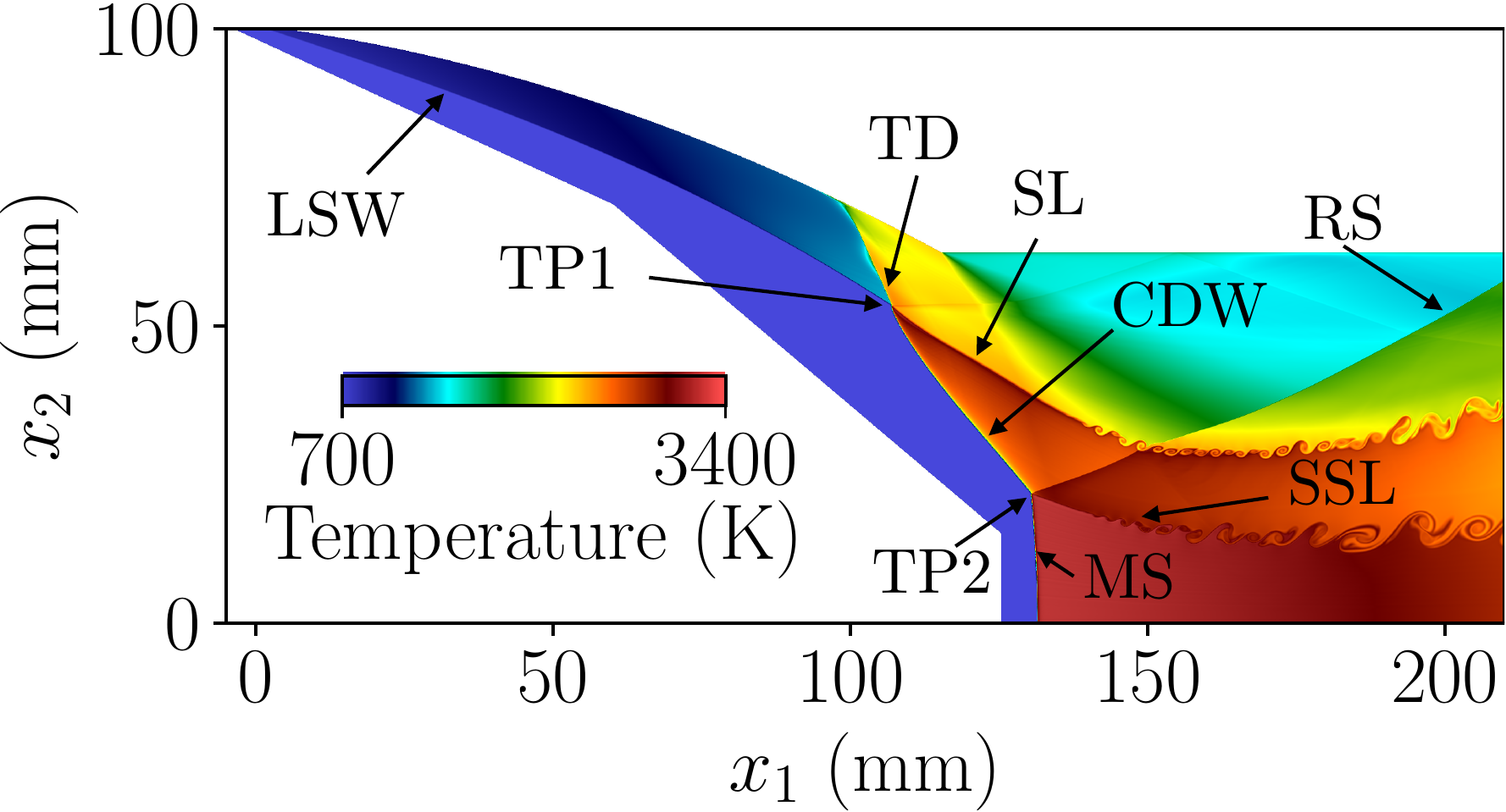}}\hfill{}\subfloat[Case 1B.]{\includegraphics[width=0.32\columnwidth]{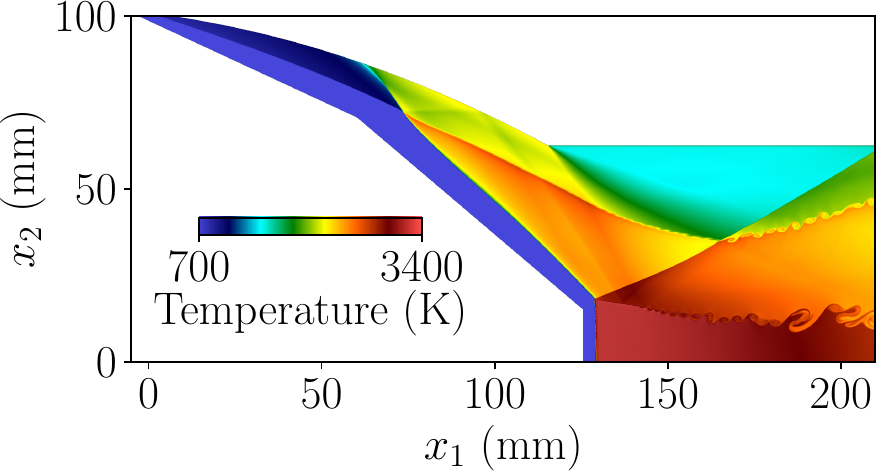}}\hfill{}\subfloat[Case 1C.]{\includegraphics[width=0.32\columnwidth]{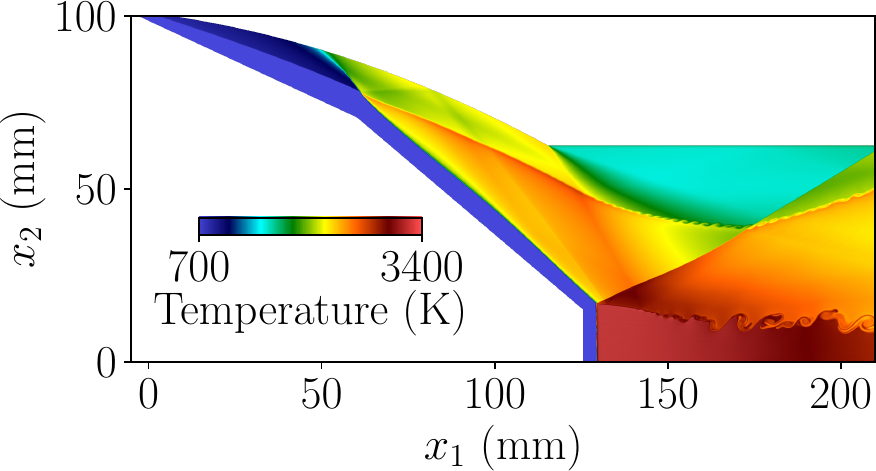}}

\caption{\label{fig:concave-case-1-temperature}Temperature fields for Cases
1A, 1B, and 1C. LSW: leading shock wave. TP1: transition point. TD:
transitional detonation. CDW: curved detonation wave. SL: slip line.
RS: reflected shock. MS: Mach stem. SSL: secondary slip line. TP2:
secondary triple point.}
\end{figure}

\begin{figure}[H]
\subfloat[Case 1A.]{\includegraphics[width=0.32\columnwidth]{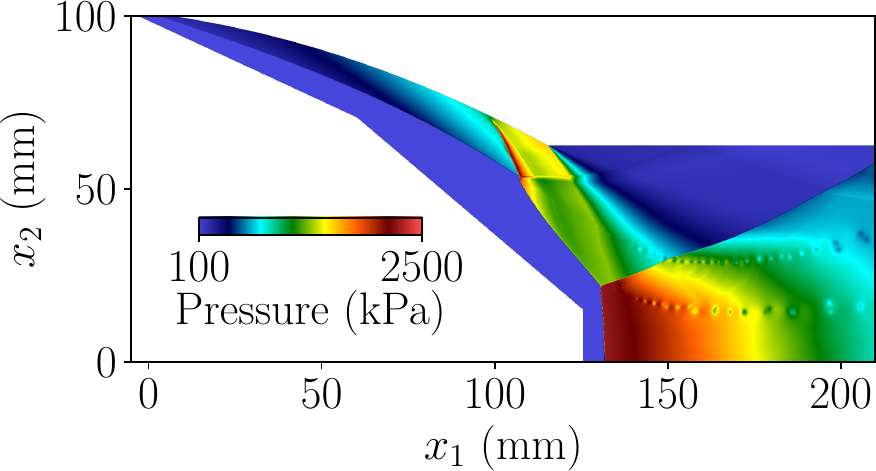}}\hfill{}\subfloat[Case 1B.]{\includegraphics[width=0.32\columnwidth]{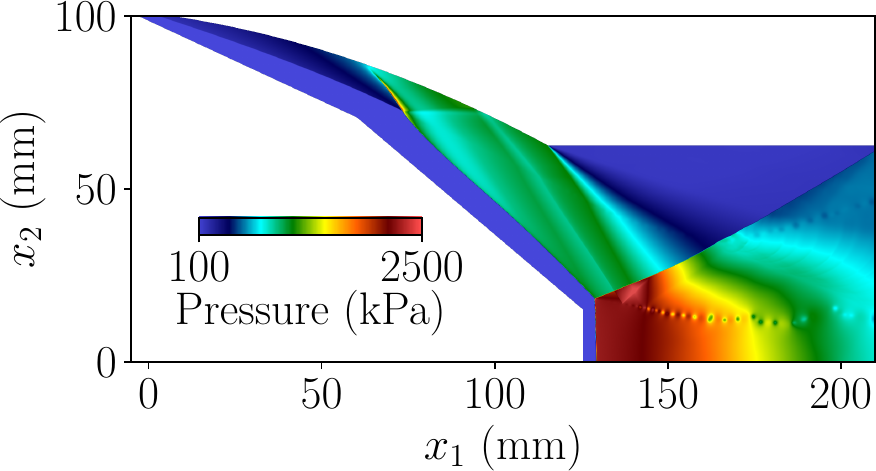}}\hfill{}\subfloat[Case 1C.]{\includegraphics[width=0.32\columnwidth]{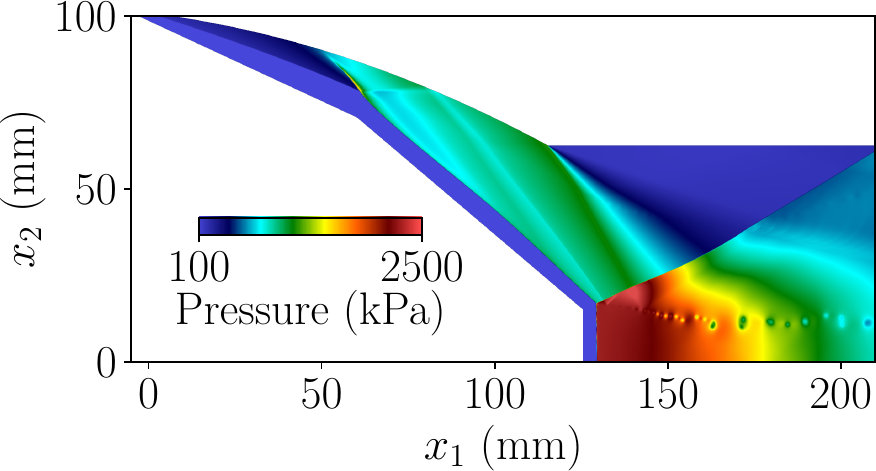}}

\caption{\label{fig:concave-case-1-pressure}Pressure fields for Cases 1A,
1B, and 1C.}
\end{figure}

Figures~\ref{fig:concave-case-1-temperature} and~\ref{fig:concave-case-1-pressure}
present the temperature and pressure fields, respectively, for Cases
1A, 1B, and 1C at $t=1.88$ ms, which is sufficient time for the CDW
to stabilize. Table~\ref{tab:concave-case-1-transition} lists the
$x_{1}$-coordinates of the points of transition from the leading
shock wave (LSW) to the CDW, as well as the percent difference with
respect to the ozone-free case. An abrupt transition occurs in all
cases, characterized by compression waves that converge into a transitional
detonation that then intersects the LSW, CDW, and primary slip line
at the transition point. The addition of ozone noticeably shortens
the initiation zone and causes the transition point to move upstream,
where the LSW is weaker as a result of the wall concavity. Consequently,
the angle of the detonation wave is shallower, leading to lower temperatures
and pressures behind the detonation and reflected shock. The height
of the Mach stem is then decreased. The transitional detonation and
associated transverse and reflected waves weaken as the ozone concentration
is increased. Note that even in the case of a straight-sided wedge,
initation-length reduction due to ozone addition can attenuate the
detonation by mitigating the coalesence of compression waves near
the reaction front~\citep{Ten24}, especially if the shock-detonation
transition changes from abrupt to smooth (not the case here); with
a concave ramp, the detonation is further attenuated by the weaker
LSW as the transition point moves upstream. The ozone-induced contraction
of the initiation zone, however, is somewhat offset by, as observed
in Figures~\ref{fig:concave-case-1-temperature} and~\ref{fig:concave-case-1-pressure},
the lower upstream temperatures and pressures ahead of the reaction
front (due to a shallower ramp angle), which normally have the competing
effect of induction-zone enlargement. This will be further confirmed
in Section~\ref{subsec:Convex-walls}, where convex ramps are considered.
In Case 1A (no ozone), a transverse wave noticeably interacts with
the expansion fan.  In Cases 1B and 1C, a transverse wave intersects
the reflected shock and reflects off the secondary slip line. The
difference between Cases 1A and 1B is greater than that between Cases
1B and 1C. 

\begin{figure}[H]
\subfloat[\label{fig:concave-case-1-line-temperature}Temperature along the
line $x_{2}=25$ mm.]{\includegraphics[width=0.32\columnwidth]{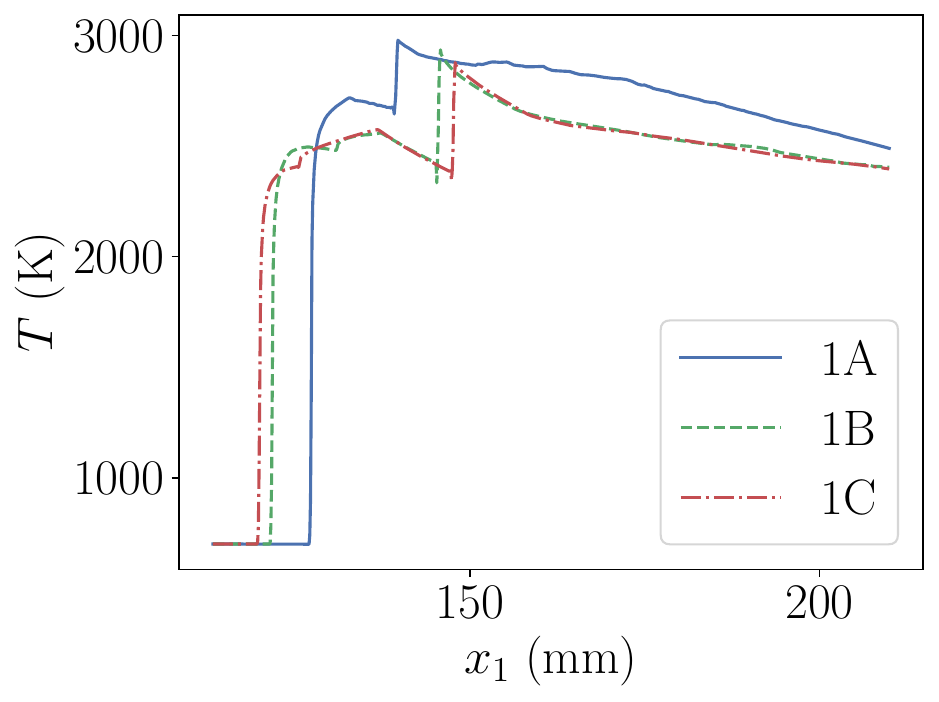}}\hfill{}\subfloat[\label{fig:concave-case-1-line-pressure}Pressure along the line $x_{2}=25$
mm.]{\includegraphics[width=0.32\columnwidth]{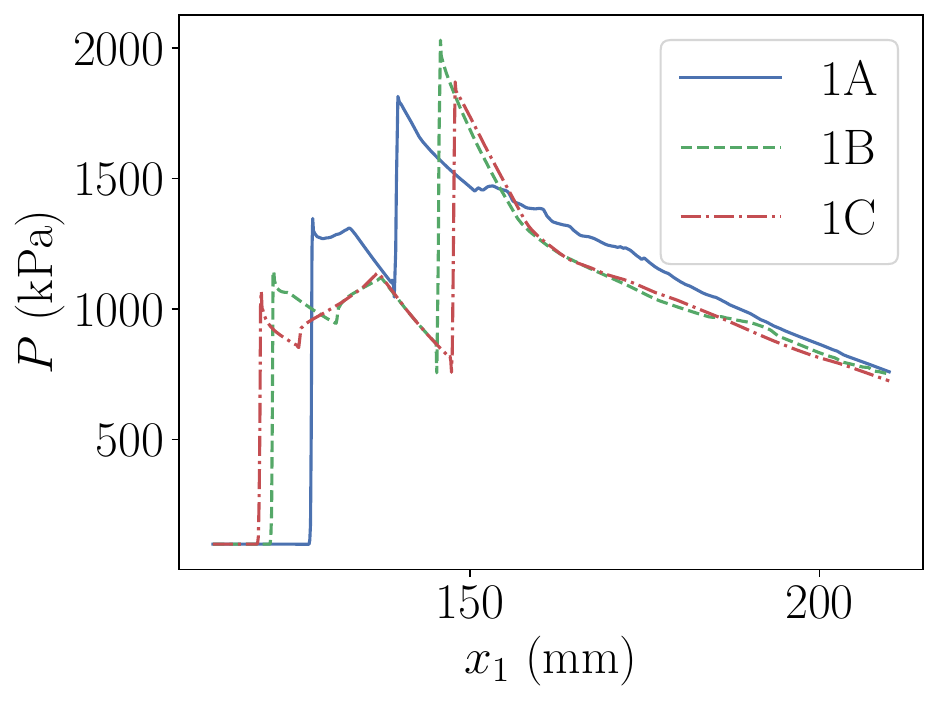}}\hfill{}\subfloat[\label{fig:concave-case-1-line-SPR}SPR along the line $x_{2}=25$
mm.]{\includegraphics[width=0.32\columnwidth]{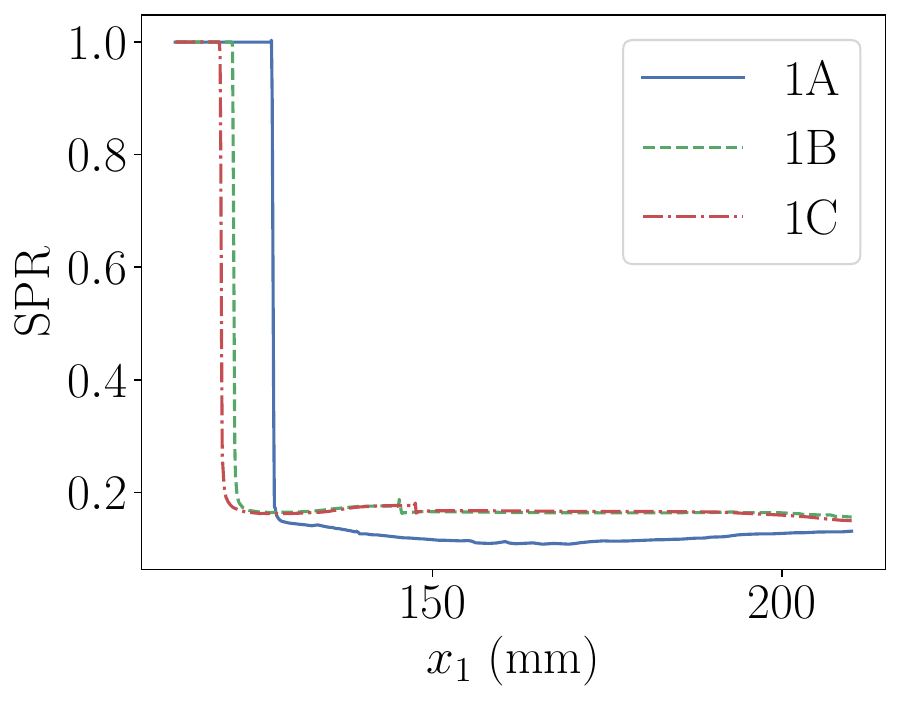}}

\caption{\label{fig:concave-case-1-lines}Variation of temperature, pressure,
and SPR along the line $x_{2}=25$ mm for Case 1 (concave).}
\end{figure}

\begin{table}[h]
\begin{centering}
\caption{$x_{1}$-coordinates of LSW-CDW transitions for Case 1 (concave).
\label{tab:concave-case-1-transition}}
\par\end{centering}
\centering{}%
\begin{tabular}{cccc}
\hline 
\noalign{\vskip\doublerulesep}
Case & Ozone addition & LSW-CDW transition & Relative change vs. Case 1A\tabularnewline[\doublerulesep]
\hline 
\noalign{\vskip\doublerulesep}
\hline 
\noalign{\vskip\doublerulesep}
1A & 0 ppm & $x_{1}\approx107$ mm & 0\%\tabularnewline[\doublerulesep]
\noalign{\vskip\doublerulesep}
\hline 
\noalign{\vskip\doublerulesep}
1B & 1000 ppm & $x_{1}\approx73$ mm & 32\%\tabularnewline[\doublerulesep]
\noalign{\vskip\doublerulesep}
\hline 
\noalign{\vskip\doublerulesep}
1C & 10,000 ppm & $x_{1}\approx61$ mm & 43\%\tabularnewline[\doublerulesep]
\hline 
\noalign{\vskip\doublerulesep}
\end{tabular}
\end{table}
Figures~\ref{fig:concave-case-1-line-temperature} and~\ref{fig:concave-case-1-line-pressure}
show the variation of temperature and pressure along the line $x_{2}=25$
mm, which intersects the CDW and reflected shock. The first large
temperature and pressure jump is due to the CDW, which is positioned
further upstream with ozone addition. Small variations in temperature
and pressure then occur as a result of transverse waves. The second
large jump in temperature and pressure is due to the reflected shock,
which is moved downstream with ozone addition. Expansion continues
to occur behind the reflected shock. The temperature increase across
the CDW is greatest in Case 1A, but the reflected shock is stronger
in the case of ozone sensitization. The combined effects of a shorter
initiation length and weaker shock at the point of transition, both
of which attenuate the CDW, lead to higher-Mach-number flow behind
the CDW and consequently a stronger reflected shock. Note that the
smaller temperature and pressure increase across the CDW with ozone
addition is in qualitative contrast with the ZND calculations discussed
in Section~\ref{subsec:ZND-calculations}, which is, as previously
mentioned, a consequence of smoothening of the shock-detonation transition
(even though it remains abrupt) and a weaker LSW as the transition
point moves upstream.

\begin{figure}[H]
\subfloat[Case 1A.]{\includegraphics[width=0.32\columnwidth]{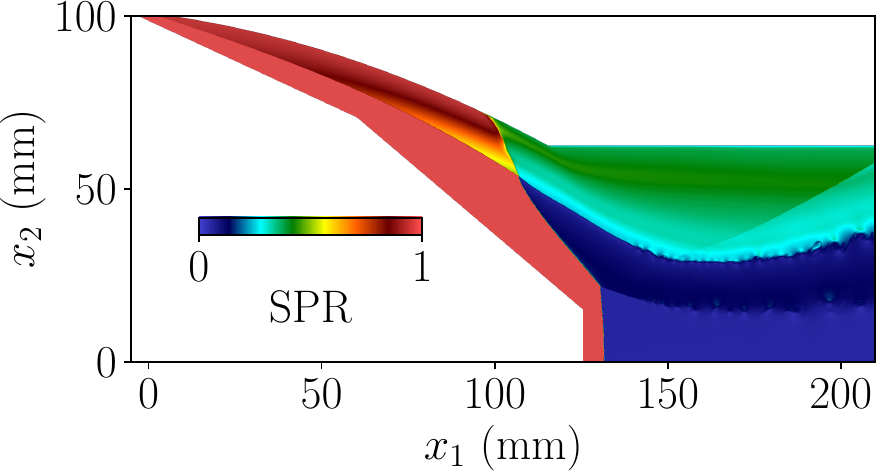}}\hfill{}\subfloat[Case 1B.]{\includegraphics[width=0.32\columnwidth]{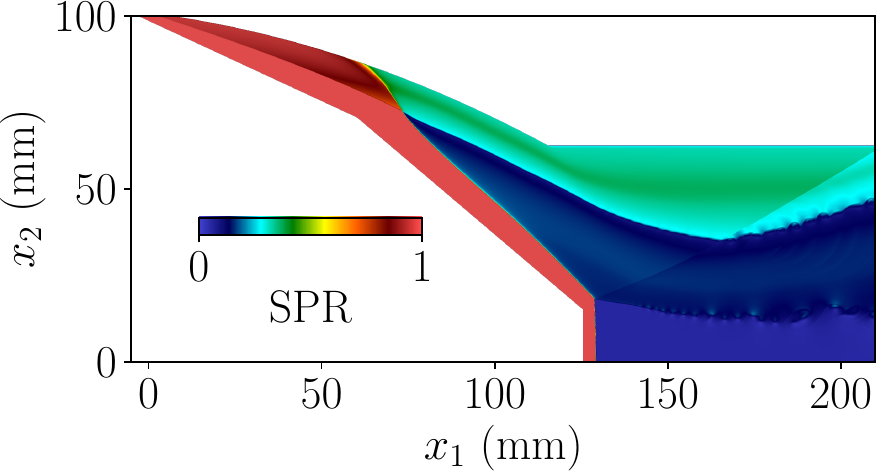}}\hfill{}\subfloat[Case 1C.]{\includegraphics[width=0.32\columnwidth]{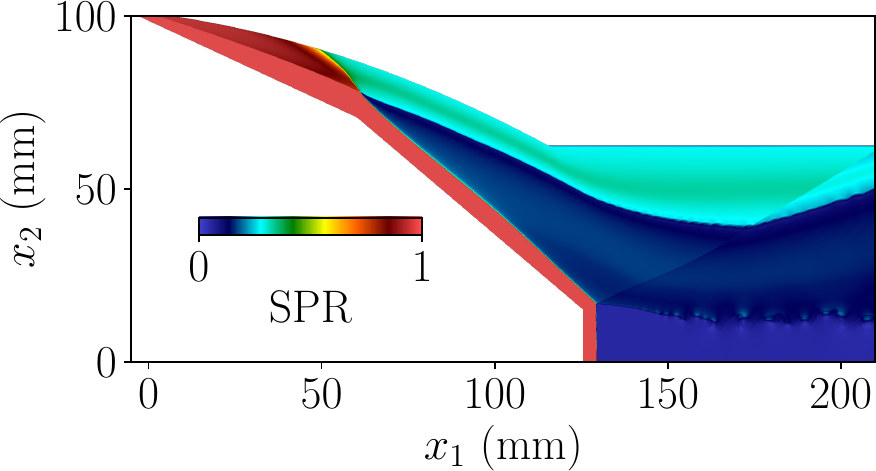}}

\caption{\label{fig:concave-case-1-TPR}SPR distributions for Cases 1A, 1B,
and 1C.}
\end{figure}

The SPR distributions for all cases are given in Figure~\ref{fig:concave-case-1-TPR}.
The stronger compression at the reaction front in Case 1A leads to
higher SPR above the slip line than in Cases 1B and 1C (note that
due to the vertically inverted combustion chamber, this region would
traditionally considered to be ``below'' the slip line). Behind
the reaction zone above the slip line, the stagnation pressure first
increases away from the wall and then decreases close to the slip
line. This initial increase in SPR is caused by coalescence of the
compression waves~\citep{Xio23}, which is then counteracted by weaker
SPR ahead of the transitional detonation caused by a steeper LSW angle.
The stagnation-pressure loss is greatest across the Mach stem. SPR
across the CDW improves as the ozone concentration increases, which
is further illustrated in the SPR variation along the line $x_{2}=25$
mm in Figure~\ref{fig:concave-case-1-line-SPR}. The SPR only slightly
decreases across the reflected shock. 

\paragraph{}

\subsubsection{Case 2}

\begin{figure}[H]
\subfloat[Case 2A.]{\includegraphics[width=0.32\columnwidth]{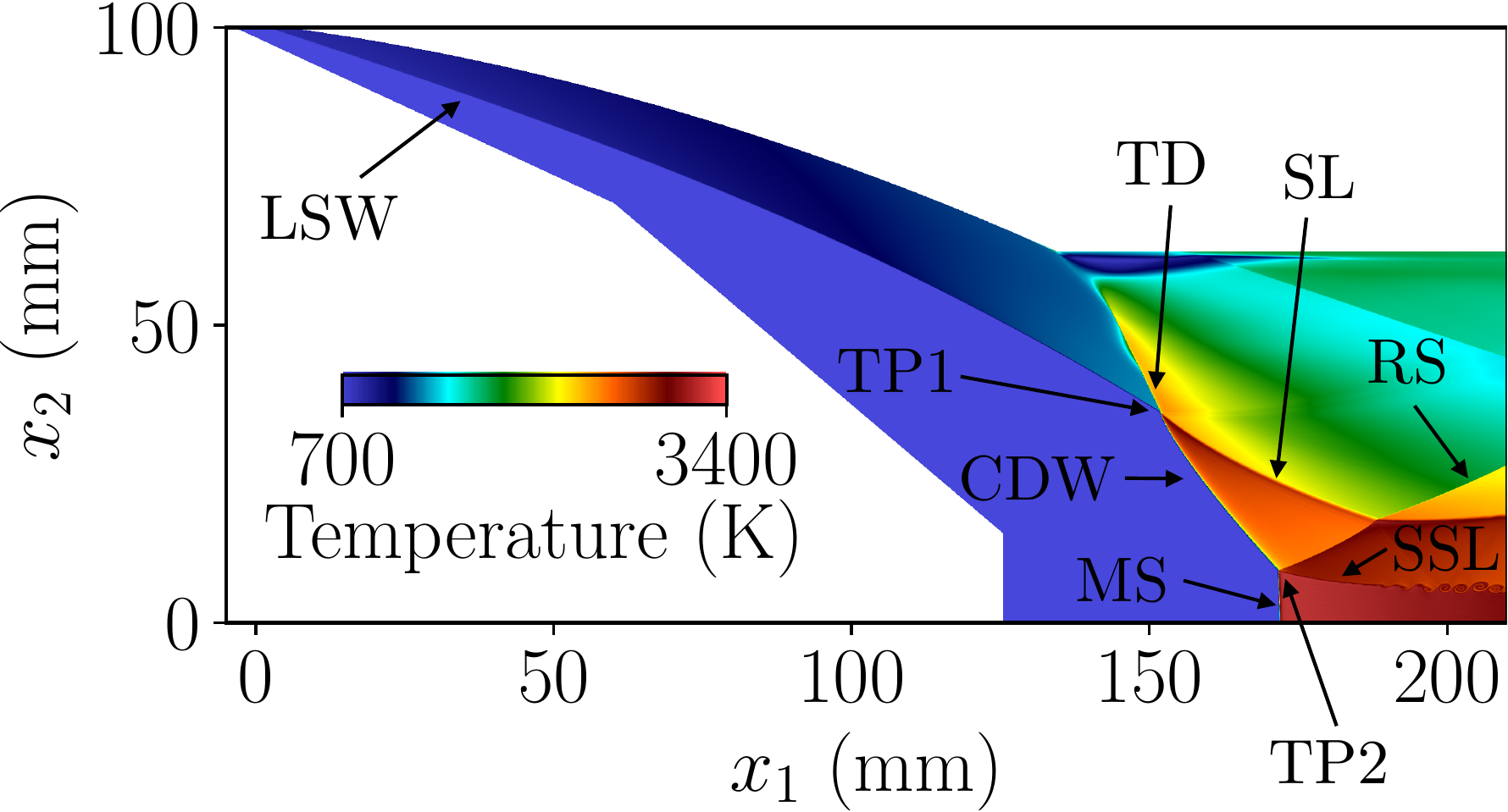}}\hfill{}\subfloat[Case 2B.]{\includegraphics[width=0.32\columnwidth]{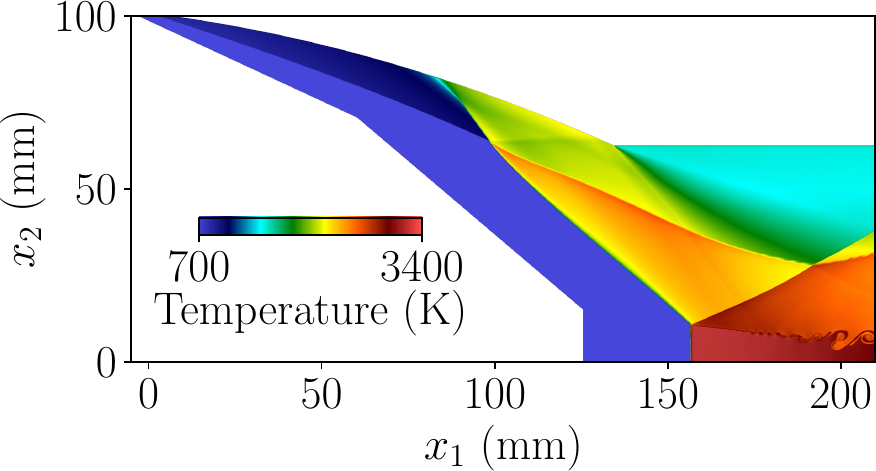}}\hfill{}\subfloat[Case 2C.]{\includegraphics[width=0.32\columnwidth]{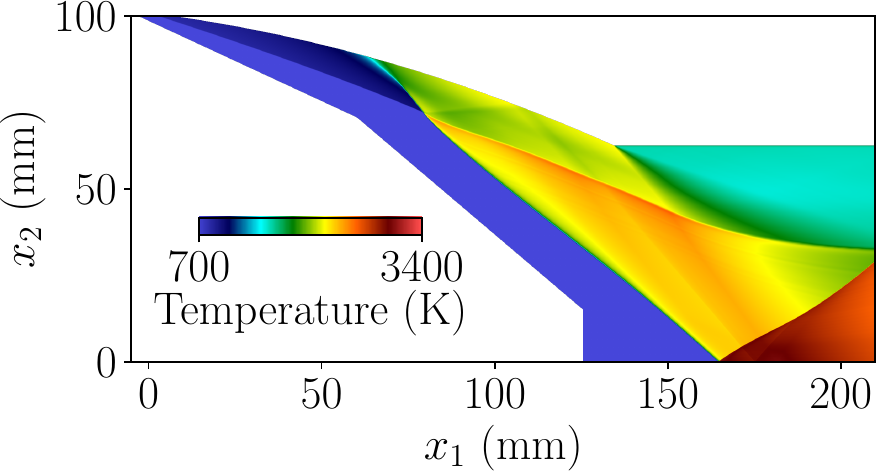}}

\caption{\label{fig:concave-case-2-temperature}Temperature fields for Cases
2A, 2B, and 2C. LSW: leading shock wave. TP1: transition point. TD:
transitional detonation. CDW: curved detonation wave. SL: slip line.
RS: reflected shock. MS: Mach stem. SSL: secondary slip line. TP2:
secondary triple point.}
\end{figure}

\begin{figure}[H]
\subfloat[Case 2A.]{\includegraphics[width=0.32\columnwidth]{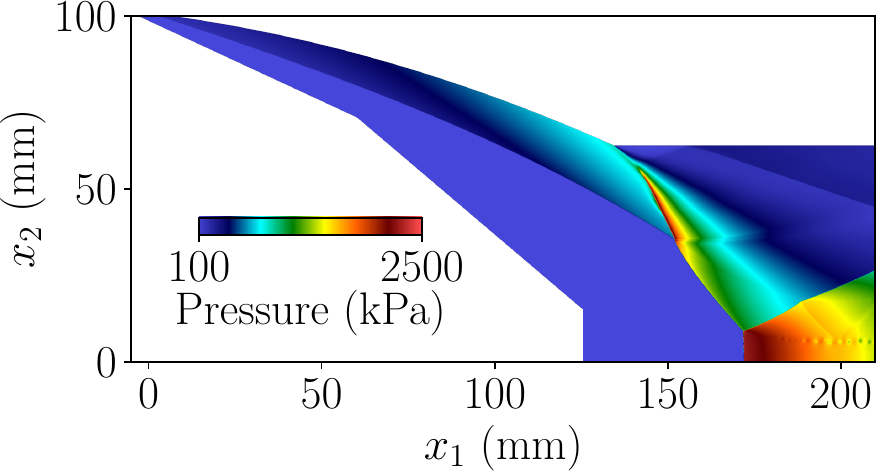}}\hfill{}\subfloat[Case 2B.]{\includegraphics[width=0.32\columnwidth]{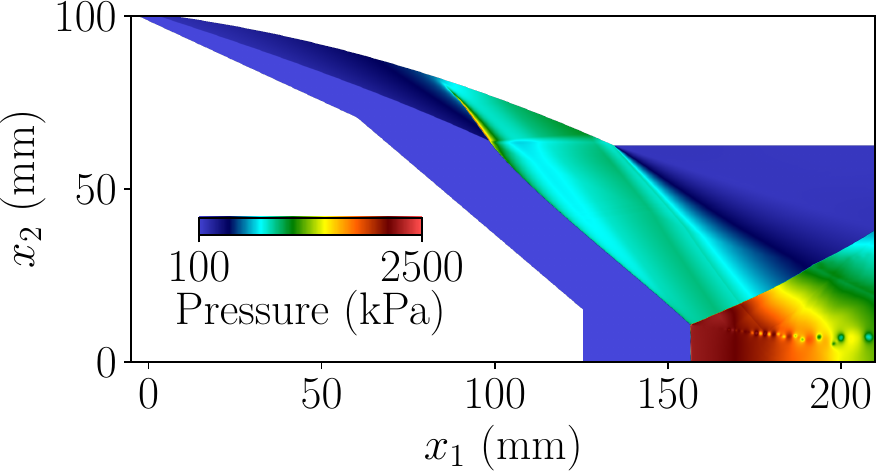}}\hfill{}\subfloat[Case 2C.]{\includegraphics[width=0.32\columnwidth]{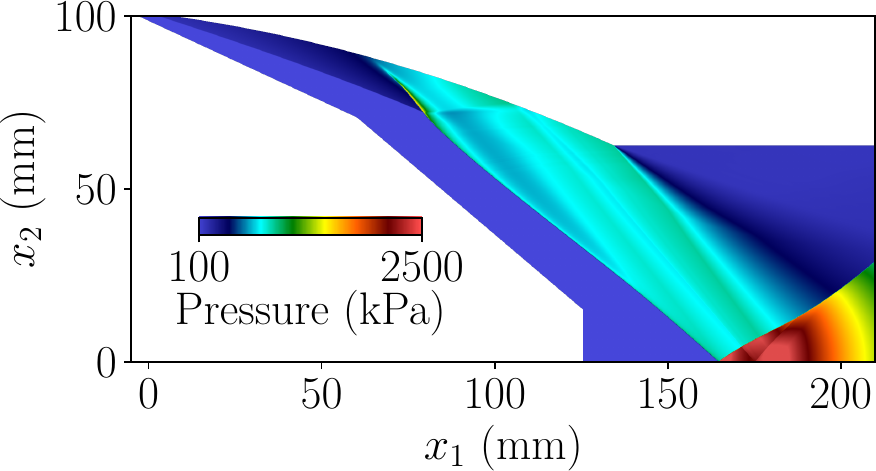}}

\caption{\label{fig:concave-case-2-pressure}Pressure fields for Cases 2A,
2B, and 2C.}
\end{figure}

Here, Point C is moved downstream, resulting in a reduced curvature
of $\kappa=2.12\;\mathrm{km}^{-1}$. The temperature and pressure
fields at $t=1.57$ ms, which is sufficiently long to reach quasi-steady
detonation, are displayed in Figures~\ref{fig:concave-case-2-temperature}
and~\ref{fig:concave-case-2-pressure}. The $x_{1}$-coordinates
of the LSW-CDW transitions for each case are given in Table~\ref{tab:concave-case-2-transition}.
A Mach reflection pattern is observed in Cases 2A and 2B, while a
regular reflection is observed in Case 2C (10,000 ppm of ozone). Abrupt
LSW-CDW transitions occur in all cases. The addition of ozone significantly
shortens the initiation zone and moves the transition point upstream,
even moreso than in Case 1, which corresponds to a larger curvature
and smaller initiation zone. Without ozone, the convergence of compression
waves into a transitional detonation takes place near the tail of
the ramp. As a result, transverse waves do not reflect off the curved
ramp, as they do in Cases 2B and 2C. Additionally, in Case 2A, the
expansion fan mitigates the coalescence of compression waves, leading
to a small low-temperature, nonreactive region immediately behind
the expansion corner. The transitional detonation is also stronger
in Case 2A. The region behind the CDW bounded by the slip line and
reflected shock is significantly smaller in the absence of ozone.
In Case 2C, the slip line and reflected shock do not intersect before
the end of the domain. Since the transition point in Case 2A is located
near the tail of the ramp, the post-CDW flow directly interacts with
the expansion fan stemming from the convex corner, almost immediately
decreasing in pressure and turning towards the $x_{1}$-direction.
In contrast, since the transition points in Cases 2B and 2C are located
further upstream, the flow behind the reaction front and CDW first
undergoes continuous compression due to the ramp concavity before
reaching the expansion fan. Furthermore, just as in Case 1, with ozone
addition, the smaller initiation zone spans shallower ramp angles
and thus reduced compression, attenuating the CDW and mitigating the
ozone-induced decrease in initiation length. The relative change in
the transition locations is greater here than in the higher-curvature
Case 1.

\begin{figure}[H]
\subfloat[\label{fig:concave-case-2-line-temperature}Temperature along the
line $x_{2}=15$ mm.]{\includegraphics[width=0.32\columnwidth]{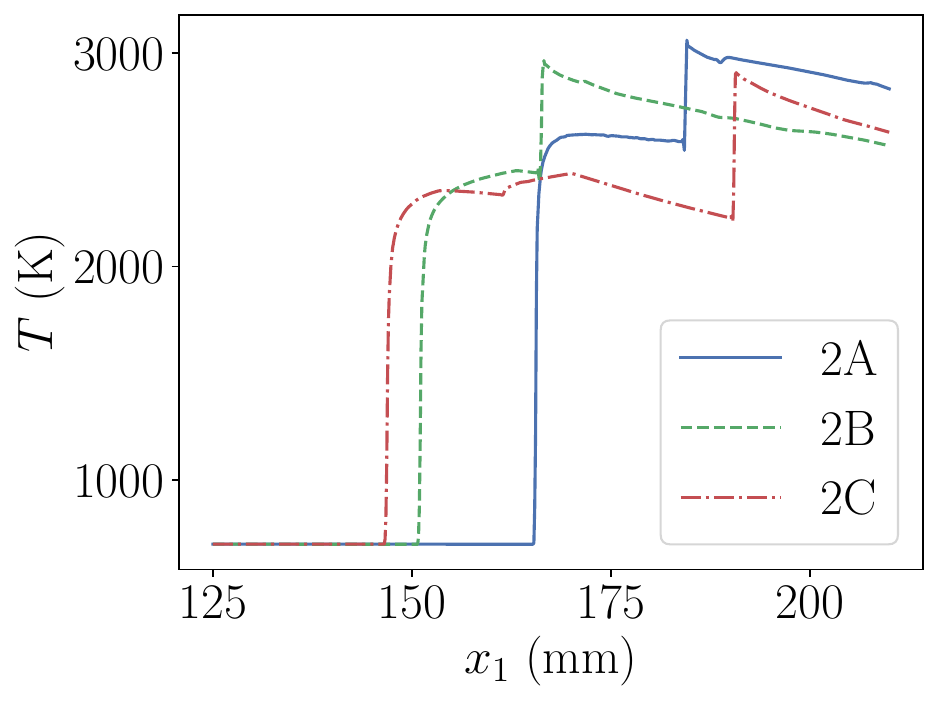}}\hfill{}\subfloat[\label{fig:concave-case-2-line-pressure}Pressure along the line $x_{2}=15$
mm.]{\includegraphics[width=0.32\columnwidth]{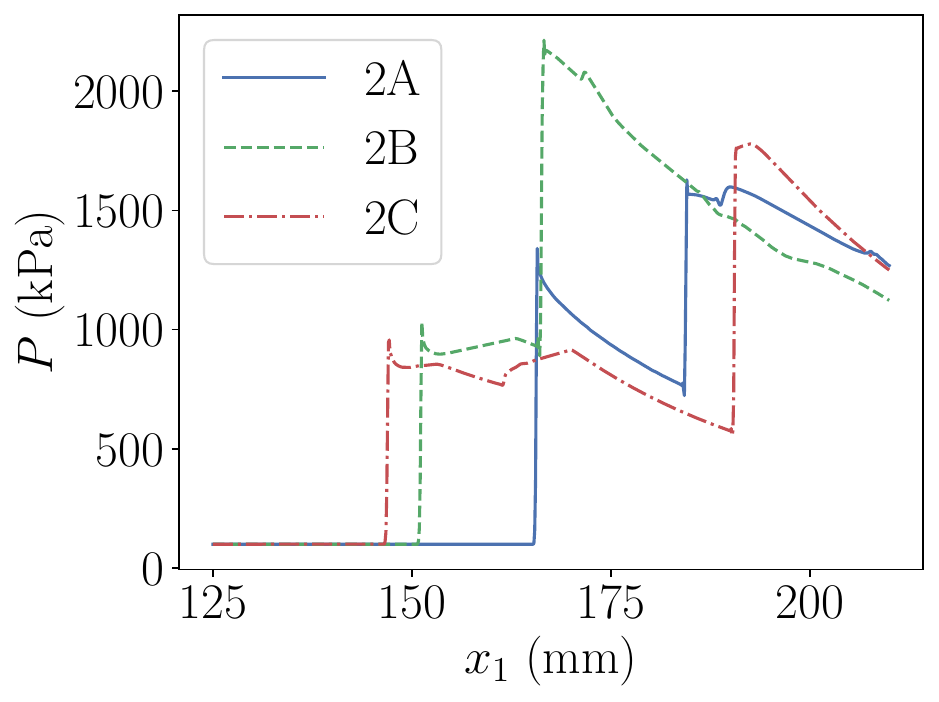}}\hfill{}\subfloat[\label{fig:concave-case-2-line-SPR}SPR along the line $x_{2}=15$
mm.]{\includegraphics[width=0.32\columnwidth]{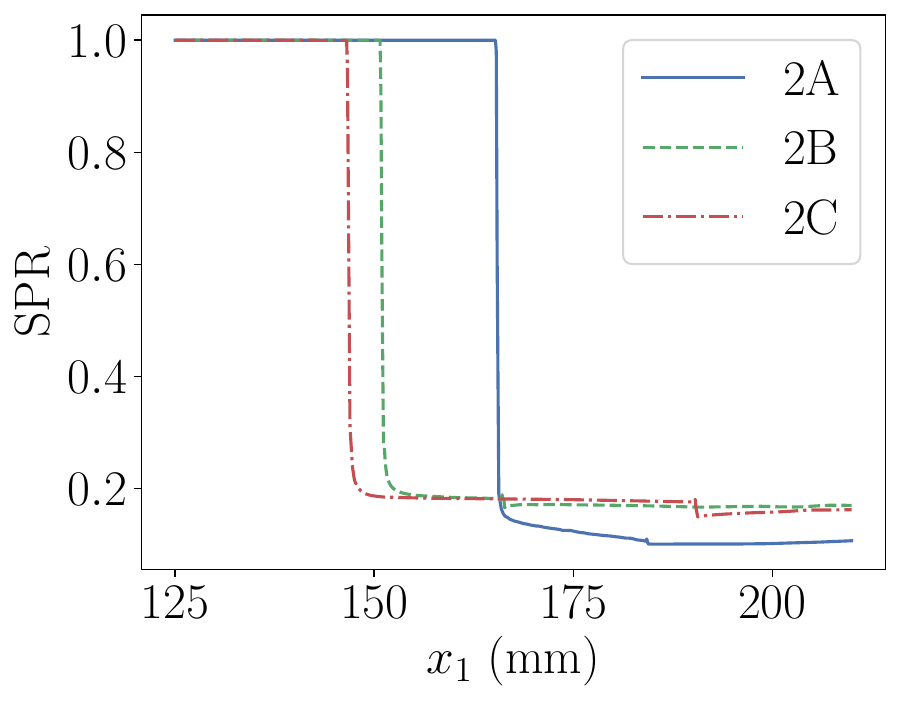}}

\caption{\label{fig:concave-case-2-lines}Variation of temperature, pressure,
and SPR along the line $x_{2}=15$ mm for Case 2 (concave).}
\end{figure}

\begin{table}[h]
\begin{centering}
\caption{$x_{1}$-coordinates of LSW-CDW transitions for Case 2 (concave).
\label{tab:concave-case-2-transition}}
\par\end{centering}
\centering{}%
\begin{tabular}{cccc}
\hline 
\noalign{\vskip\doublerulesep}
Case & Ozone addition & LSW-CDW transition & Relative change vs. Case 2A\tabularnewline[\doublerulesep]
\hline 
\noalign{\vskip\doublerulesep}
\hline 
\noalign{\vskip\doublerulesep}
2A & 0 ppm & $x_{1}\approx152$ mm & 0\%\tabularnewline[\doublerulesep]
\noalign{\vskip\doublerulesep}
\hline 
\noalign{\vskip\doublerulesep}
2B & 1000 ppm & $x_{1}\approx98$ mm & 36\%\tabularnewline[\doublerulesep]
\noalign{\vskip\doublerulesep}
\hline 
\noalign{\vskip\doublerulesep}
2C & 10,000 ppm & $x_{1}\approx79$ mm & 48\%\tabularnewline[\doublerulesep]
\hline 
\noalign{\vskip\doublerulesep}
\end{tabular}
\end{table}
Figures~\ref{fig:concave-case-2-line-temperature} and~\ref{fig:concave-case-2-line-pressure}
show the temperature and pressure along the line $x_{2}=15$ mm, which
intersects the CDW and reflected shock. The temperature and pressure
jumps across the CDW are lowest in Case 2C. This is again due to the
smaller initiation zone, which smoothens the LSW-CDW transition (even
though it is still abrupt), combined with the reduced ramp angle and
therefore weaker LSW as the transition is moved upstream. This weaker
CDW in Case 2C leads to faster, less deflected flow behind the CDW,
resulting in a stronger reflected shock, as indicated by the greater
temperature and pressure jumps near the outflow boundary, and suppressing
the formation of a Mach stem. The CDW is positioned furthest upstream
in Case 2C, and the lack of a Mach stem causes the reflected shock
to intersect the line $x_{2}=15$ mm further downstream than Cases
2A and 2B. In all cases, significant flow expansion is observed behind
the reflected shock, and the small variations in pressure are caused
by transverse and reflected waves. Overall, the influence of ozone
sensitization is qualitatively similar to that in the higher-curvature
Case 1, although magnified here due to the very large initiation zone
in the absence of ozone.

\begin{figure}[H]
\subfloat[Case 2A.]{\includegraphics[width=0.32\columnwidth]{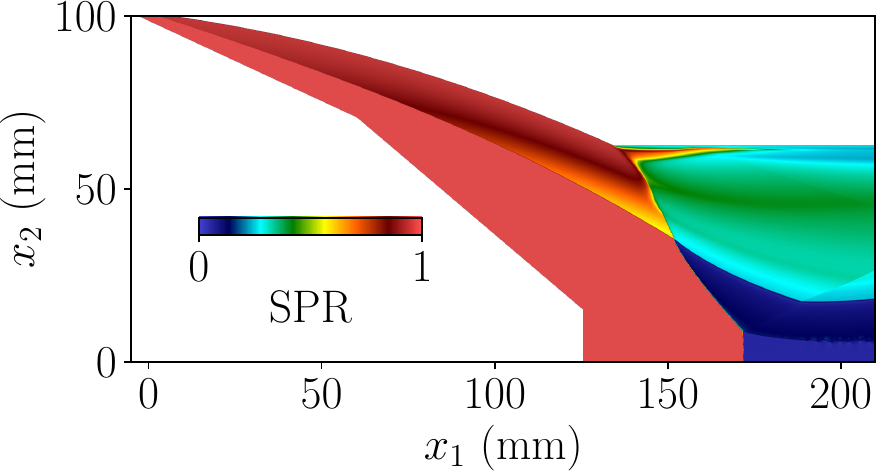}}\hfill{}\subfloat[Case 2B.]{\includegraphics[width=0.32\columnwidth]{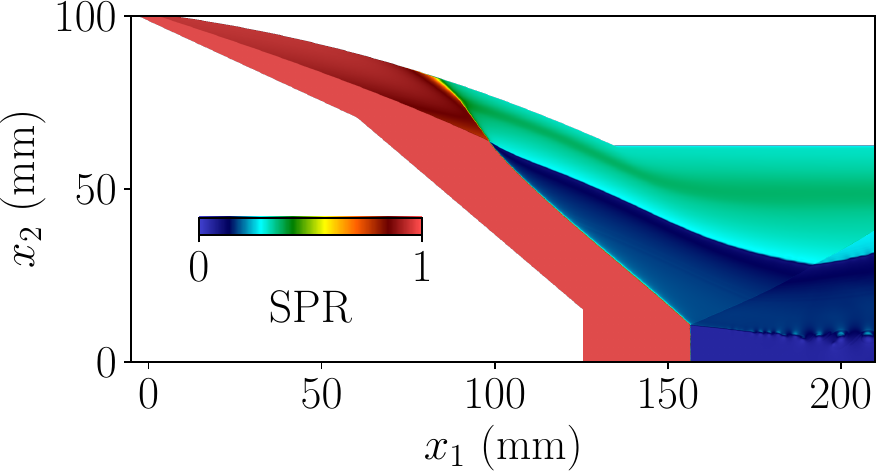}}\hfill{}\subfloat[Case 2C.]{\includegraphics[width=0.32\columnwidth]{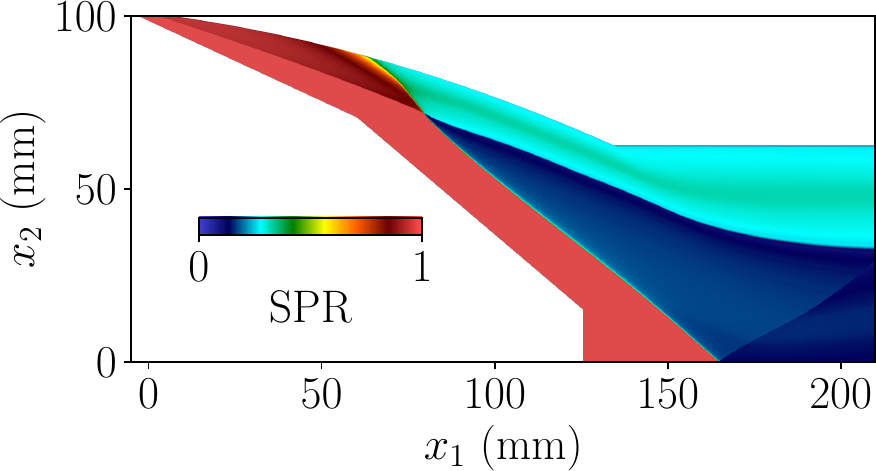}}

\caption{\label{fig:concave-case-2-TPR}SPR distributions for Cases 2A, 2B,
and 2C.}
\end{figure}

As indicated in the SPR distributions presented in Figure~\ref{fig:concave-case-2-TPR},
the SPR behind the reaction zone above the slip line is overall higher
in the absence of ozone, just as in Case 1. The greatest stagnation-pressure
losses are due to the Mach stem, which is completely avoided in Case
2C (10,000 ppm of ozone). The SPR across the CDW is improved with
the addition of ozone, as further illustrated in Figure~\ref{fig:concave-case-2-line-SPR}.
The stronger reflected shock in Case 2C causes a greater loss in stagnation
pressure than in Case 2B, such that the SPR in Case 2C is slightly
lower at the outflow boundary in Figure~\ref{fig:concave-case-2-line-SPR}.

\subsubsection{Case 3}

\label{subsec:Convex-case-3}

\begin{figure}[H]
\subfloat[Case 3A.]{\includegraphics[width=0.32\columnwidth]{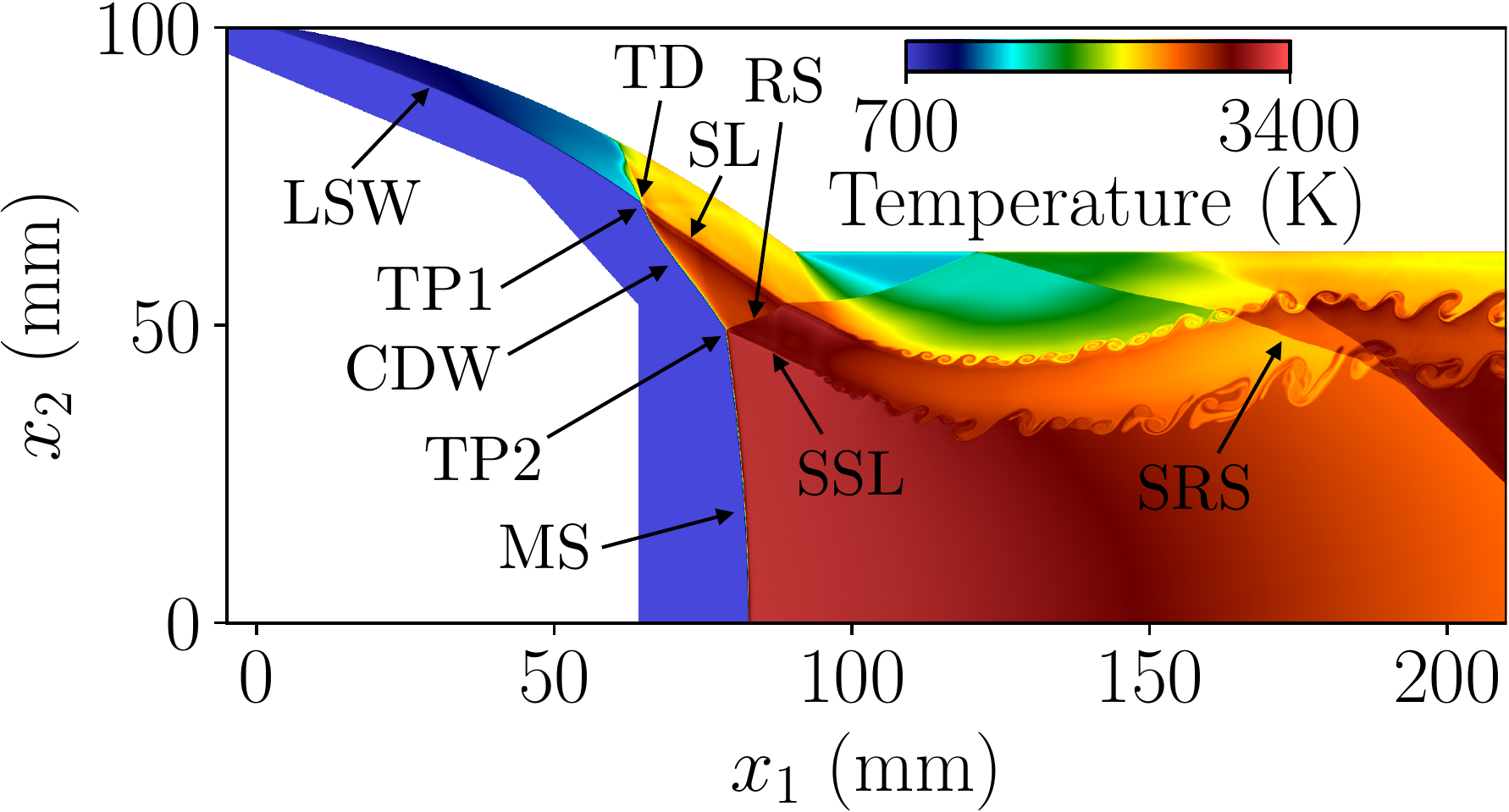}}\hfill{}\subfloat[Case 3B.]{\includegraphics[width=0.32\columnwidth]{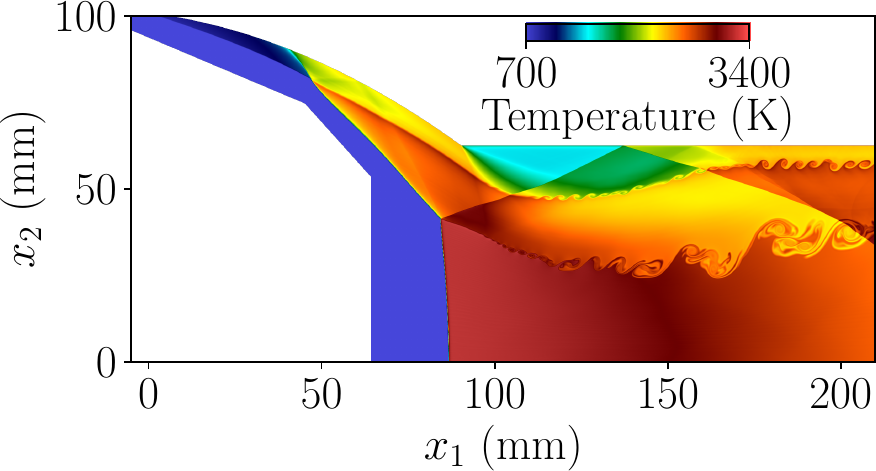}}\hfill{}\subfloat[Case 3C.]{\includegraphics[width=0.32\columnwidth]{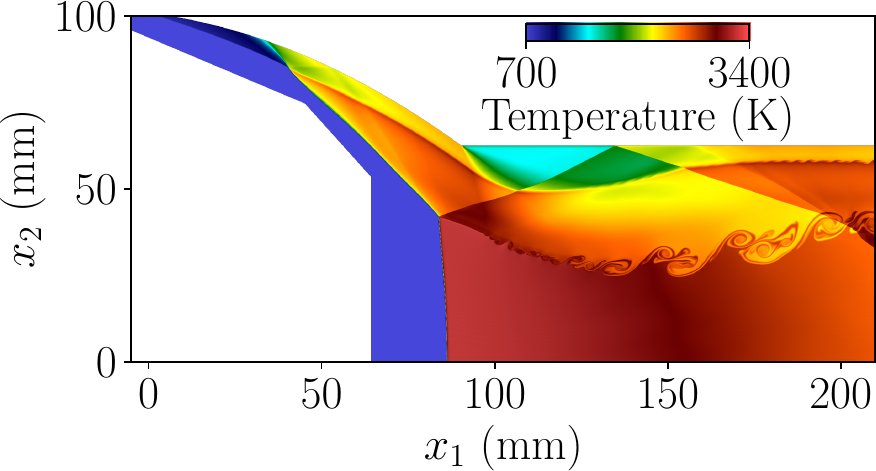}}

\caption{\label{fig:concave-case-3-temperature}Temperature fields for Cases
3A, 3B, and 3C. LSW: leading shock wave. TP1: transition point. TD:
transitional detonation. CDW: curved detonation wave. SL: slip line.
RS: reflected shock. MS: Mach stem. SSL: secondary slip line. TP2:
secondary triple point. SRS: secondary reflected shock.}
\end{figure}

\begin{figure}[H]
\subfloat[Case 3A.]{\includegraphics[width=0.32\columnwidth]{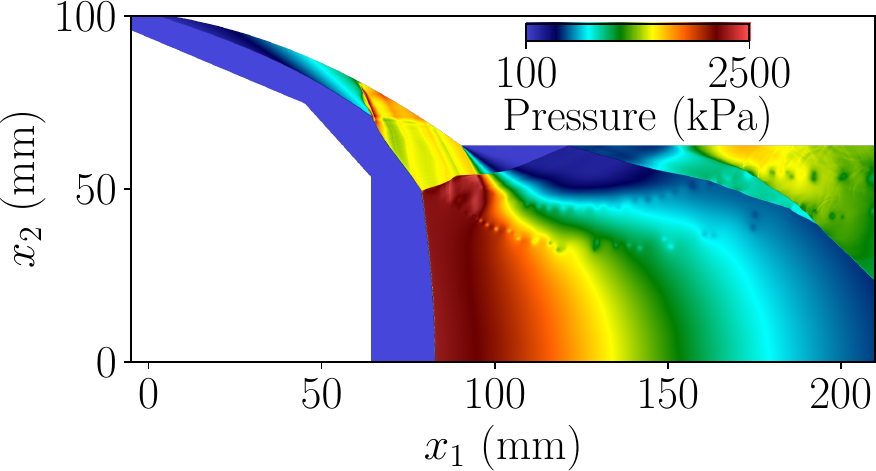}}\hfill{}\subfloat[Case 3B.]{\includegraphics[width=0.32\columnwidth]{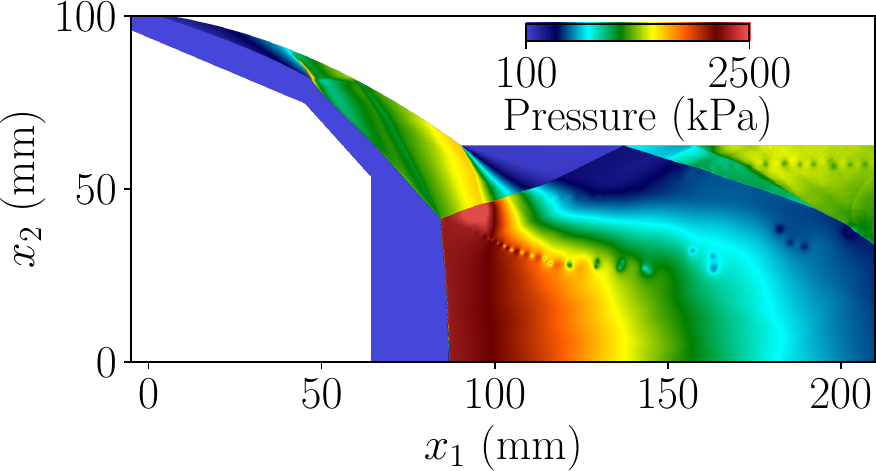}}\hfill{}\subfloat[Case 3C.]{\includegraphics[width=0.32\columnwidth]{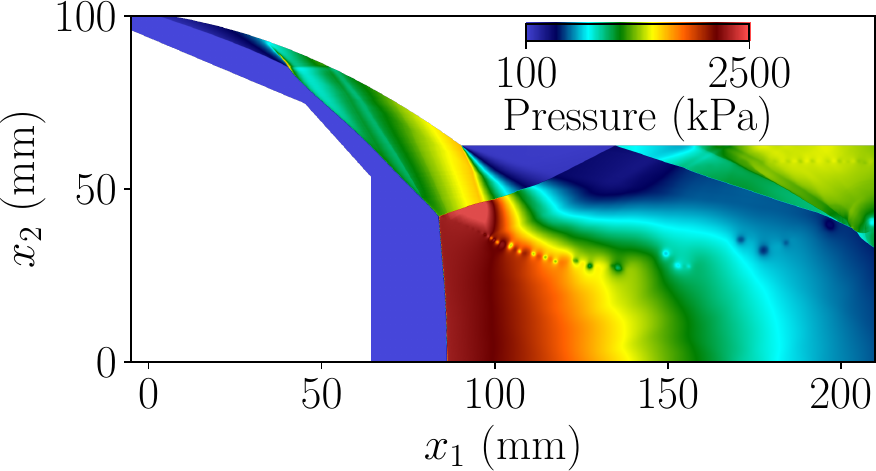}}

\caption{\label{fig:concave-case-3-pressure}Pressure fields for Cases 3A,
3B, and 3C.}
\end{figure}

This case corresponds to a higher curvature of $\kappa=5.09\;\mathrm{km}^{-1}$,
with Point C in Figure~\ref{fig:concave-domain} moved upstream.
In addition, because a larger domain is required, Point F is moved
to $(-5,0)$ mm, and Points G and H are removed (such that Points
A and F are now connected via a line segment). Figures~\ref{fig:concave-case-3-temperature}
and~\ref{fig:concave-case-3-pressure} display the temperature and
pressure fields, respectively, for Cases 3A, 3B, and 3C at $t=4.08$
ms, which is sufficiently long to reach a quasi-steady state. To reduce
computational cost, each simulation is first run on a coarser mesh
with characteristic element size $h=0.2$ mm until approximately $t=2.20$
ms. After uniform refinement of the mesh, such that the element size
becomes $h=0.1$ mm, the simulation is then continued until the final
time is reached. An abrupt LSW-CDW transition and a Mach reflection
pattern is observed in all cases. The height of the Mach stem is significantly
greater than in Cases 1 and 2.  The reflected shock originating from
the secondary triple point reflects off the top horizontal wall. 
These shocks interact with the Kelvin-Helmholtz instabilities. Upon
crossing the Mach stem, the flow becomes subsonic and then accelerates
as the secondary slip line is deflected downwards in the $x_{2}$-direction;
after reaching sonic conditions, the flow accelerates further to supersonic
speeds as the secondary slip line is deflected upwards.

Table~\ref{tab:concave-case-3-transition} presents the LSW-CDW locations
for each case. The addition of ozone shortens the initiation zone
and moves the transition point upstream, though to a lesser degree
than in the previous (lower-curvature) cases. Just as previously observed,
the transitional detonation and transverse and reflected waves, as
well as the CDW, are attenuated. Behind the combustion front and the
CDW, the curvature of the wall induces noticeable compression in the
ozonated cases. Such compression behind the combustion front and CDW
is more prominent here than in Cases 1 and 2 because of the greater
ramp angle, especially near the tail. In the ozone-free case, the
region between the slip lines is significantly narrower than in Cases
3B and 3C, as a result of the taller Mach stem and larger initiation
zone. There is almost no visible difference between Cases 3B and 3C
(apart from the discrepancy in initiation length), which will be further
discussed below.

\begin{figure}[H]
\subfloat[\label{fig:concave-case-3-line-temperature}Temperature along the
line $x_{2}=60$ mm.]{\includegraphics[width=0.32\columnwidth]{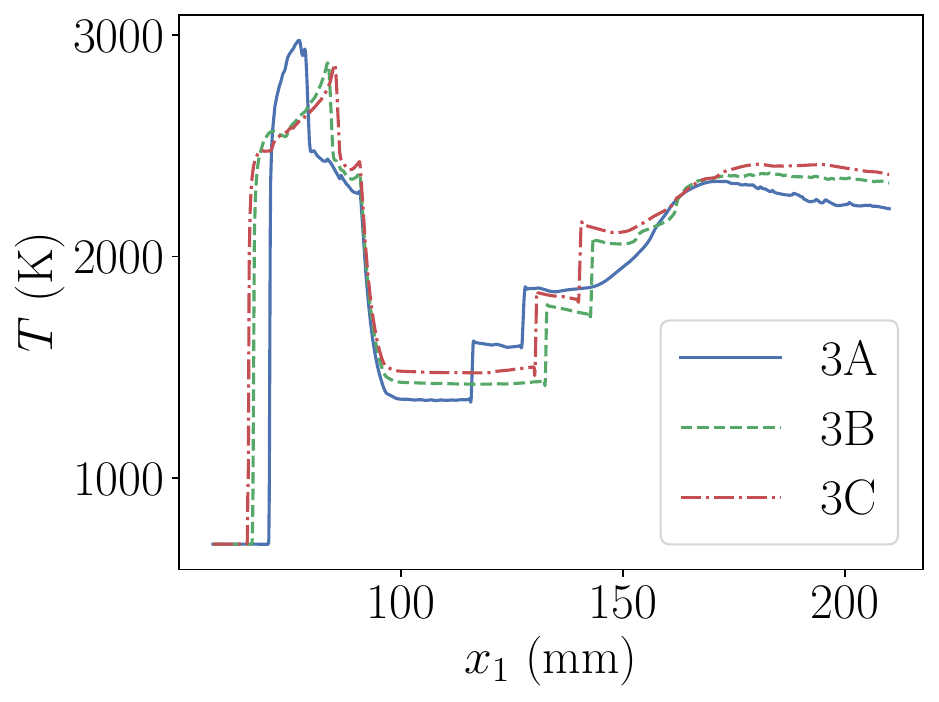}}\hfill{}\subfloat[\label{fig:concave-case-3-line-pressure}Pressure along the line $x_{2}=60$
mm.]{\includegraphics[width=0.32\columnwidth]{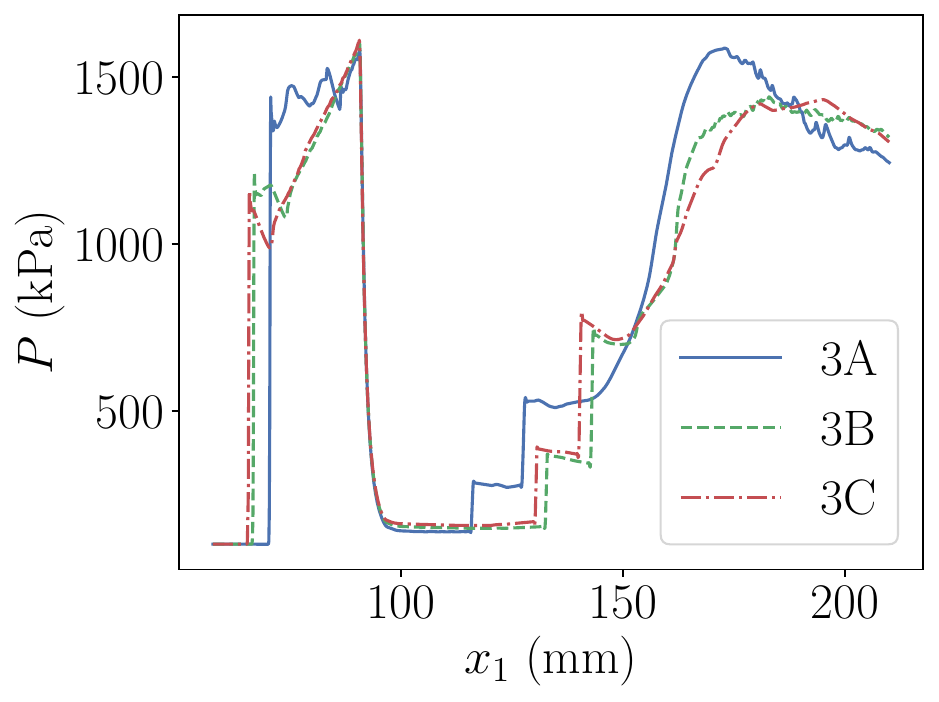}}\hfill{}\subfloat[\label{fig:concave-case-3-line-TPR}SPR along the line $x_{2}=60$
mm.]{\includegraphics[width=0.32\columnwidth]{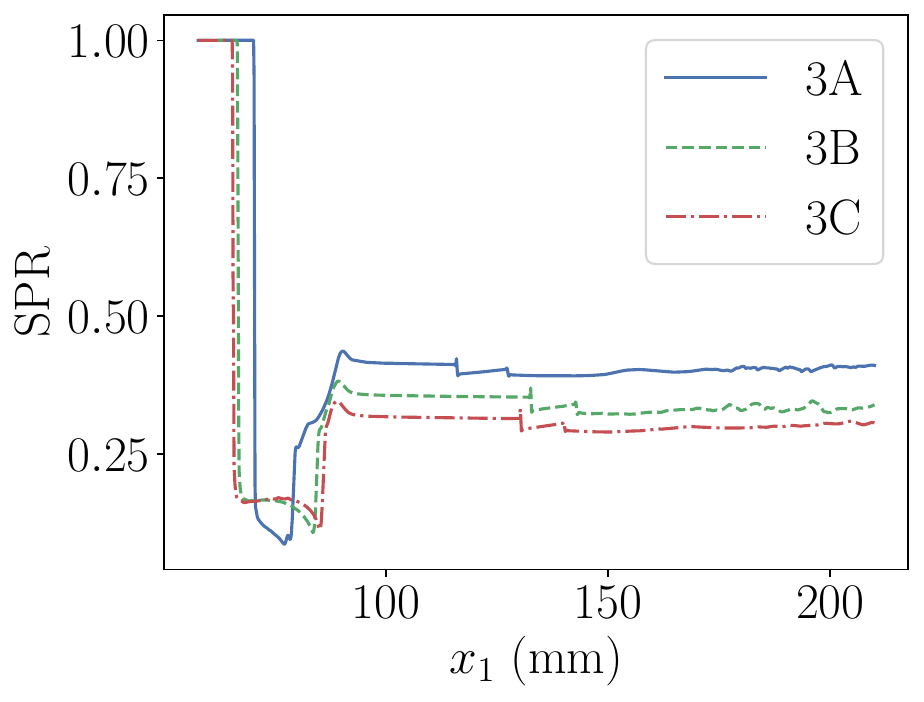}}

\caption{\label{fig:concave-case-3-lines}Variation of temperature, pressure,
and SPR along the line $x_{2}=60$ mm for Case 3 (concave).}
\end{figure}

\begin{table}[h]
\begin{centering}
\caption{$x_{1}$-coordinates of LSW-CDW transitions for Case 3 (concave).
\label{tab:concave-case-3-transition}}
\par\end{centering}
\centering{}%
\begin{tabular}{cccc}
\hline 
\noalign{\vskip\doublerulesep}
Case & Ozone addition & LSW-CDW transition & Relative change vs. Case 3A\tabularnewline[\doublerulesep]
\hline 
\noalign{\vskip\doublerulesep}
\hline 
\noalign{\vskip\doublerulesep}
3A & 0 ppm & $x_{1}\approx64$ mm & 0\%\tabularnewline[\doublerulesep]
\noalign{\vskip\doublerulesep}
\hline 
\noalign{\vskip\doublerulesep}
3B & 1000 ppm & $x_{1}\approx47$ mm & 27\%\tabularnewline[\doublerulesep]
\noalign{\vskip\doublerulesep}
\hline 
\noalign{\vskip\doublerulesep}
3C & 10,000 ppm & $x_{1}\approx40$ mm & 38\%\tabularnewline[\doublerulesep]
\hline 
\noalign{\vskip\doublerulesep}
\end{tabular}
\end{table}
Figures~\ref{fig:concave-case-3-line-temperature} and~\ref{fig:concave-case-3-line-pressure}
present the variation of temperature and pressure, respectively, along
the line $x_{2}=60$ mm, which intersects the CDW, slip line, reflected
shock, and secondary reflected shock. The initial pressure and temperature
jump is due to the CDW. Immediately behind the CDW, the overall increase
in pressure is a result of compression induced by the concave wall,
with the non-monotonicity of the pressure profile caused by waves
propagating from the initial reaction zone, which are especially noticeable
in Case 3A.  The expansion fan then leads to a large, rapid decrease
in pressure. The next two jumps in pressure and temperature are caused
by the reflected shock originating from the secondary triple point
and the secondary reflected shock. 

The difference between Cases 3A and 3B is considerably larger than
the difference between Cases 3B and 3C. In contrast, in the lower-curvature
Case 1 and Case 2, the difference between 1000 ppm of ozone and 10,000
ppm of ozone addition is still appreciable. To help explain this observation,
we first compare Cases 3A and 3B. As shown in the pressure fields
in Figure~\ref{fig:concave-case-3-pressure}, even though the location
of the transition does not change as significantly with 1000 ppm ozone
addition as in Cases 1 and 2, there is a very noticeable difference
in the LSW strengths at the transition point. This is because of the
relatively large change in ramp angle in this high-curvature case
(compared to Cases 1 and 2). When comparing Cases 3B and 3C, a difference
in LSW strength at the transition point can still be discerned. However,
due to the smaller reduction in initiation length, this difference
is not as large and is eventually offset by the subsequent rapid continuous
compression induced by the high-curvature ramp. On the other hand,
this continuous compression behind the transition point is not sufficient
to offset the difference between Cases 3A and 3B. These results illustrate
the complex interactions among various competing effects in the case
of curved walls (as opposed to straight-sided walls).

\begin{figure}[H]
\subfloat[Case 3A.]{\includegraphics[width=0.32\columnwidth]{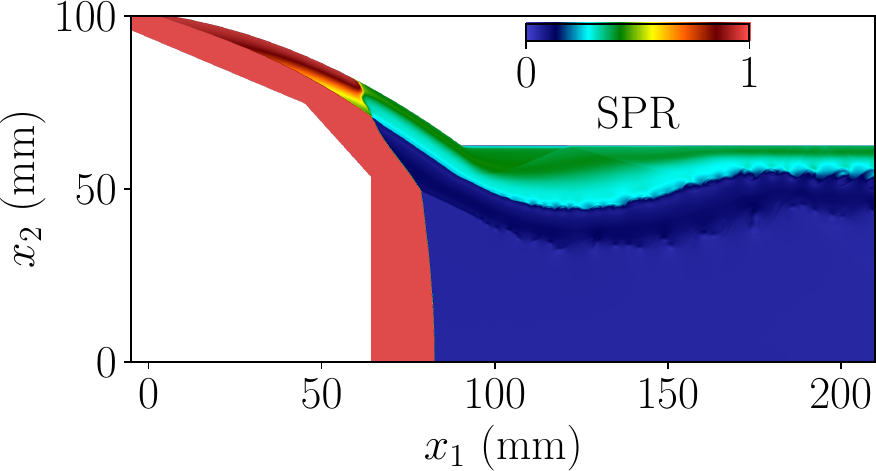}}\hfill{}\subfloat[Case 3B.]{\includegraphics[width=0.32\columnwidth]{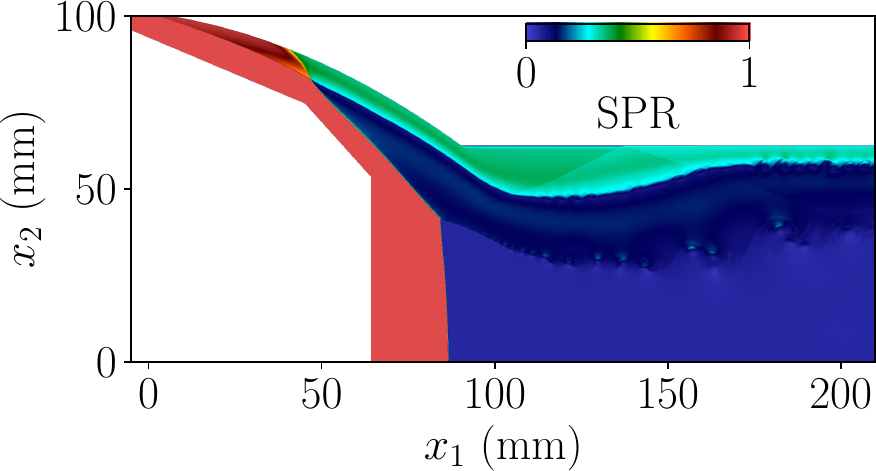}}\hfill{}\subfloat[Case 3C.]{\includegraphics[width=0.32\columnwidth]{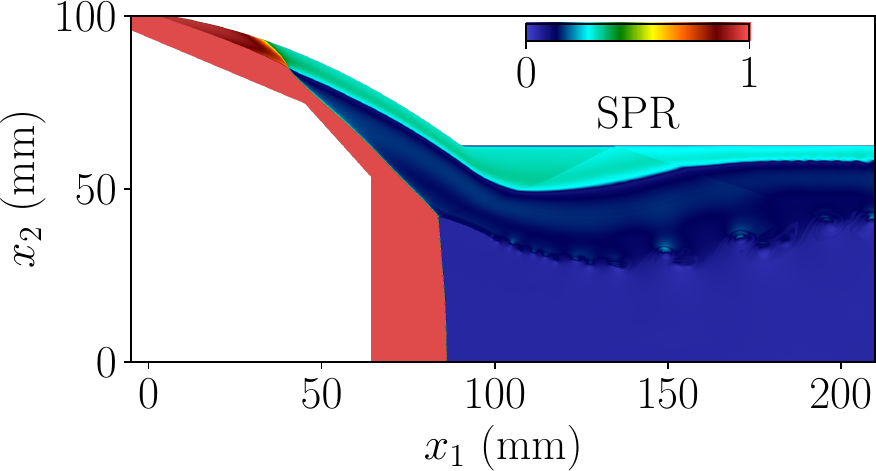}}

\caption{\label{fig:concave-case-3-SPR}SPR distributions for Cases 3A, 3B,
and 3C.}
\end{figure}

The SPR distributions are presented in Figure~\ref{fig:concave-case-3-SPR}.
In all cases, the greatest loss in stagnation pressure occurs across
the Mach stem. Consistent with previous results, the addition of ozone
improves SPR across the CDW but reduces it behind the reaction zone
above the slip line. This is further illustrated in Figure~\ref{fig:concave-case-3-line-TPR},
which shows the variation of SPR along the line $x_{2}=60$ mm. 

\subsubsection{Case 4}

\LyXZeroWidthSpace{}

\begin{figure}[H]
\subfloat[Case 4A, $t=4.15$ ms.]{\includegraphics[width=0.32\columnwidth]{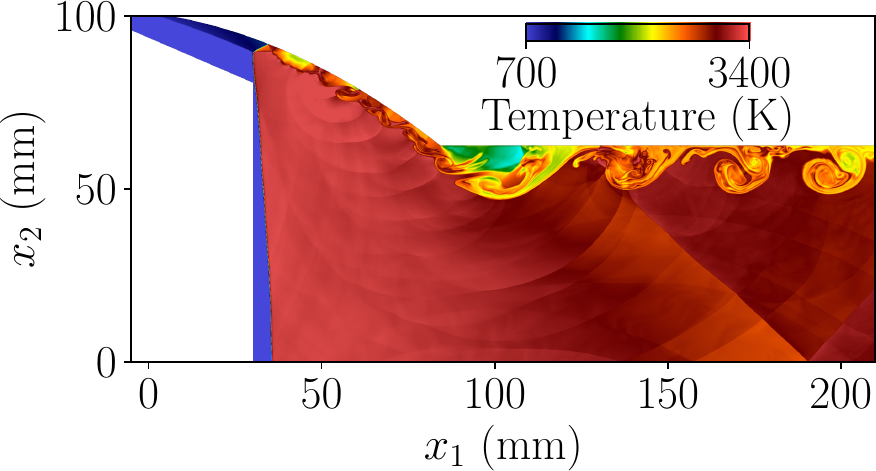}}\hfill{}\subfloat[Case 4B, $t=4.24$ ms.]{\includegraphics[width=0.32\columnwidth]{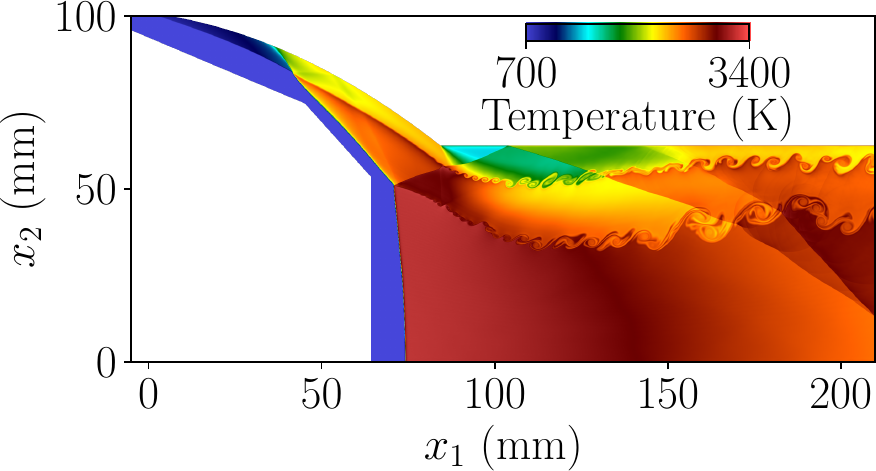}}\hfill{}\subfloat[Case 4C, $t=4.24$ ms.]{\includegraphics[width=0.32\columnwidth]{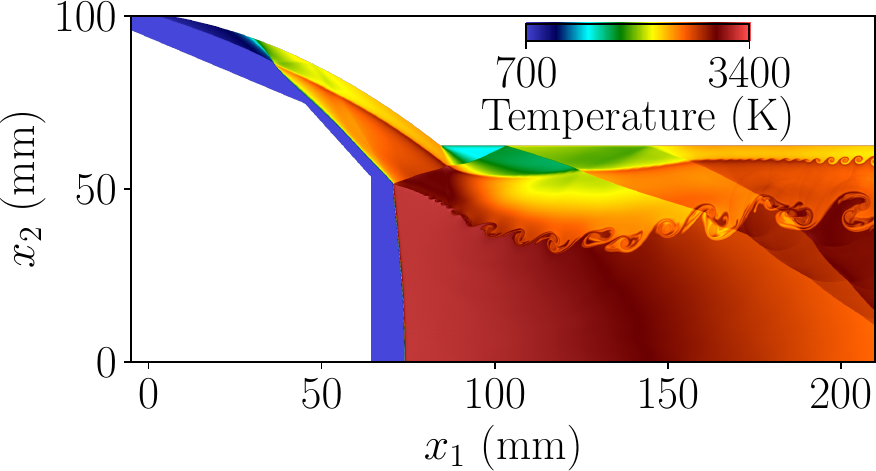}}

\caption{\label{fig:concave-case-4-temperature}Temperature fields for Cases
4A, 4B, and 4C.}
\end{figure}

\begin{figure}[H]
\subfloat[$t=2.51$ ms.]{\includegraphics[width=0.32\columnwidth]{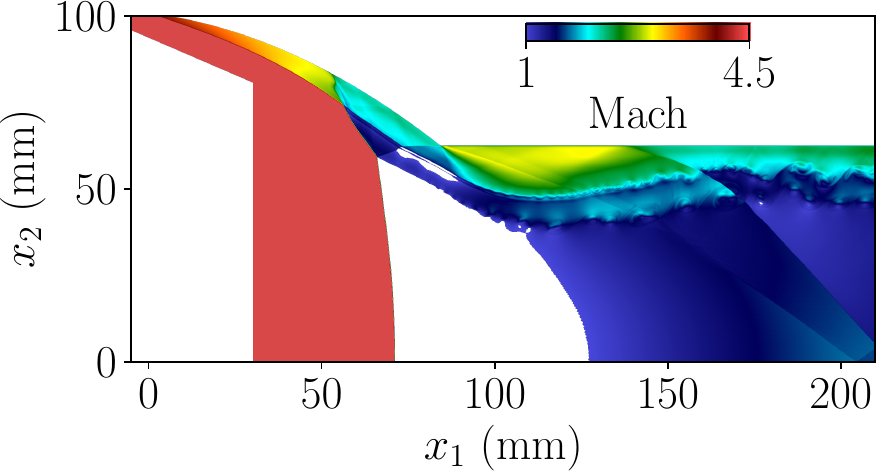}}\hfill{}\subfloat[$t=3.77$ ms.]{\includegraphics[width=0.32\columnwidth]{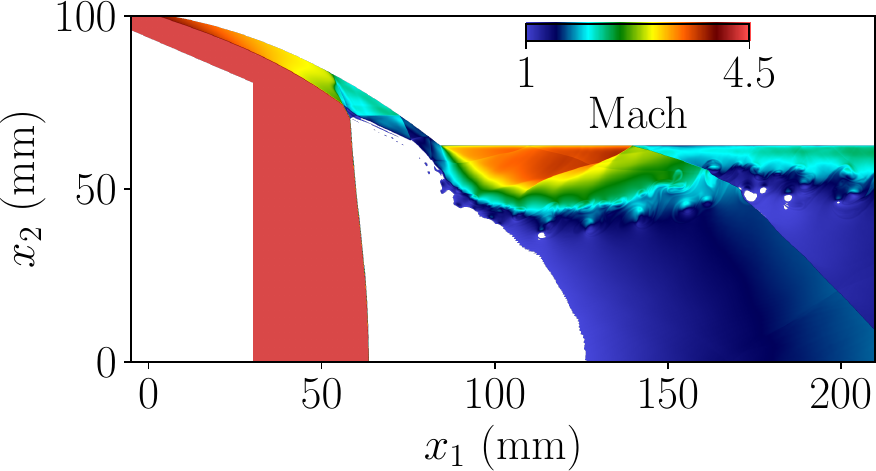}}\hfill{}\subfloat[$t=4.15$ ms.]{\includegraphics[width=0.32\columnwidth]{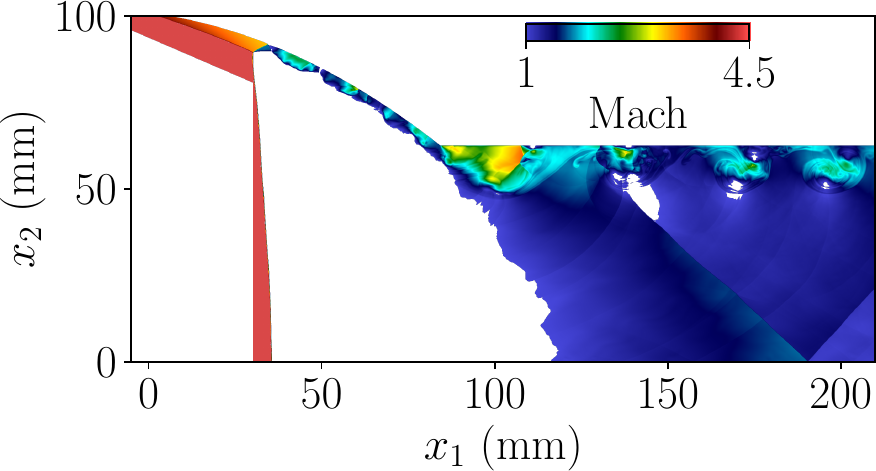}}

\caption{\label{fig:concave-case-4A-mach}Mach-number distributions for Case
4A. Subsonic regions are left uncolored.}
\end{figure}

\begin{figure}[H]
\subfloat[$t=2.36$ ms.]{\includegraphics[width=0.32\columnwidth]{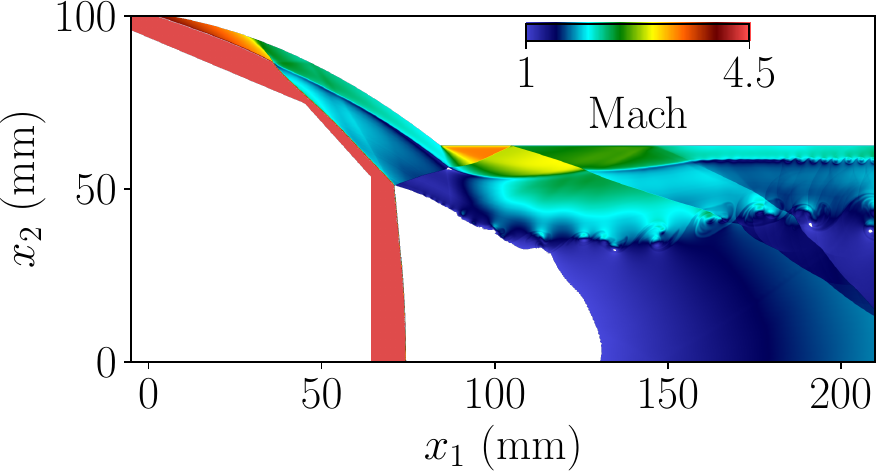}}\hfill{}\subfloat[$t=3.30$ ms.]{\includegraphics[width=0.32\columnwidth]{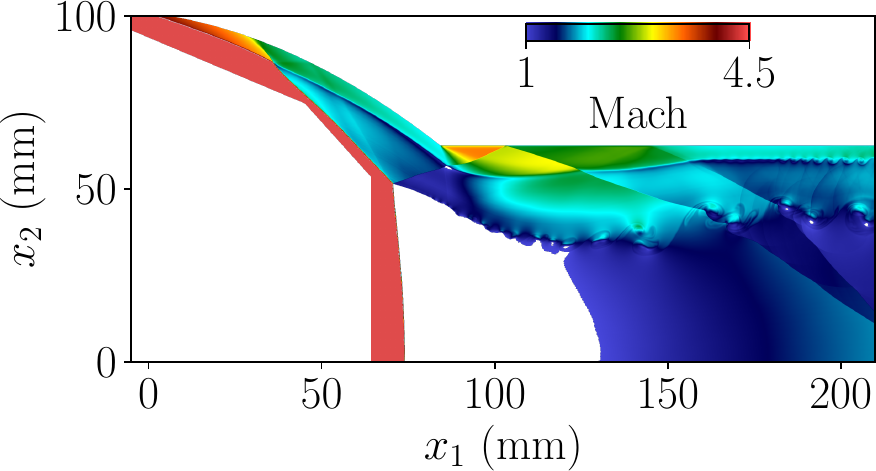}}\hfill{}\subfloat[$t=4.24$ ms.]{\includegraphics[width=0.32\columnwidth]{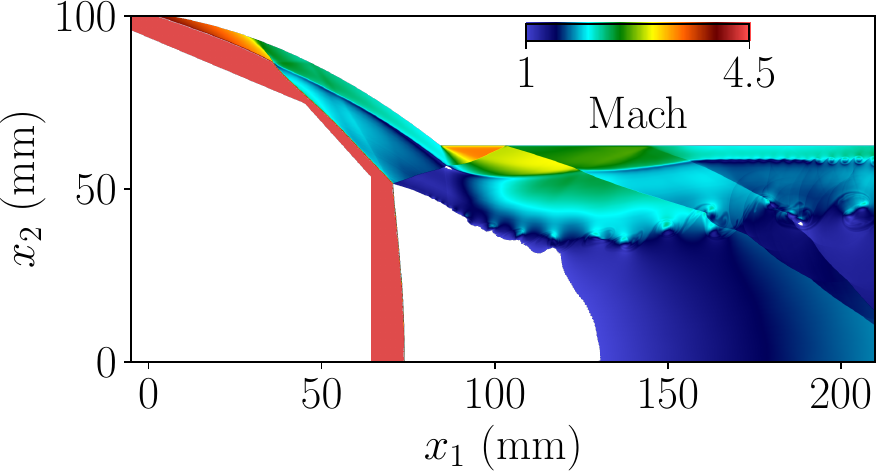}}

\caption{\label{fig:concave-case-4C-mach}Mach-number distributions for Case
4C. Subsonic regions are left uncolored.}
\end{figure}

The purpose of this case is to present a geometry for which the detonation
is non-stationary in the absence of ozone but can be stabilized with
ozone addition. Point C is moved upstream to $\left(84,62\right)$
mm, leading to a mean curvature of $\kappa=5.80\;\mathrm{km}^{-1}$,
which is slightly larger than the mean curvature in Case 3. Each case
is first computed on a coarser mesh with $h=0.2$ mm until approximately
$t=2.20$ ms. The mesh is then uniformly refined, and the simulation
is restarted. Figure~\ref{fig:concave-case-4-temperature} shows
the temperature fields at $t=4.15$ ms for Case 4A and $t=4.24$ ms
for Cases 4B and 4C. The temperature fields for Case 4B and 4C resemble
those for Case 3B and 3C due to the similar geometries. However, in
the absence of ozone, the detonation fails to stabilize. The Mach
stem grows and accelerates towards the inflow boundary, and the CDW
disappears. The reason for the non-stationary behavior is interaction
between large high-pressure, subsonic regions in close proximity~\citep{Yan24}.
To illustrate, Figure~\ref{fig:concave-case-4A-mach} presents Mach-number
distributions for Case 4A at three different points in time, where
the subsonic regions are left uncolored. At earlier times, there is
a subsonic region behind the reflected shock that eventually merges
with the subsonic region behind the Mach stem, causing a non-stationary
condition. Conversely, Figure~\ref{fig:concave-case-4C-mach} shows
Mach-number distributions for Case 4C at three different points in
time, where it is observed that the flow behind the reflected shock
is nearly fully supersonic as a result of the detonation attenuation,
preventing non-stationary behavior. 

It should be noted, however, that the ability of ozone addition to
stabilize a normally non-stationary detonation is limited, at least
at the considered conditions. If Point C is moved marginally upstream
to $\left(80,62\right)$ mm, resulting in a mean curvature of $\kappa=6.33\;\mathrm{km}^{-1}$,
then ozone addition at concentrations of 1000 ppm and 10,000 ppm fail
to stabilize the detonation (not shown for brevity). One potential
reason is the very high ramp angle near the tail, which, as previously
discussed, causes rapid post-transition compression that can appreciably
counteract the benefits of a shortened initation zone. In the case
of a non-stationary detonation with reduced continuous compression
behind the combustion front and CDW, the ability of ozone addition
to stabilize the detonation may be improved.


\subsection{Convex walls}

\label{subsec:Convex-walls}

Table~\ref{tab:convex-cases} lists the two convex-wall cases considered
in this study. $\alpha_{2}$ is fixed across all cases, and Point
C in Figure~\ref{fig:convex-domain} is varied in the $x_{1}$-direction
to obtain different curvatures. Case 1 corresponds to the highest
curvature (in magnitude). At higher curvatures, the initiation length
in the absence of ozone becomes very small, such that the effect of
ozone addition is significantly reduced; therefore, higher curvatures
are not considered. As indicated in the fifth column of Table~\ref{tab:convex-cases},
the letter refers to the ozone amount (e.g., Case 1C corresponds to
$\kappa=-1.04\;\mathrm{km}^{-1}$ and 10,000 ppm ozone). A regular
reflection pattern is observed in all cases apart from Case 1A, which
corresponds to a (stationary) Mach reflection, and Case 2A, in which
the LSW fails to transition to a detonation. 

\begin{table}[h]
\begin{centering}
\caption{Geometric information for convex ramp. Units are in mm unless otherwise
specified.\label{tab:convex-cases}}
\par\end{centering}
\centering{}%
\begin{tabular}{cccccc}
\hline 
\noalign{\vskip\doublerulesep}
Case & Point C & Wall profile & $\kappa\;\left(\mathrm{km}^{-1}\right)$ & Ozone & Reflection pattern\tabularnewline[\doublerulesep]
\hline 
\noalign{\vskip\doublerulesep}
\hline 
\noalign{\vskip\doublerulesep}
1 & $\left(150,62\right)$ & $x_{2,c}=0.0005750x_{1}^{2}-0.3396x_{1}+100$ & $-1.04$ & %
\begin{tabular}{c}
1A: 0 ppm\tabularnewline
1B: 1000 ppm\tabularnewline
1C: 10,000 ppm\tabularnewline
\end{tabular} & %
\begin{tabular}{c}
1A: Mach reflection\tabularnewline
1B: Regular reflection\tabularnewline
1C: Regular reflection\tabularnewline
\end{tabular}\tabularnewline[\doublerulesep]
\noalign{\vskip\doublerulesep}
\hline 
\noalign{\vskip\doublerulesep}
2 & $\left(180,52\right)$ & $x_{2,c}=0.0002446x_{1}^{2}-0.2551x_{1}+100$ & $-0.46$ & %
\begin{tabular}{c}
2A: 0 ppm\tabularnewline
2B: 1000 ppm\tabularnewline
2C: 10,000 ppm\tabularnewline
\end{tabular} & %
\begin{tabular}{c}
2A: No detonation\tabularnewline
2B: Regular reflection\tabularnewline
2C: Regular reflection\tabularnewline
\end{tabular}\tabularnewline[\doublerulesep]
\hline 
\noalign{\vskip\doublerulesep}
\end{tabular}
\end{table}

\subsubsection{Case 1}

\begin{figure}[H]
\subfloat[Case 1A.]{\includegraphics[width=0.32\columnwidth]{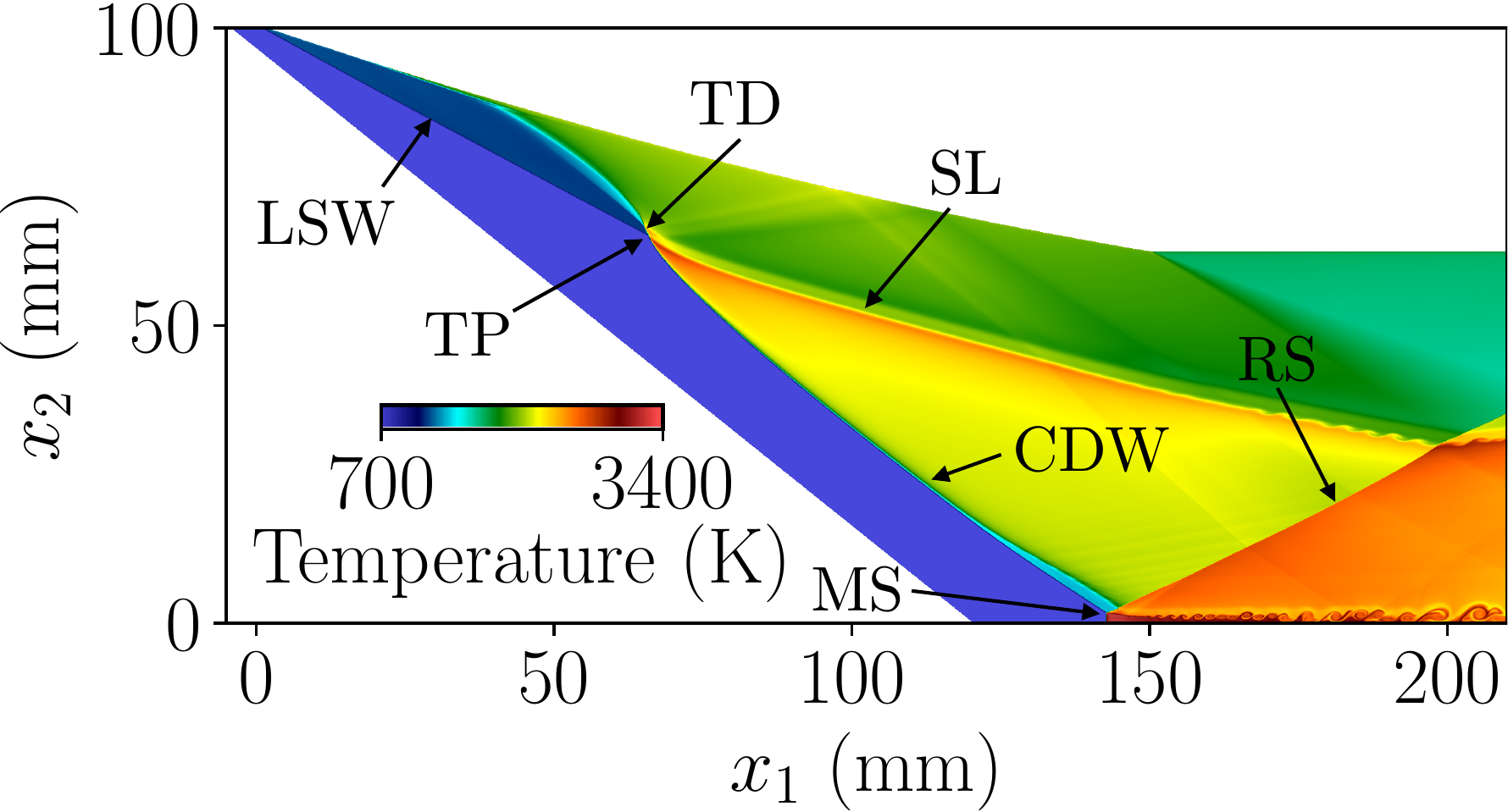}}\hfill{}\subfloat[Case 1B.]{\includegraphics[width=0.32\columnwidth]{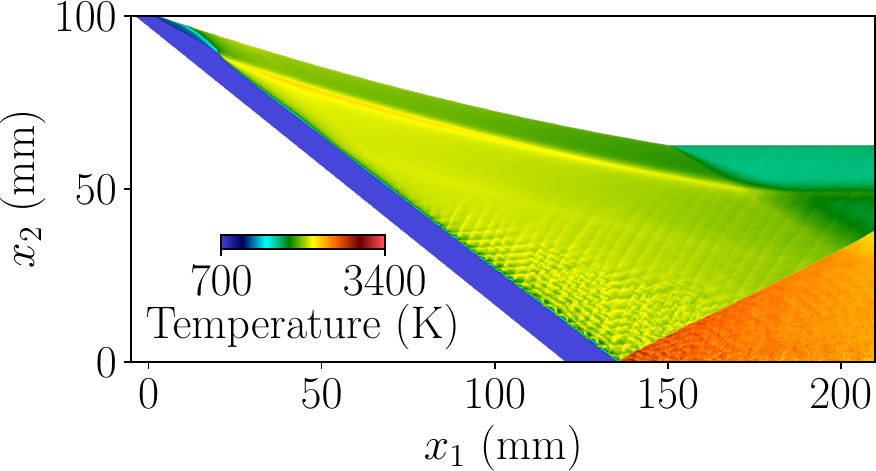}}\hfill{}\subfloat[Case 1C.]{\includegraphics[width=0.32\columnwidth]{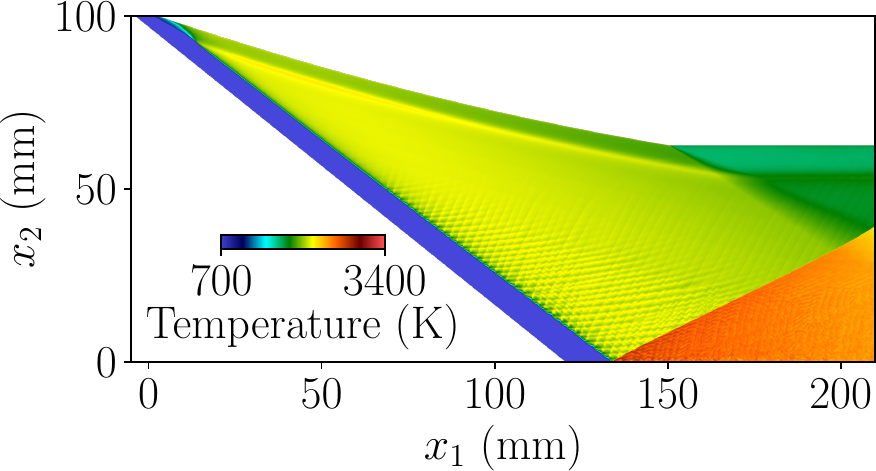}}

\caption{\label{fig:convex-case-1-temperature-1}Temperature fields for Cases
1A, 1B, and 1C. LSW: leading shock wave. TP: transition point. TW:
transverse wave. CDW: curved detonation wave. SL: slip line. RS: reflected
shock.}
\end{figure}

\begin{figure}[H]
\subfloat[Case 1A.]{\includegraphics[width=0.32\columnwidth]{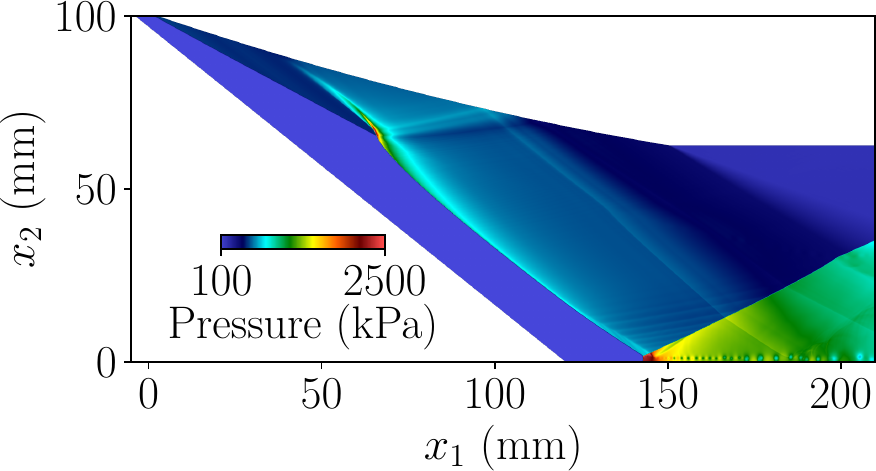}}\hfill{}\subfloat[Case 1B.]{\includegraphics[width=0.32\columnwidth]{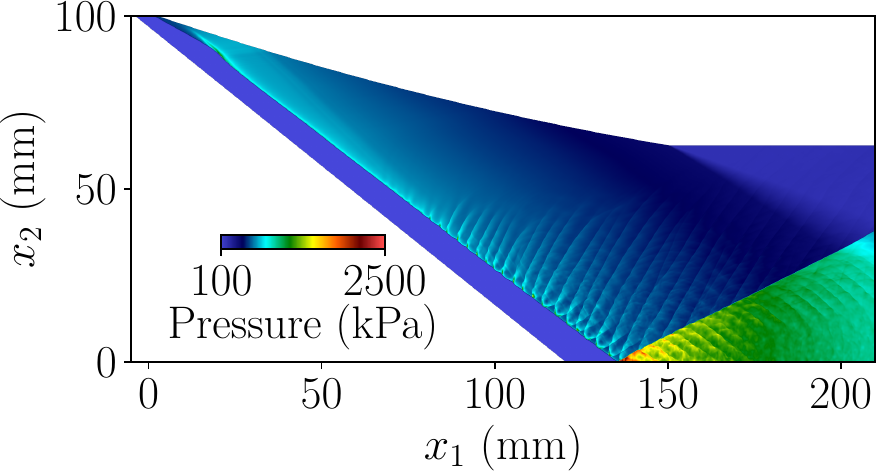}}\hfill{}\subfloat[Case 1C.]{\includegraphics[width=0.32\columnwidth]{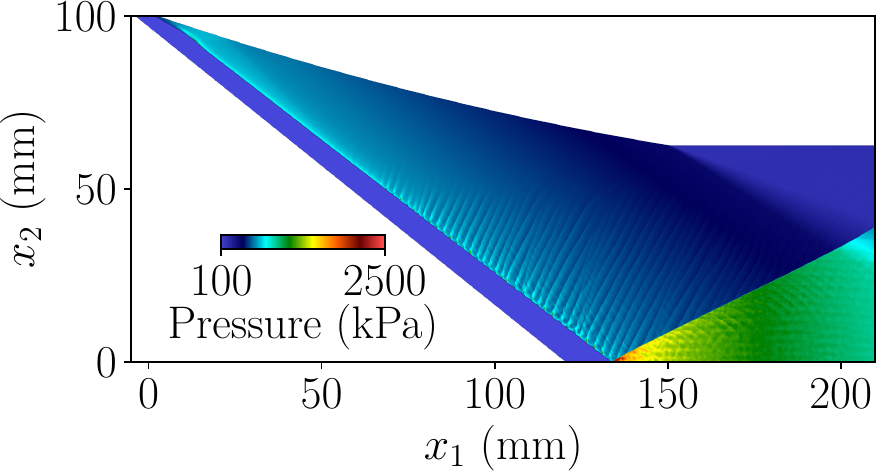}}

\caption{\label{fig:convex-case-1-pressure-1}Pressure fields for Cases 1A,
1B, and 1C.}
\end{figure}

Figures~\ref{fig:convex-case-1-temperature-1} and~\ref{fig:convex-case-1-pressure-1}
display the quasi-steady temperature and pressure fields, respectively,
for Case 1A at $t=1.57$ ms and Cases 1B and 1C at $t=0.628$ ms.
Case 1A is first computed on a coarser mesh with characteristic element
size $h=0.2$ mm until $t=0.628$ ms, after which the mesh is uniformly
refined (which yields $h=0.1$ mm) and the simulation is continued
until the final time. A regular reflection pattern is observed Case
1B and Case 1C, in which the LSW transitions to a CDW that then reflects
off the bottom wall. Case 1A is instead characterized by a Mach reflection
pattern, where the Mach stem is extremely small. Cellular structures
are clearly observed in Cases 2B and 2C, while hints of a cellular
structure can be discerned in Case 2A. Furthermore, in Case 2A, the
induction zone ahead of the CDW noticeably increases in size downstream.
In the absence of ozone, the LSW-CDW transition is abrupt. The addition
of ozone reduces the initiation length and leads to a smooth transition,
along with weakening or disappearance of the transverse waves originating
from the LSW-CDW transition. In the vicinity of the transition point,
the CDW is attenuated with ozone addition; however, as will be discussed
below, the effect of ozone sensitization on the CDW away from the
transition point (near the bottom wall) is more complicated. Due to
the smaller initiation zone, the slip line is closer to the curved
ramp. In Case 1A, the slip line intersects the reflected shock before
the end of the domain. The difference between 0 ppm and 1000 ppm of
ozone is greater than that between 1000 ppm and 10,000 ppm of ozone. 

The $x_{1}$-coordinates of the LSW-CDW transitions, as well as the
relative change with respect to the ozone-free case, are listed in
Table~\ref{tab:convex-case-1-transition}. Despite the smaller initiation
zone in the absence of ozone addition, the upstream relocation of
the transition point is comparatively greater here than in the concave-ramp
cases. The reason is that as the onset of combustion is moved upstream
with greater ozone concentrations, the ramp angle increases, which
has the amplifying effect of accelerating combustion. It should be
noted, however, that this amplifying effect is not expected to be
significant in this relatively low-curvature case. In the case of
concave ramps, as previously discussed, the opposite is observed since
the ramp angle instead decreases, thus mitigating combustion.

\begin{figure}[H]
\subfloat[\label{fig:convex-case-1-line-temperature-1}Temperature along the
line $x_{2}=15$ mm.]{\includegraphics[width=0.32\columnwidth]{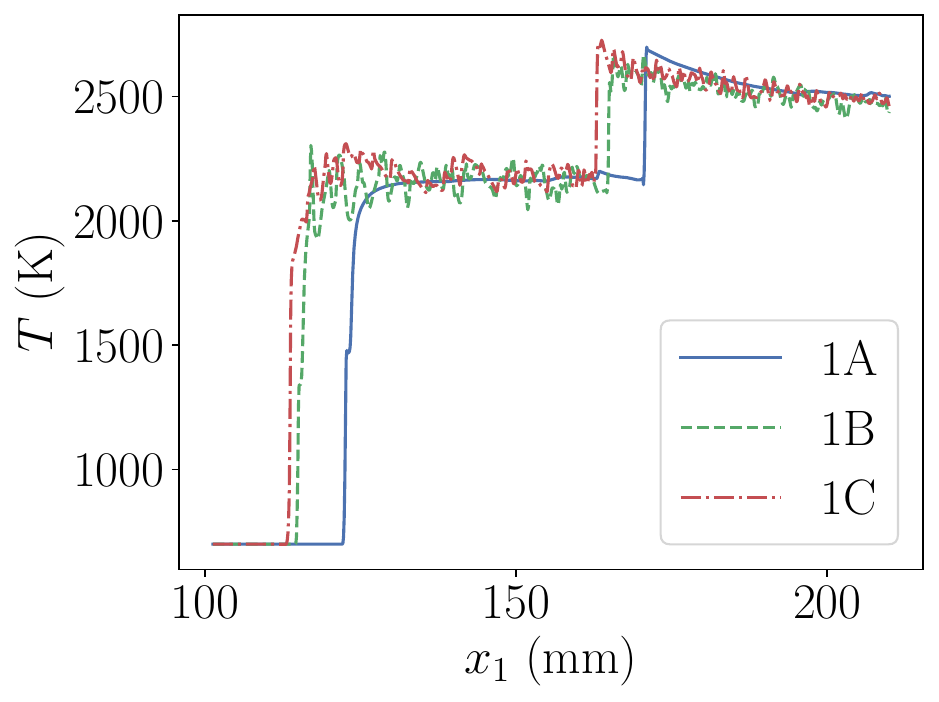}}\hfill{}\subfloat[\label{fig:convex-case-1-line-pressure-1}Pressure along the line
$x_{2}=15$ mm.]{\includegraphics[width=0.32\columnwidth]{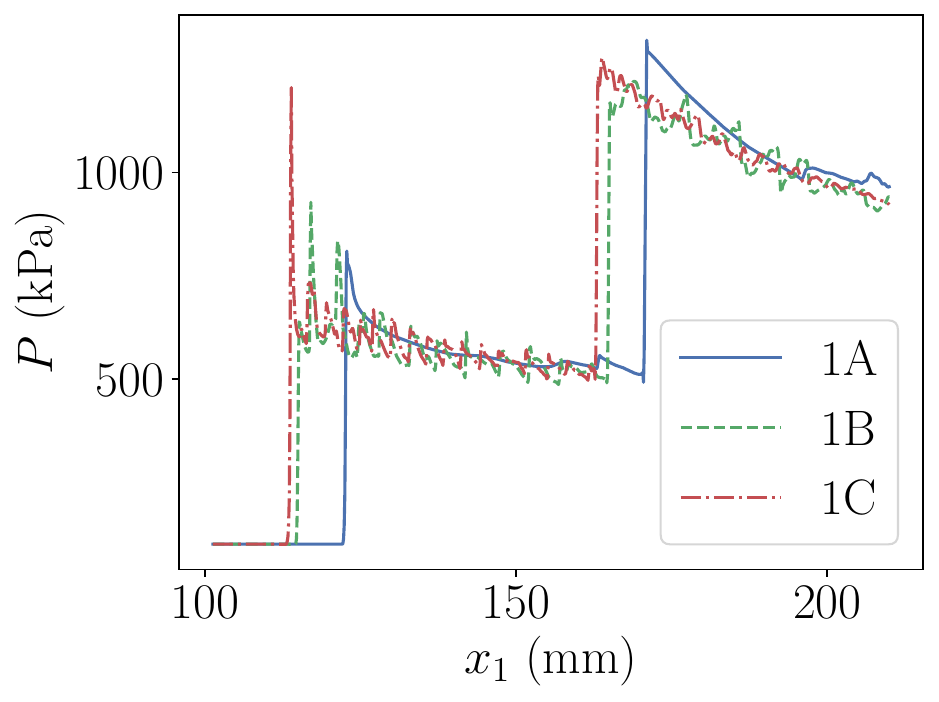}}\hfill{}\subfloat[\label{fig:convex-case-1-line-SPR-1}SPR along the line $x_{2}=15$
mm.]{\includegraphics[width=0.32\columnwidth]{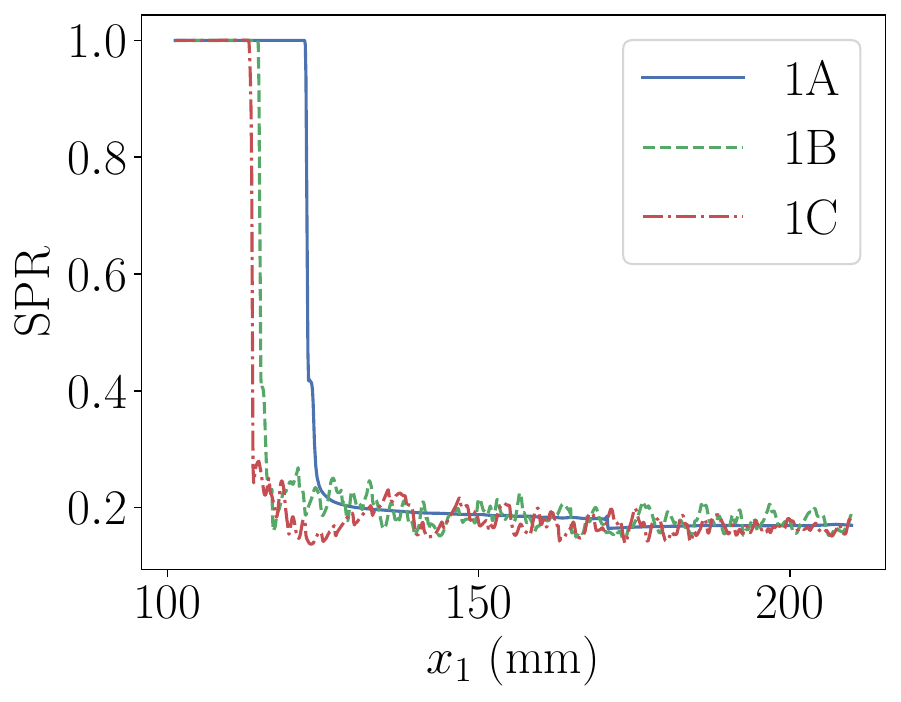}}

\caption{\label{fig:convex-case-1-lines-1}Variation of temperature, pressure,
and SPR along the line $x_{2}=15$ mm for Case 1 (convex).}
\end{figure}

\begin{table}[h]
\begin{centering}
\caption{$x_{1}$-coordinates of LSW-CDW transitions for Case 1 (convex). \label{tab:convex-case-1-transition}}
\par\end{centering}
\centering{}%
\begin{tabular}{cccc}
\hline 
\noalign{\vskip\doublerulesep}
Case & Ozone addition & LSW-CDW transition & Relative change vs. Case 1A\tabularnewline[\doublerulesep]
\hline 
\noalign{\vskip\doublerulesep}
\hline 
\noalign{\vskip\doublerulesep}
1A & 0 ppm & $x_{1}\approx66$ mm & 0\%\tabularnewline[\doublerulesep]
\noalign{\vskip\doublerulesep}
\hline 
\noalign{\vskip\doublerulesep}
1B & 1000 ppm & $x_{1}\approx21$ mm & 68\%\tabularnewline[\doublerulesep]
\noalign{\vskip\doublerulesep}
\hline 
\noalign{\vskip\doublerulesep}
1C & 10,000 ppm & $x_{1}\approx14$ mm & 79\%\tabularnewline[\doublerulesep]
\hline 
\noalign{\vskip\doublerulesep}
\end{tabular}
\end{table}
Figures~\ref{fig:convex-case-1-line-temperature-1} and~\ref{fig:convex-case-1-line-pressure-1}
presents the variation of temperature and pressure along the line
$x_{2}=15$ mm, which intersects the CDW and the reflected shock.
The oscillations in temperature and pressure are due to transverse
waves associated with the cellular structure. The addition of ozone
moves the CDW and reflected shock upstream. The first increase in
temperature and pressure is a result of the CDW, and the second increase
is due to the reflected shock. The gradual decreases in pressure behind
the CDW and the reflected shock are caused by continuous expansion
induced by the convex wall.  

In the case of ODWs, ozone addition often attenuates the detonation,
especially if the transition type changes from abrupt to smooth~\citep{Ten24,Vas22}
(which is the case here), leading to improved propulsion performance.
For convex ramps, however, there is a competing effect wherein a shorter
initiation zone leads to a stronger shock at the transition point
(due to a greater ramp angle). Here, this competing effect slightly
offsets the smoothening of the LSW-CDW transition, such that far away
from the transition point, the temperature behind the CDW in Case
1C (10,000 ppm of ozone) is overall slightly greater than in Case
1A (no ozone). However, given the relatively low curvature, this competing
effect is not expected to be significant, especially since another
contributing factor to the higher overall post-CDW temperature
in Case 1C is that ozone addition already leads to slightly greater
post-detonation temperatures in the one-dimensional ZND setting (as
in Section~\ref{subsec:ZND-calculations}). 

\begin{figure}[H]
\subfloat[Case 1A.]{\includegraphics[width=0.32\columnwidth]{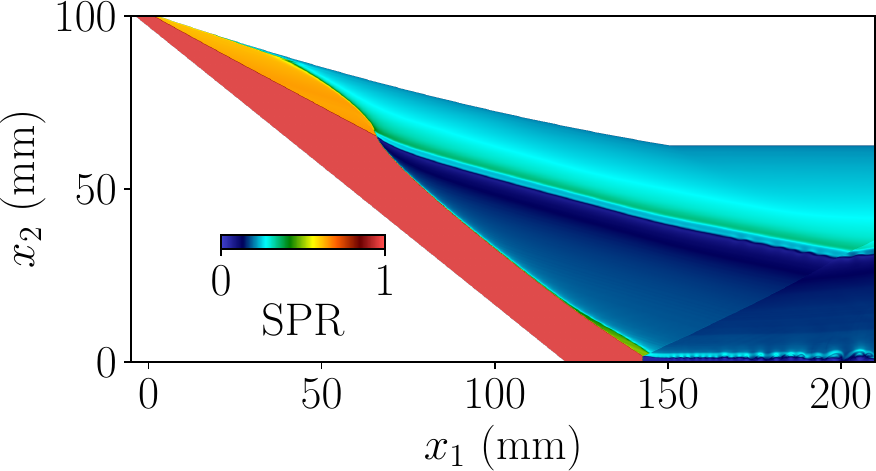}}\hfill{}\subfloat[Case 1B.]{\includegraphics[width=0.32\columnwidth]{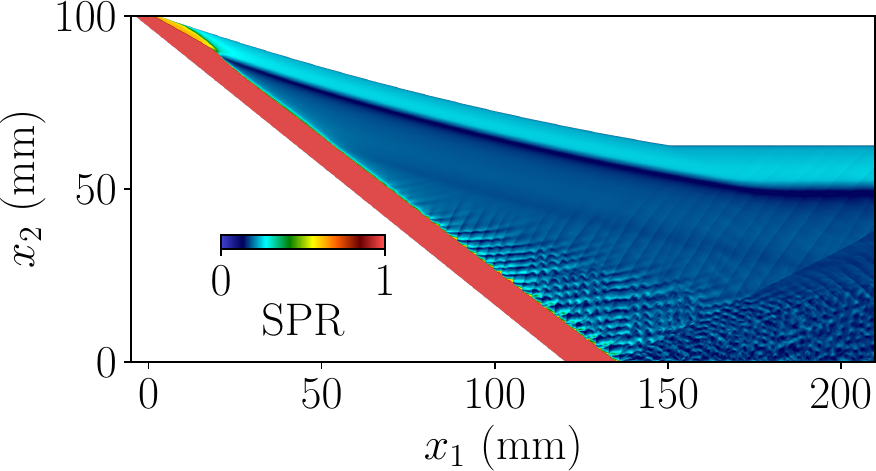}}\hfill{}\subfloat[Case 1C.]{\includegraphics[width=0.32\columnwidth]{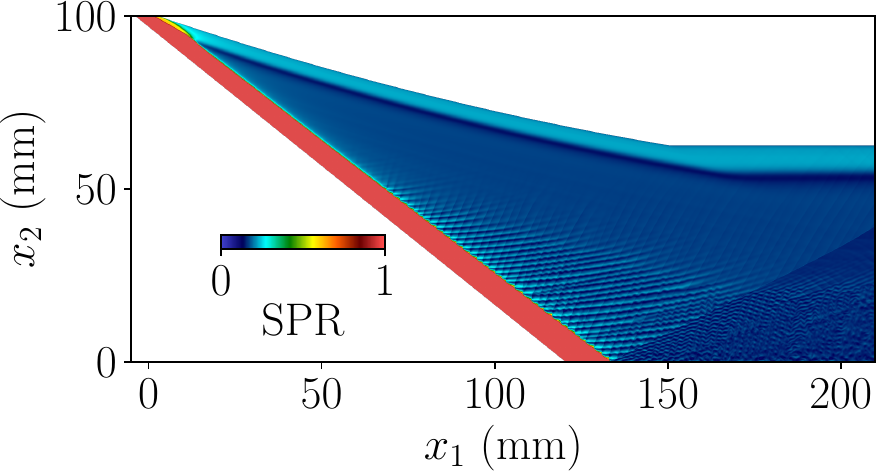}}

\caption{\label{fig:convex-case-1-TPR}SPR distributions for Cases 1A, 1B,
and 1C.}
\end{figure}

Figure~\ref{fig:convex-case-1-TPR} shows the distributions of SPR.
The post-reaction region above the slip line exhibits higher SPR in
the absence of ozone additive. This is because ozone sensitization,
by way of shortening the initiation zone, mitigates the coalescence
of compression waves along the combustion front~\citep{Ten24}, and
stronger compression in the reaction zone improves the SPR~\citep{Xio23}.
Greater stagnation-pressure losses are observed across the CDW, especially
near the transition point where the detonation is strongest. Additional
loss in stagnation pressure occurs across the reflected shock. Near
the transition point, ozone addition improves SPR across the CDW,
which is consistent with the ODW setting~\citep{Ten24}. However,
away from the transition point (near the bottom wall), stagnation-pressure
loss across the CDW is highest in Case 1C (10,000 ppm of ozone), as
Figure~\ref{fig:convex-case-1-line-SPR-1} shows.  The overall greatest
stagnation-pressure loss is observed behind the Mach stem in Case
1A. 

\subsubsection{Case 2}

\begin{figure}[H]
\subfloat[Case 2A.]{\includegraphics[width=0.32\columnwidth]{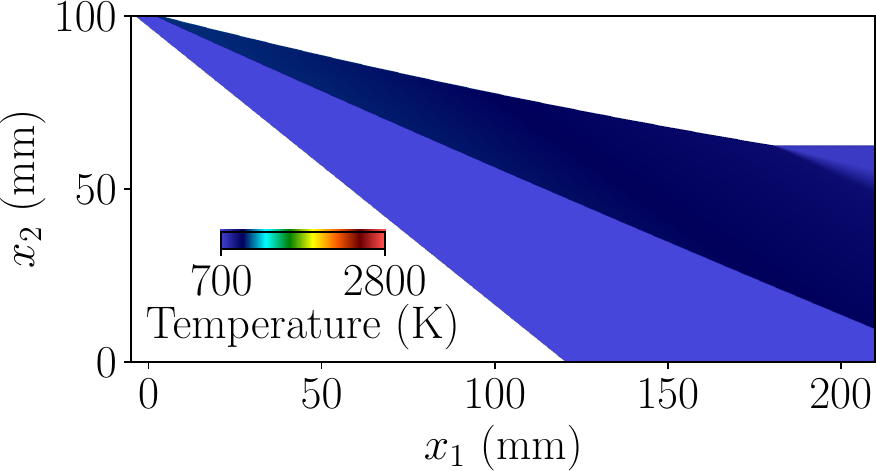}}\hfill{}\subfloat[Case 2B.]{\includegraphics[width=0.32\columnwidth]{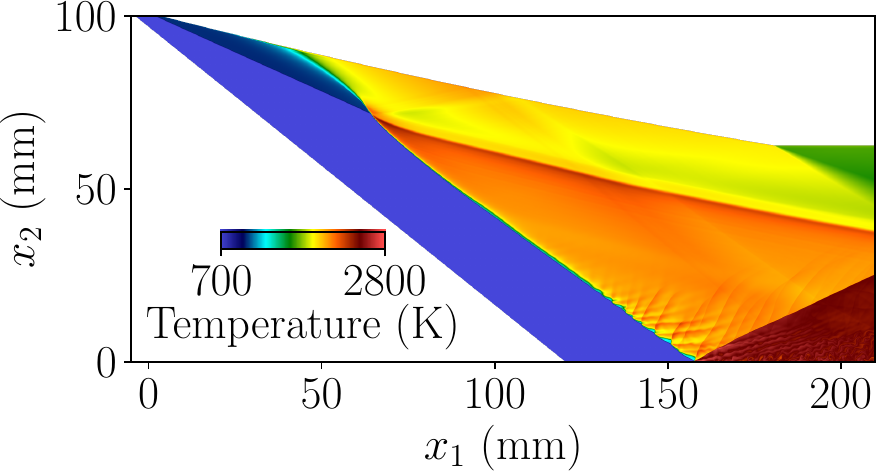}}\hfill{}\subfloat[Case 2C.]{\includegraphics[width=0.32\columnwidth]{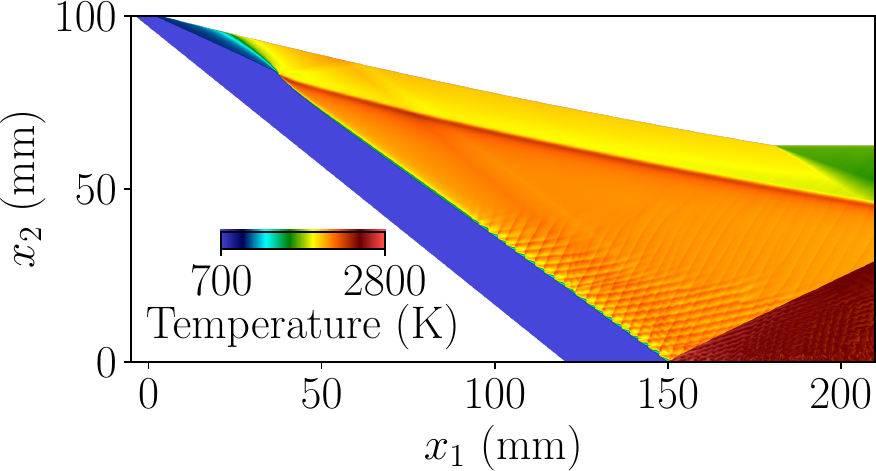}}

\caption{\label{fig:convex-case-2-temperature}Temperature fields for Cases
2A, 2B, and 2C.}
\end{figure}

\begin{figure}[H]
\subfloat[Case 2A.]{\includegraphics[width=0.32\columnwidth]{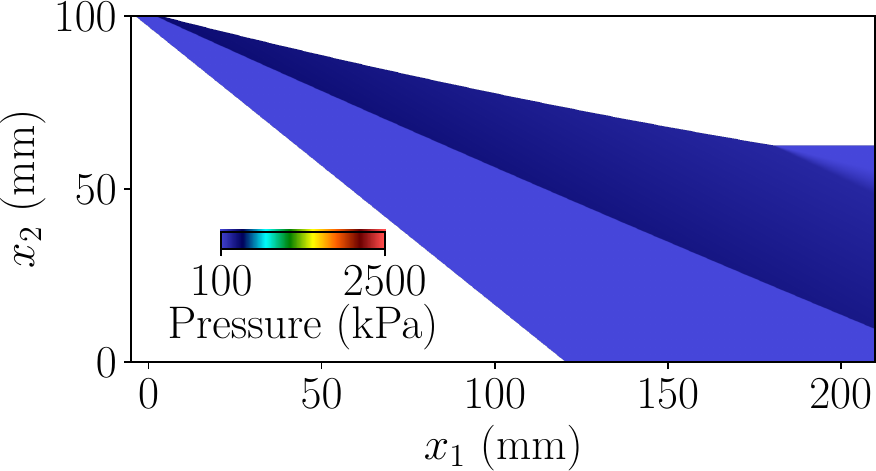}}\hfill{}\subfloat[Case 2B.]{\includegraphics[width=0.32\columnwidth]{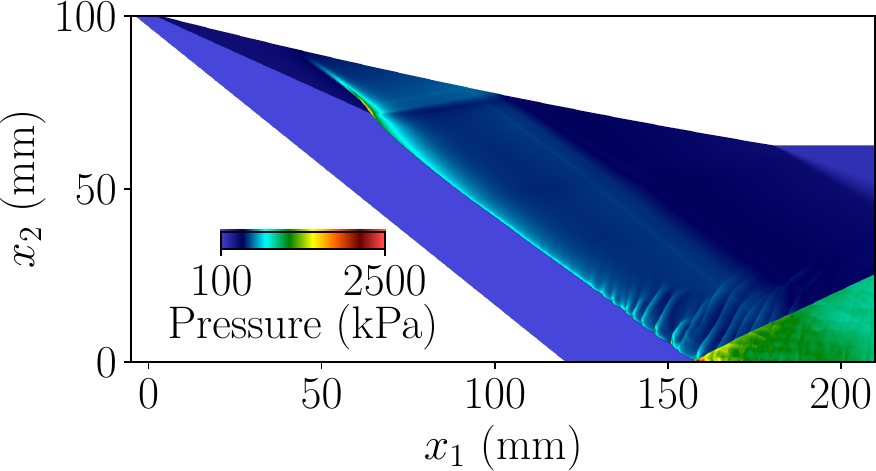}}\hfill{}\subfloat[Case 2C.]{\includegraphics[width=0.32\columnwidth]{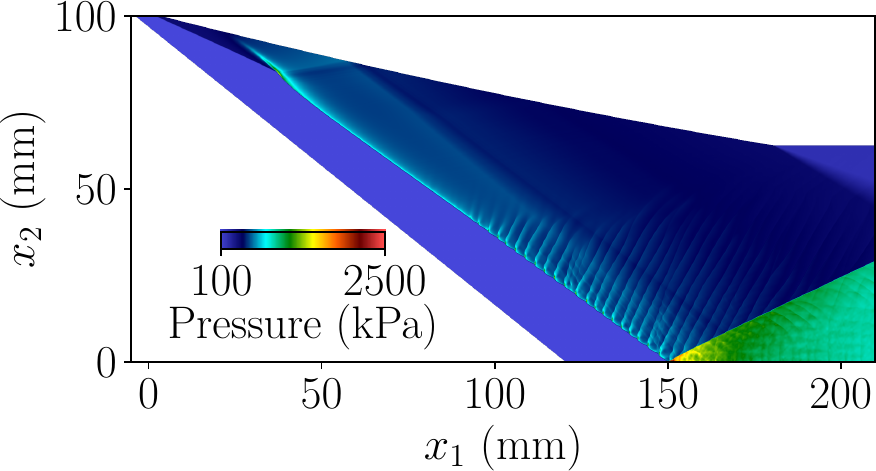}}

\caption{\label{fig:convex-case-2-pressure}Pressure fields for Cases 2A, 2B,
and 2C.}
\end{figure}

This case corresponds to a very low mean curvature of $\kappa=-0.46\;\mathrm{km}^{-1}$,
such that the ramp is nearly linear. Figures~\ref{fig:convex-case-2-temperature}
and~\ref{fig:convex-case-2-pressure} present the temperature and
pressure fields, respectively, at $t=0.628$ ms. In the absence of
ozone, the LSW fails to transition to a detonation. With ozone addition,
shock-detonation transition occurs, and a regular reflection pattern
is observed. Table~\ref{tab:convex-case-2-transition} provides the
transition locations. Cellular structures are apparent in Cases 2B
and 2C. It is worth noting that the difference between 1000 ppm ozone
and 10,000 ppm ozone is greater here than in the concave-ramp cases
(at least in terms of transition-point relocation), again due to additional
effects induced by the ramp concavity. Compared to Case 2B, the transition
point in Case 2C is moved upstream, the transverse waves are weaker,
and the slip line is closer to the convex ramp. 


\begin{figure}[H]
\subfloat[\label{fig:convex-case-2-line-temperature}Temperature along the line
$x_{2}=15$ mm.]{\includegraphics[width=0.32\columnwidth]{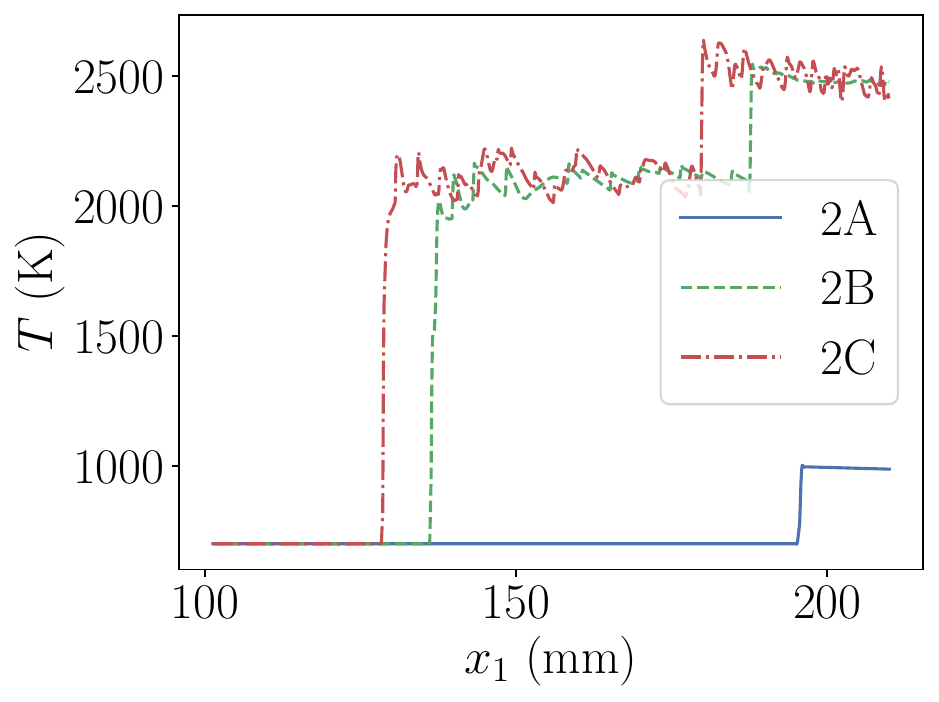}}\hfill{}\subfloat[\label{fig:convex-case-2-line-pressure}Pressure along the line $x_{2}=15$
mm.]{\includegraphics[width=0.32\columnwidth]{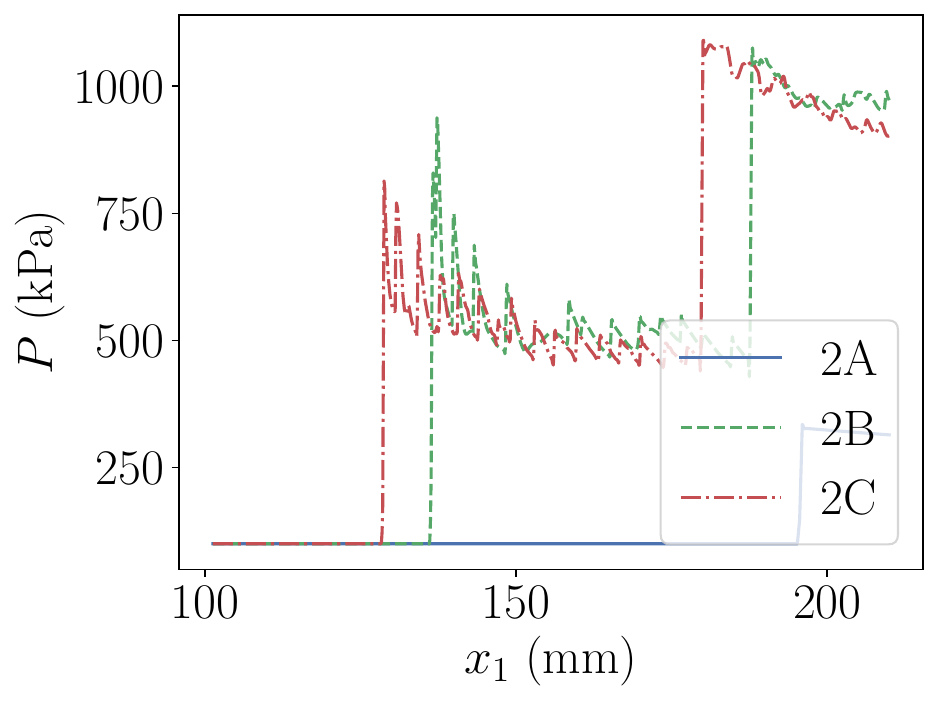}}\hfill{}\subfloat[\label{fig:convex-case-2-line-SPR}SPR along the line $x_{2}=15$
mm.]{\includegraphics[width=0.32\columnwidth]{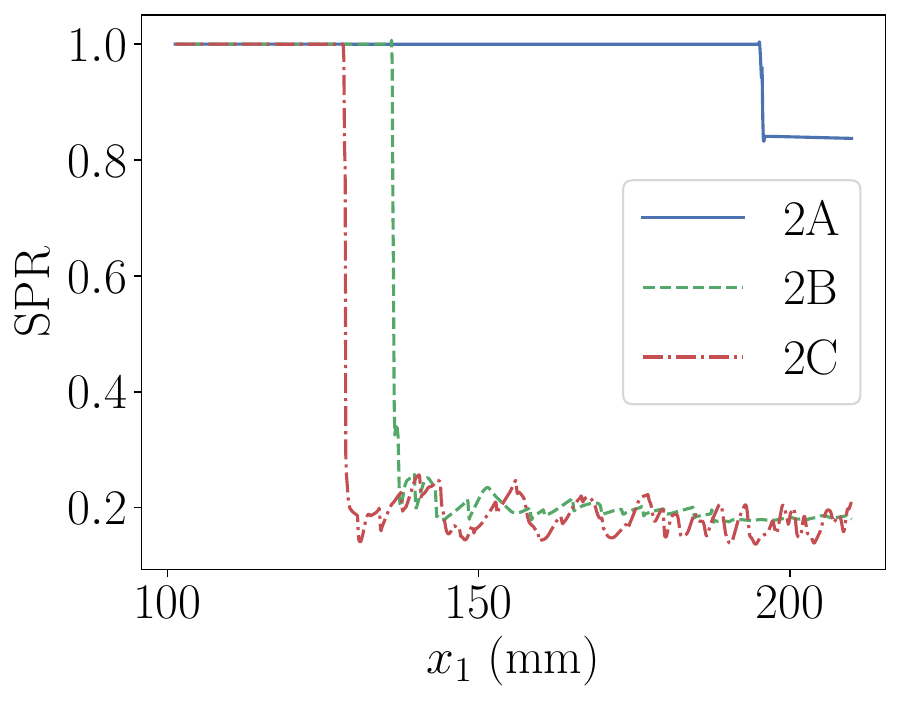}}

\caption{\label{fig:convex-case-1-lines}Variation of temperature, pressure,
and SPR along the line $x_{2}=15$ mm for Case 2 (convex).}
\end{figure}

\begin{table}[h]
\begin{centering}
\caption{$x_{1}$-coordinates of LSW-CDW transitions for Case 1 (convex). \label{tab:convex-case-2-transition}}
\par\end{centering}
\centering{}%
\begin{tabular}{cccc}
\hline 
\noalign{\vskip\doublerulesep}
Case & Ozone addition & LSW-CDW transition & Relative change vs. Case 2B\tabularnewline[\doublerulesep]
\hline 
\noalign{\vskip\doublerulesep}
\hline 
\noalign{\vskip\doublerulesep}
2A & 0 ppm & n/a & n/a\tabularnewline[\doublerulesep]
\noalign{\vskip\doublerulesep}
\hline 
\noalign{\vskip\doublerulesep}
2B & 1000 ppm & $x_{1}\approx64$ mm & 0\%\tabularnewline[\doublerulesep]
\noalign{\vskip\doublerulesep}
\hline 
\noalign{\vskip\doublerulesep}
2C & 10,000 ppm & $x_{1}\approx37$ mm & 42\%\tabularnewline[\doublerulesep]
\hline 
\noalign{\vskip\doublerulesep}
\end{tabular}
\end{table}
Figures~\ref{fig:convex-case-2-line-temperature} and~\ref{fig:convex-case-2-line-pressure}
show the variation of temperature and pressure, respectively, along
the line $x_{2}=15$ mm. In Case 2A, this line sampler intersects
the LSW, while in Cases 2B and 2C, it intersects the CDW and reflected
shock. The temperature and pressure oscillations are again due to
the transverse waves associated with the cellular structure. In Cases
2B and 2C, the first large jump in temperature and pressure is due
to the CDW, and the second is due to the reflected shock. Expansion
due to the ramp convexity and the convex corner is also observed.
 Compared to Case 2B, the CDW and reflected shock in Case 2C are
located further upstream. Apart from the different CDW and reflected-shock
locations, the temperature, pressure, and SPR profiles are overall
similar between Cases 2B and 2C.  SPR is of course highest in Case
2A since the LSW fails to even transition to a detonation. 

\section{Concluding remarks}

CDWs induced by curved ramps represent a promising, relatively new
standing-detonation-engine concept. We performed two-dimensional simulations
of CDWs in a confined combustion chamber and examined the effect of
ozone sensitization on their reflection patterns and flow characteristics.
The main findings are as follows:
\begin{itemize}
\item Just as for ODWs, ozone addition reduces the size of the initiation
zone and can change the LSW-CDW transition type from abrupt to smooth.
The overall effect of ozone addition is typically greater in cases
with large initiation zones in the absence of ozone.  The use of
curved ramps introduces complexities that can lead to additional competing
effects not present in the ODW setting.
\item In the case of concave walls, the reduction of the initiation length
causes the LSW-CDW transition to move upstream, where the LSW is shallower
and weaker. This somewhat offsets the ozone-induced decrease in initiation
length since the region with greater ramp angles (and thus greater
compression) is no longer spanned by the initiation zone. At the same
time, the detonation-attenuation effect associated with smoothening
of the LSW-CDW transition is magnified, again due to the shallower
ramp angle at the transition point. Ozone addition can thus reduce
stagnation-pressure losses across the CDW and improve detonation stabilization.
In particular, it was observed that ozone addition can change a non-stationary
Mach reflection pattern to a stationary Mach reflection pattern and
a stationary Mach reflection pattern to a regular reflection pattern.
However, for high mean curvatures, the large ramp angle near the tail
of the ramp can cause significant continuous compression behind the
LSW-CDW transition that then mitigates the detonation-attenuation
effect. Finally, the higher-Mach-number flow behind the CDW can reduce
or eliminate the Mach stem, thus further reducing stagnation-pressure
losses.
\item In the case of convex walls, the reduction of the initiation length
causes the LSW-CDW transition to similarly move upstream, where the
LSW is instead steeper and stronger. The initiation zone is thus shortened
to a greater extent than in the case of concave walls. The CDW can
possibly be strengthened, which is in direct competition with smoothening
of the LSW-CDW transition. If strengthening of the CDW is the dominant
effect, then propulsion performance and detonation stabilization may
be detrimentally affected. However, it should be noted that the overall
effect on SPR, which is a measure of propulsion performance, is not
observed to be significant (also the case for concave ramps).
\end{itemize}
This study contributes to improved understanding of standing CDWs
and ozone sensitization. Future work may investigate the effect of
ozone addition on CDWs at a wider range of conditions (e.g., additional
flight Mach numbers, different chamber dimensions/configurations,
and more complex ramp shapes, such as ramps that change in convexity~\citep{Xio23,Yan24}).
 Viscous effects will also be considered.

\section*{Acknowledgments}

This work is sponsored by the Office of Naval Research through the
Naval Research Laboratory 6.1 Computational Physics Task Area. 

\bibliographystyle{elsarticle-num}
\bibliography{../../JCP_submission/citations}

\appendix

\section{Grid convergence}

\label{sec:grid-convergence}

\subsection{Concave walls}

\label{subsec:Concave-walls-grid-convergence}

\begin{figure}[H]
\subfloat[\label{fig:concave-convergence-study-temperature-coarse}Mesh 1.]{\includegraphics[width=0.32\columnwidth]{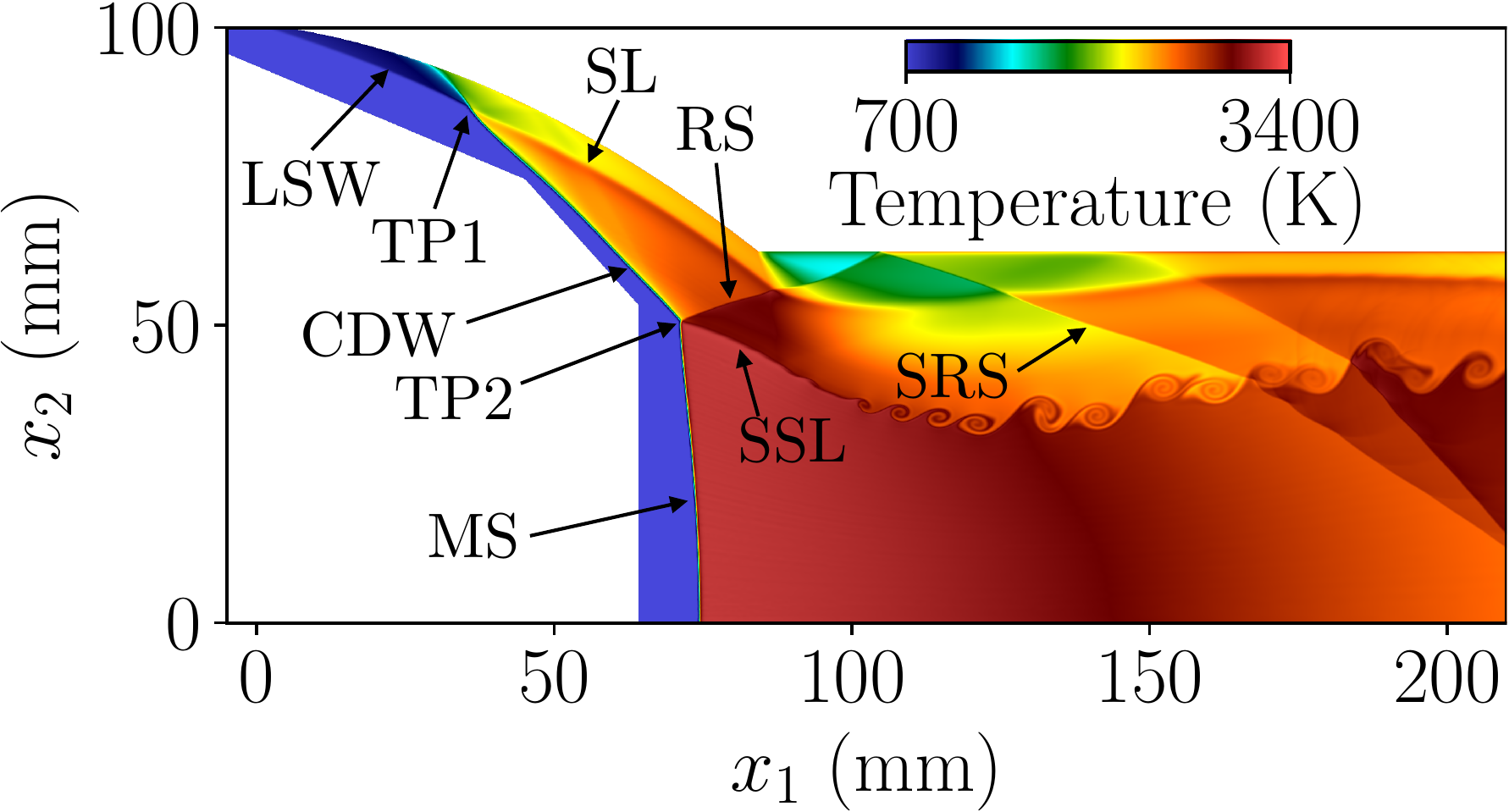}}\hfill{}\subfloat[\label{fig:concave-convergence-study-temperature-fine}Mesh 2.]{\includegraphics[width=0.32\columnwidth]{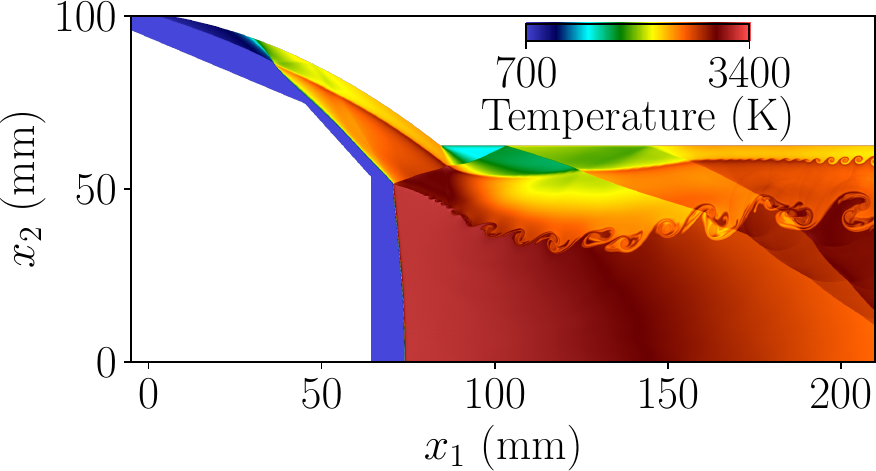}}\hfill{}\subfloat[\label{fig:concave-convergence-study-temperature-finest}Mesh 3.]{\includegraphics[width=0.32\columnwidth]{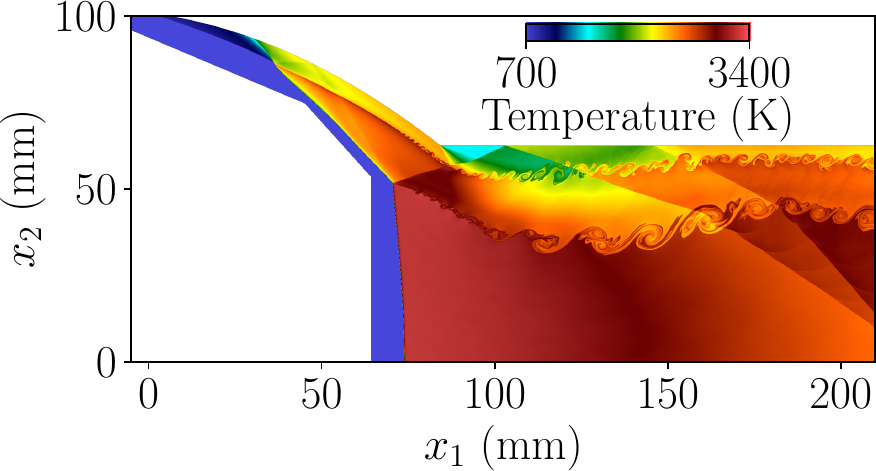}}

\caption{\label{fig:concave-convergence-study-temperature}Temperature fields
for Case 1C with different meshes. Mesh 1: $h=0.2$ mm. Mesh 2: $h=0.1$
mm. Mesh 3: $h=0.05$ mm. LSW: leading shock wave. TP1: transition
point. CDW: curved detonation wave. SL: slip line. RS: reflected shock.
MS: Mach stem. SSL: secondary slip line. TP2: secondary triple point.
SRS: secondary reflected shock.}
\end{figure}

\begin{figure}[H]
\subfloat[Mesh 1.]{\includegraphics[width=0.32\columnwidth]{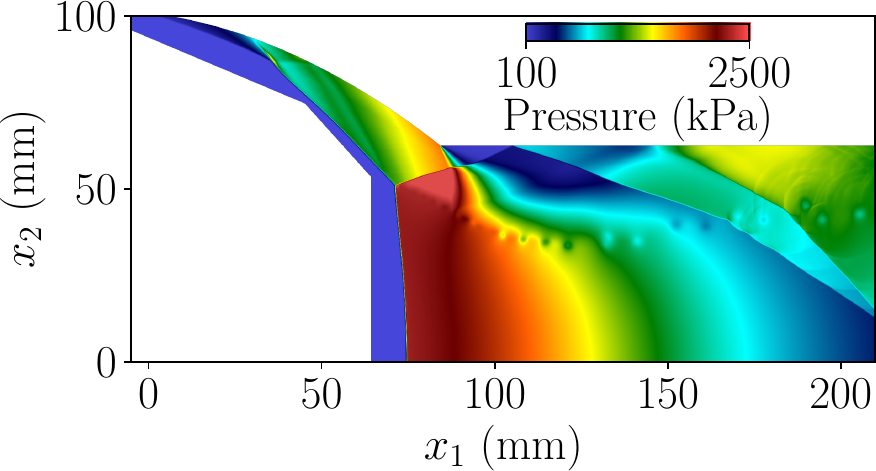}}\hfill{}\subfloat[Mesh 2.]{\includegraphics[width=0.32\columnwidth]{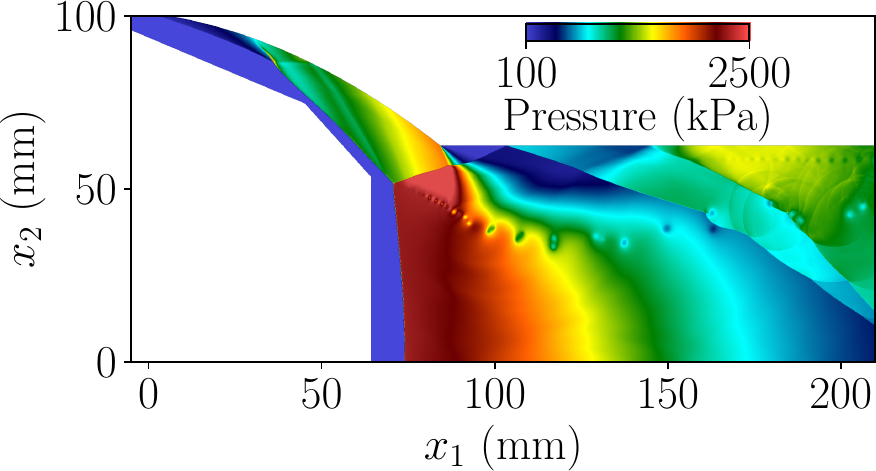}}\hfill{}\subfloat[Mesh 3.]{\includegraphics[width=0.32\columnwidth]{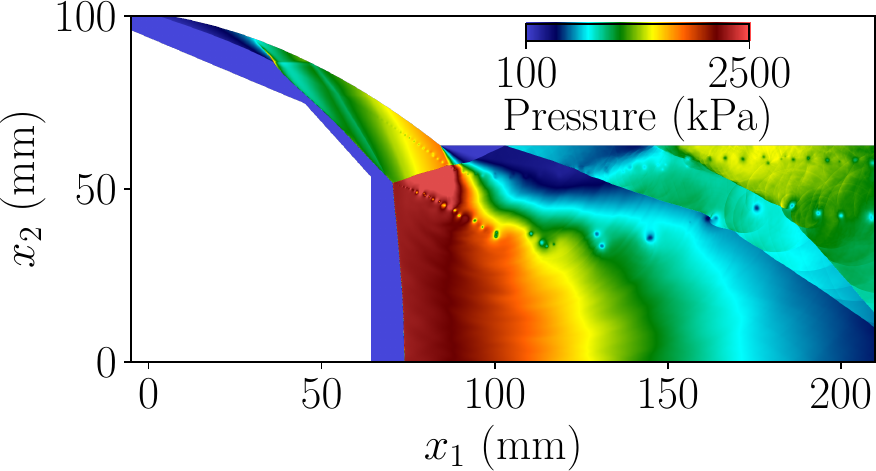}}

\caption{\label{fig:concave-convergence-study-pressure}Pressure fields for
Case 1C with different meshes. Mesh 1: $h=0.2$ mm. Mesh 2: $h=0.1$
mm. Mesh 3: $h=0.05$ mm.}
\end{figure}

Grid convergence is assessed on Case 4C, which corresponds to the
smallest initiation zone across all (stationary) concave-ramp cases.
Three mesh sizes are considered: $h=0.2$ mm (Mesh 1), $h=0.1$ mm
(Mesh 2), and $h=0.05$ mm (Mesh 3). Note that with a $p=1$ solution
approximation, each triangular element has three degrees of freedom
(per state variable). We first compute this case on Mesh 1 up to $t=2.20$
ms. The solution is uniformly refined (which corresponds to Mesh 2)
and then computed up to $t=4.24$ ms. Finally, after another level
of uniform refinement (Mesh 3), the simulation is continued until
$t=7.92$ ms. A quasi-steady state is achieved on each mesh.  Figures~\ref{fig:concave-convergence-study-temperature}
and~\ref{fig:concave-convergence-study-pressure} provide the final
temperature and pressure fields, respectively. The Kelvin-Helmholtz
instabilities along the lower (secondary) slip line exhibit smaller-scale
features with the finer meshes. Along the upper slip line, the Kelvin-Helmholtz
instabilities are most noticeable on Mesh 3; on Mesh 1, they do not
appear at all. Nevertheless, the overall reflection pattern is consistent
across all meshes.


\begin{figure}[H]
\subfloat[\label{fig:concave-convergence-study-line-upper}$x_{2}=68$ mm.]{\includegraphics[width=0.24\columnwidth]{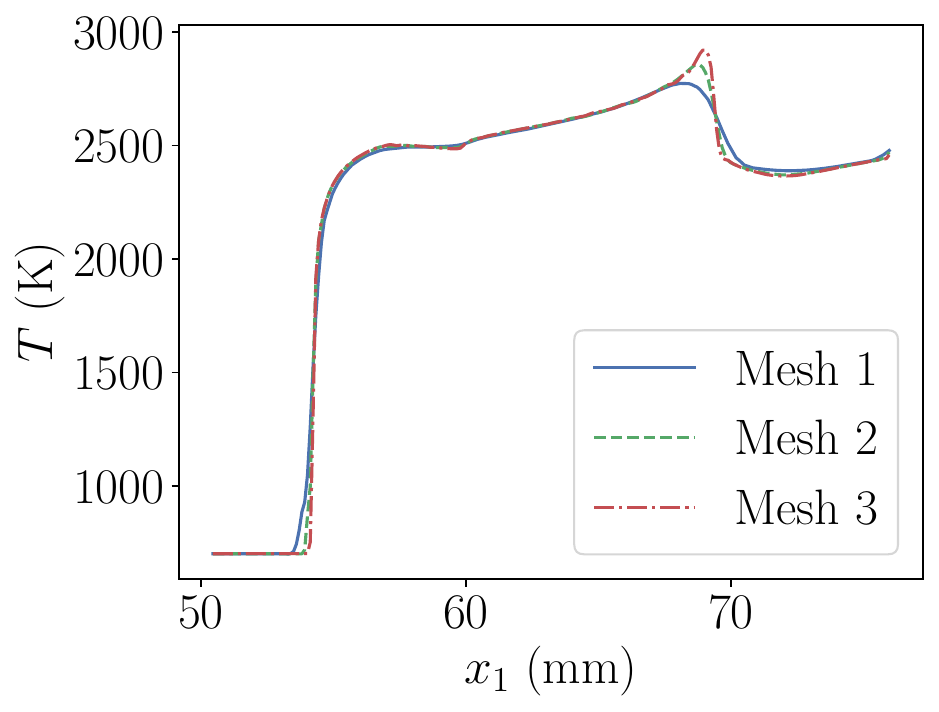}}\hfill{}\subfloat[\label{fig:concave-convergence-study-line-upper-zoom}$x_{2}=68$
mm, zoomed.]{\includegraphics[width=0.24\columnwidth]{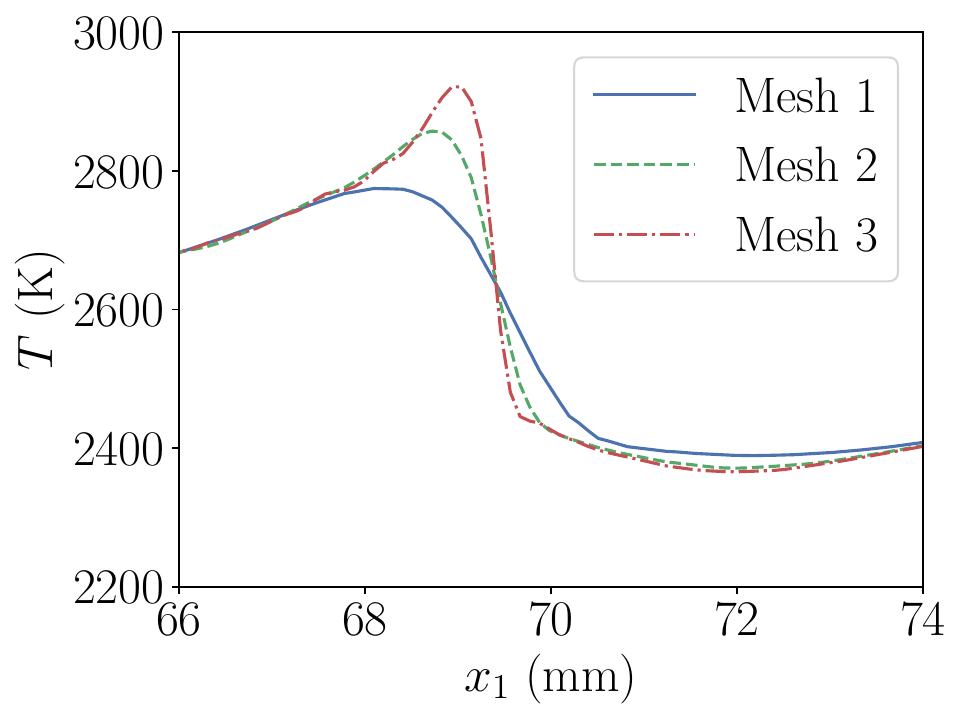}}\hfill{}\subfloat[\label{fig:concave-convergence-study-line-lower}$x_{2}=8$ mm.]{\includegraphics[width=0.24\columnwidth]{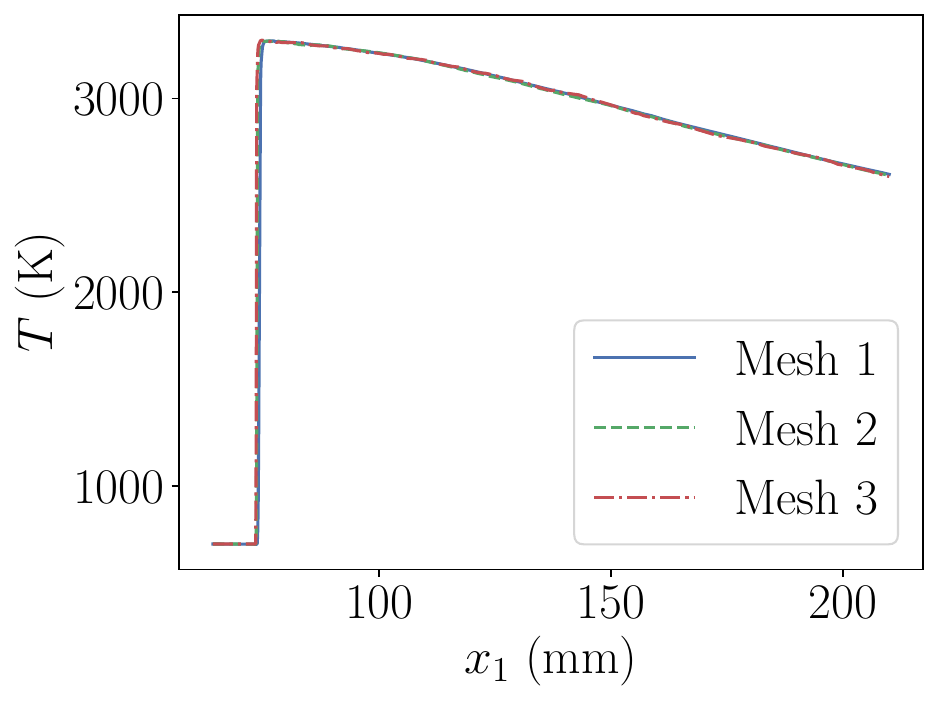}}\hfill{}\subfloat[\label{fig:concave-convergence-study-line-lower-zoom}$x_{2}=8$ mm,
zoomed.]{\includegraphics[width=0.24\columnwidth]{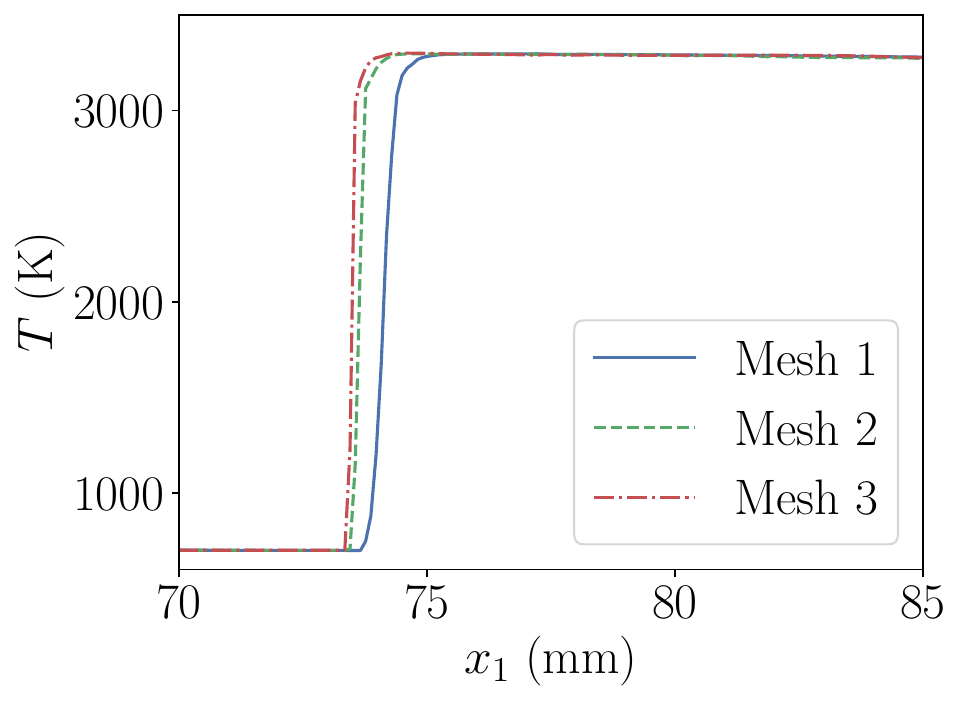}}

\caption{\label{fig:concave-convergence-study-lines}Variation of temperature
along the lines $x_{2}=68$ mm and $x_{2}=8$ mm. Zoomed-in views
are also provided. Mesh 1: $h=0.2$ mm. Mesh 2: $h=0.1$ mm. Mesh
3: $h=0.05$ mm.}
\end{figure}

Figure~\ref{fig:concave-convergence-study-line-upper} presents the
variation of temperature along the line $x_{2}=68$ mm, which intersects
the CDW front and the slip line. Smearing of the CDW front and slip
line is most evident on Mesh 1. Figure~\ref{fig:concave-convergence-study-line-upper-zoom}
zooms in on the vicinity of the slip line, across which a sharp decrease
in temperature occurs. The peak temperature immediately before the
slip line increases as the mesh is refined. Figure~\ref{fig:concave-convergence-study-line-lower}
shows the temperature variation along $x_{2}=8$ mm, which intersects
the Mach stem. As illustrated in the close-up view in Figure~\ref{fig:concave-convergence-study-line-lower-zoom},
the Mach-stem location is moved upstream as the resolution is increased.
On the whole, the temperature profiles for Mesh 2 and Mesh 3 are very
similar. Despite a slight underprediction of the peak temperature
in Figure~\ref{fig:concave-convergence-study-line-upper}, Mesh 2
($h=0.1$ mm) is deemed to be sufficient to capture the overall reflection
patterns and flow characteristics of interest and is therefore used
for the simulations in Section~\ref{subsec:Concave-walls}.

\subsection{Convex walls}

\label{subsec:Convex-walls-grid-convergence}

\begin{figure}[H]
\subfloat[\label{fig:convex-convergence-study-temperature-coarse}Mesh 1.]{\includegraphics[width=0.32\columnwidth]{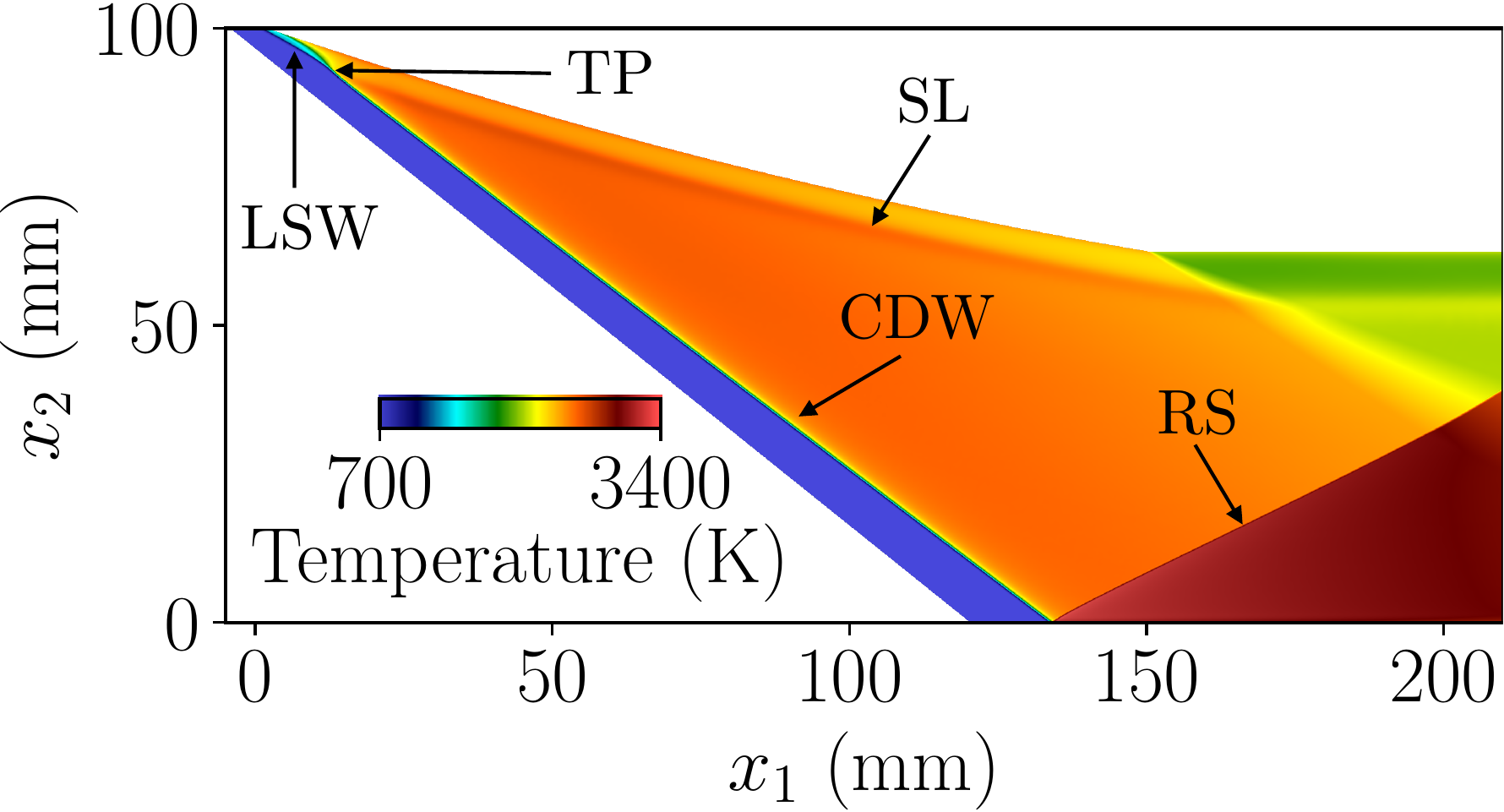}}\hfill{}\subfloat[\label{fig:convex-convergence-study-temperature-fine}Mesh 2.]{\includegraphics[width=0.32\columnwidth]{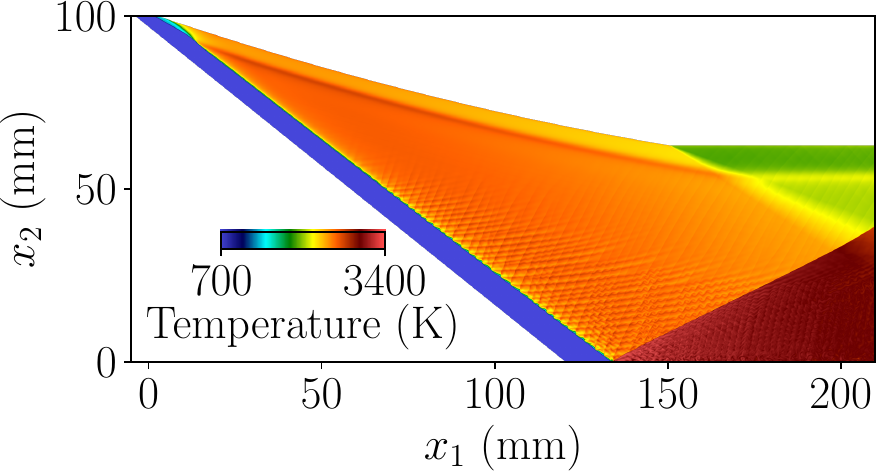}}\hfill{}\subfloat[\label{fig:convex-convergence-study-temperature-finest}Mesh 3.]{\includegraphics[width=0.32\columnwidth]{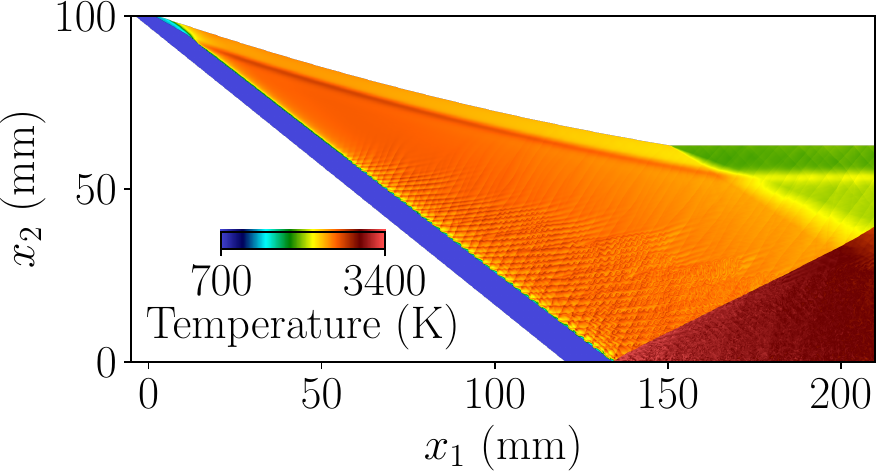}}

\caption{\label{fig:convex-convergence-study-temperature}Temperature fields
for Case 2C with different meshes. Mesh 1: $h=0.2$ mm. Mesh 2: $h=0.1$
mm. Mesh 3: $h=0.05$ mm. LSW: leading shock wave. CDW: curved detonation
wave. SL: slip line. RS: reflected shock. TP: transition point.}
\end{figure}

\begin{figure}[H]
\subfloat[Mesh 1.]{\includegraphics[width=0.32\columnwidth]{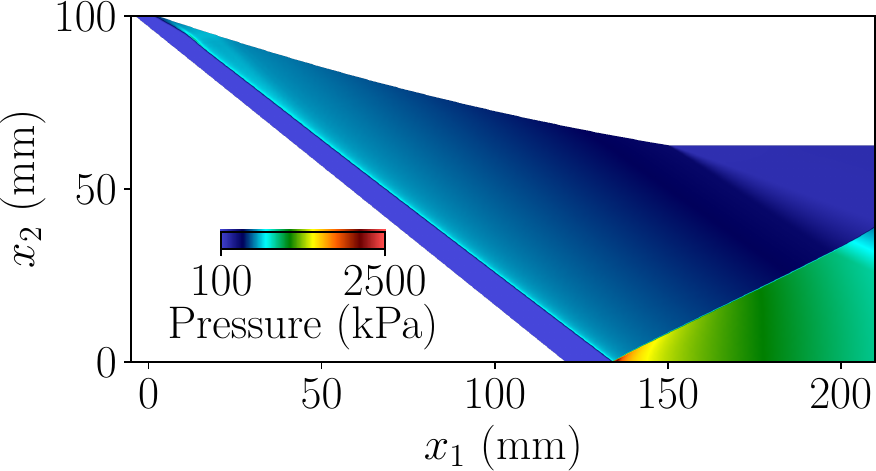}}\hfill{}\subfloat[Mesh 2.]{\includegraphics[width=0.32\columnwidth]{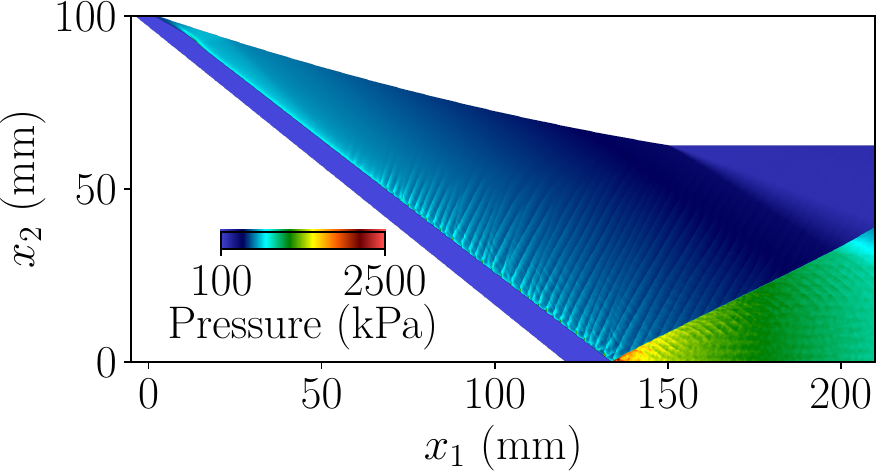}}\hfill{}\subfloat[Mesh 3.]{\includegraphics[width=0.32\columnwidth]{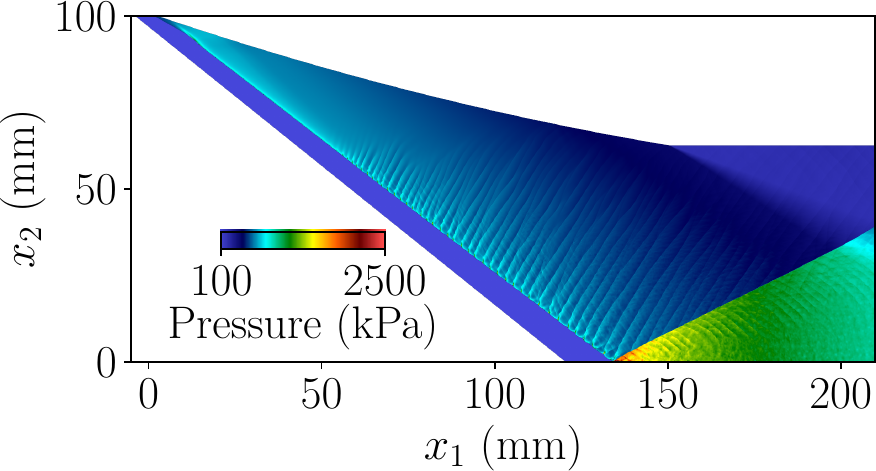}}

\caption{\label{fig:convex-convergence-study-pressure}Pressure fields for
Case 2C with different meshes. Mesh 1: $h=0.2$ mm. Mesh 2: $h=0.1$
mm. Mesh 3: $h=0.05$ mm.}
\end{figure}

Grid convergence is assessed on Case 1C, which corresponds to the
smallest initiation zone across all convex-ramp cases. Three characteristic
mesh sizes are considered: $h=0.2$ mm (Mesh 1), $h=0.1$ mm (Mesh
2), and $h=0.05$ mm (Mesh 3). The final time for the Mesh 1 and Mesh
2 solutions is $t=0.627$ ms. The Mesh 3 solution is obtained by uniformly
refining the Mesh 2 solution and restarting the simulation, with a
final time of $t=0.785$ ms. A quasi-steady state is achieved in all
cases. Figures~\ref{fig:convex-convergence-study-temperature} and~\ref{fig:convex-convergence-study-pressure}
present the final temperature and pressure fields, respectively. In
all cases, the LSW transitions into a CDW that reflects off the bottom
wall. A cellular structure along the CDW front appears with Mesh 2
and especially Mesh 3, but not with Mesh 1. It should be noted that
the grid is likely not converged with respect to the cellular structure
since mesh sizes approaching the mean free path may be required~\citep{Ram24}.
Nevertheless, this study is concerned with the overall reflection
patterns, which are similar across all meshes, so accurate resolution
of the cellular structure is not required; indeed, in related studies,
the mesh resolution often is not sufficiently fine for a cellular
structure to appear~\citep{Ten24,Vas22,Yan24}.


\begin{figure}[H]
\subfloat[\label{fig:convex-convergence-study-line-upper}$x_{2}=80$ mm.]{\includegraphics[width=0.24\columnwidth]{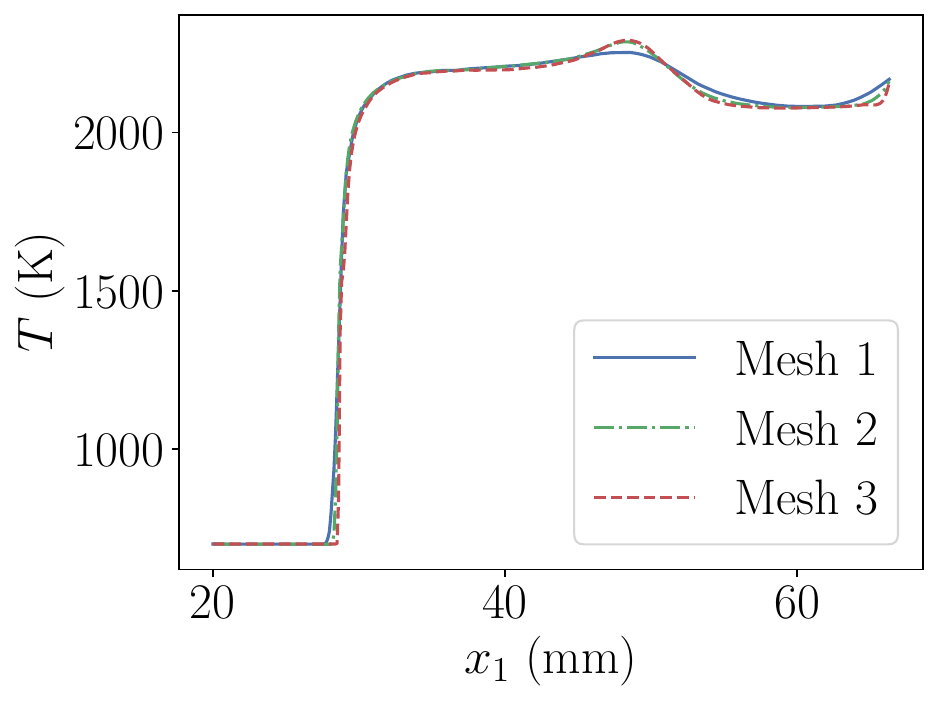}}\hfill{}\subfloat[\label{fig:convex-convergence-study-line-upper-zoom}$x_{2}=80$ mm,
zoomed.]{\includegraphics[width=0.24\columnwidth]{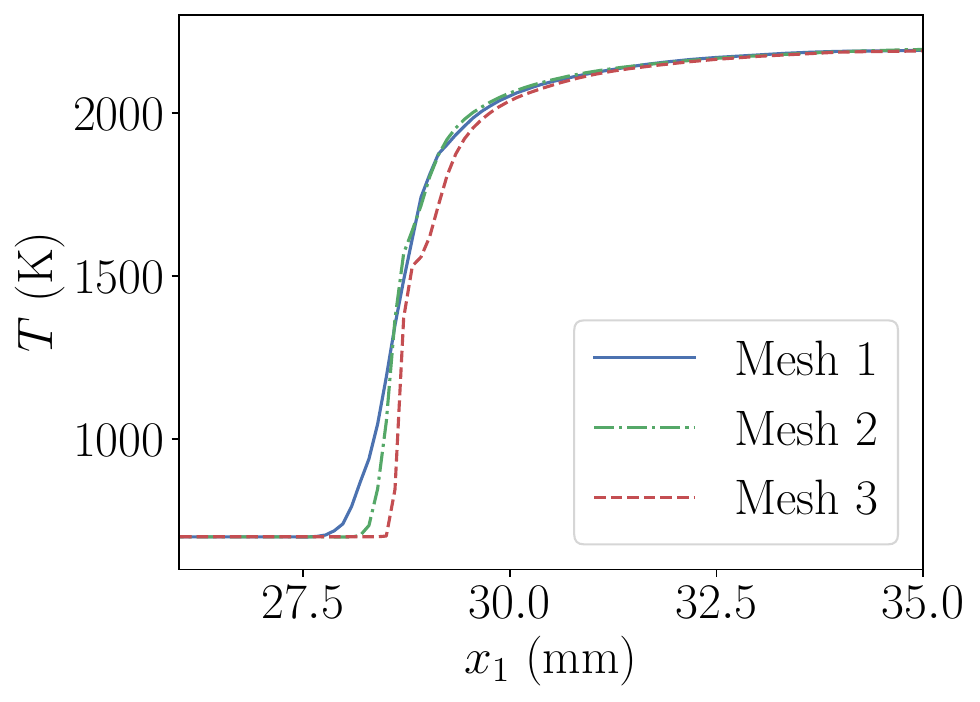}}\hfill{}\subfloat[\label{fig:convex-convergence-study-line-lower}$x_{2}=59$ mm.]{\includegraphics[width=0.24\columnwidth]{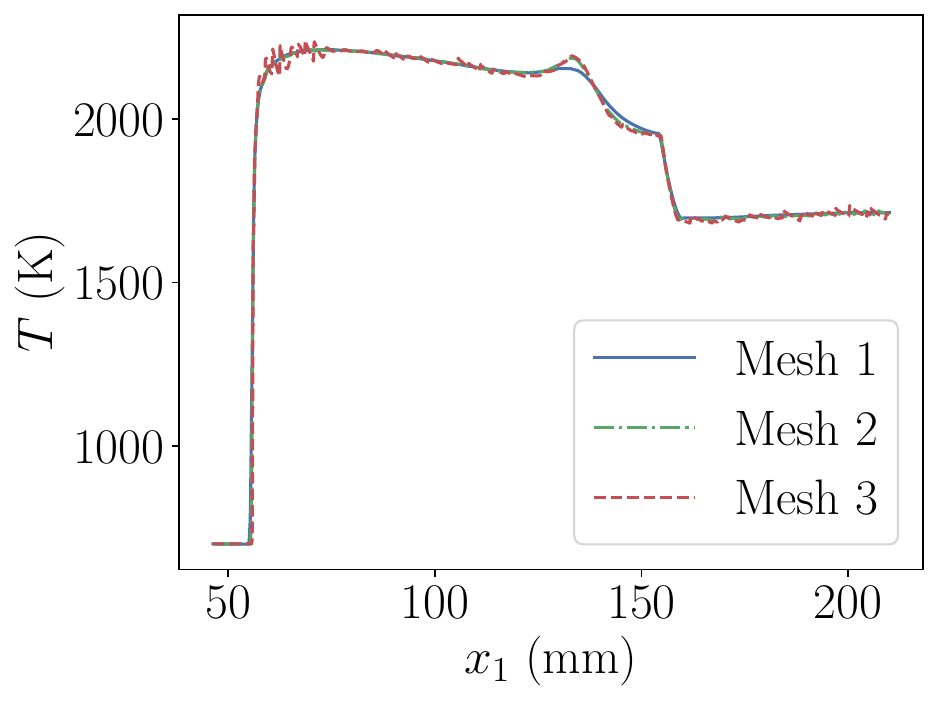}}\hfill{}\subfloat[\label{fig:convex-convergence-study-line-lower-zoom}$x_{2}=59$ mm,
zoomed.]{\includegraphics[width=0.24\columnwidth]{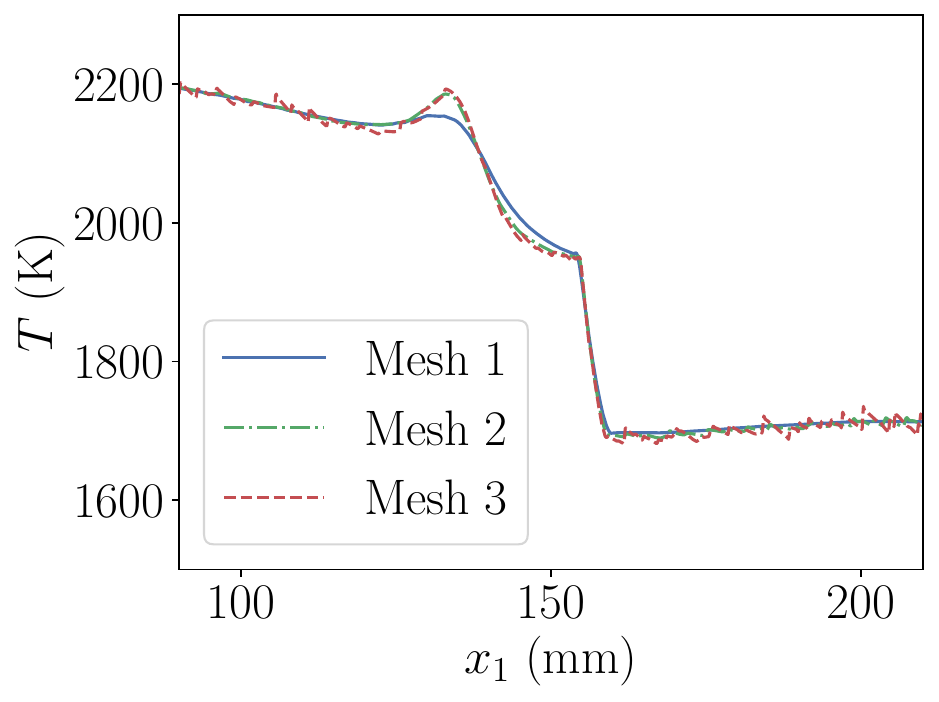}}

\caption{\label{fig:convex-convergence-study-lines}Variation of temperature
along the lines $x_{2}=80$ mm and $x_{2}=59$ mm. Zoomed-in views
are also provided. Mesh 1: $h=0.2$ mm. Mesh 2: $h=0.1$ mm. Mesh
3: $h=0.05$ mm.}
\end{figure}

For a more detailed assessment of grid convergence (with respect to
the global wave structure), Figure~\ref{fig:convex-convergence-study-line-upper}
presents the variation of temperature along the line $x_{2}=80$ mm,
which intersects the CDW and the slip line. The temperature profile
in the vicinity of the slip line is diffused with Mesh 1, and the
peak temperature is somewhat underpredicted. Figure~\ref{fig:convex-convergence-study-line-upper-zoom}
zooms in on the CDW front, where the finer meshes lead to a sharper
profile. In addition, with Mesh 3, an induction zone can be discerned.
Nevertheless, there is good agreement among the temperatures behind
the CDW front and the CDW locations. Figure~\ref{fig:convex-convergence-study-line-lower}
shows the temperature variation along $x_{2}=59$ mm, which additionally
intersects the expansion fan around the convex corner. The temperature
oscillations for Mesh 3 are due to the transverse waves associated
with the cellular structure. As illustrated in the zoomed-in view
presented in Figure~\ref{fig:convex-convergence-study-line-lower-zoom},
the finer meshes yield a sharper temperature profile and slightly
higher peak temperature in the vicinity of the slip line. Given these
results, Mesh 2 ($h=0.1$ mm) is deemed to be sufficient to capture
the overall reflection patterns and flow characteristics of interest
and is therefore used for the simulations in Section~\ref{subsec:Convex-walls}.

\end{document}